\documentclass[12pt,preprintnumbers,nofootinbib,letterpaper]{article}
\pdfoutput=1
\usepackage{amsmath,amssymb,amsfonts,cite,xcolor,epsf,epsfig,epstopdf,graphics,graphicx,mathtools,subcaption,physics,verbatim}
\usepackage{float}

\def\thefootnote{*\arabic{footnote}}
\definecolor{ultramarine}{rgb}{0.07, 0.04, 0.56}
\definecolor{cadmiumgreen}{rgb}{0.0, 0.42, 0.24}
\definecolor{indigodye}{rgb}{0.0, 0.25, 0.42}
\usepackage[linktocpage=true,breaklinks]{hyperref}
\hypersetup{
    colorlinks=true,
    citecolor=ultramarine,
    linkcolor=cadmiumgreen,
    urlcolor=indigodye,
}

\usepackage{orcidlink}

\usepackage[utf8]{inputenc}
\usepackage{newunicodechar}
\newunicodechar{、}{,}

\usepackage{pgfplots}
\pgfplotsset{compat=newest}

\numberwithin{equation}{section}

\usepackage{array}
\newcolumntype{P}[1]{>{\centering\arraybackslash}p{#1}}

\usepackage{dcolumn}
\newcolumntype{M}[1]{>{\centering\arraybackslash}m{#1}}
\newcolumntype{N}{@{}m{0pt}@{}}

\newcommand{\Mpl}{M_{\rm Pl}}

\newcommand{\wh}{\omega_h}
\newcommand{\wt}{\omega_{t,\lambda}}

\newcommand{\ncpl}{{\alpha}}
\newcommand{\pvc}{\theta}

\newcommand{\D}{{\rm d}}

\newcommand{\be}{\begin{equation}}
\newcommand{\ee}{\end{equation}}

\usepackage{tikz}

\begin{document}

\begin{flushright} {\footnotesize
YITP-25-90}  \end{flushright}
\vspace{0.3cm}

\begin{center}

\def\thefootnote{\fnsymbol{footnote}}

{\Large {\bf Oscillations and parity violation in gravitational wave background from extra tensor modes
}}
\\[1cm]

{Jaume Garriga\,\orcidlink{0000-0002-7315-3599}$^{1}$,
Mohammad Ali Gorji\,\orcidlink{0000-0001-8848-2283}$^{2}$,
Fazlollah Hajkarim\,\orcidlink{0000-0002-5935-882X}$^{3}$,
Misao Sasaki\,\orcidlink{0000-0001-5924-0664}$^{4,5,6,7}$}
\\[.7cm]

{\small \textit{$^1$Departament de F\'{i}sica Qu\`{a}ntica i Astrof\'{i}sica, i Institut de Ci\`encies del Cosmos, Universitat de Barcelona, Mart\'{i} i Franqu\`{e}s 1, 08028 Barcelona, Spain
}}\\

{\small \textit{$^2$Cosmology, Gravity, and Astroparticle Physics Group, Center for Theoretical Physics of the Universe, Institute for Basic Science (IBS), Daejeon, 34126, Korea
}}\\

{\small \textit{$^3$ Department of Physics and Astronomy, University of Oklahoma, Norman, OK 73019, USA
}}\\

{\small \textit{$^{4}$Kavli Institute for the Physics and Mathematics of the Universe (WPI), The University of Tokyo, \\ Chiba 277-8583, Japan}}\\

{\small \textit{$^{5}$Asia Pacific Center for Theoretical Physics (APCTP),
Pohang 37673, Korea}}

{\small \textit{$^{6}$
Center for Gravitational Physics and Quantum Information, Yukawa Institute for Theoretical Physics, \\ Kyoto University, Kyoto 606-8502, Japan}}\\

{\small \textit{$^{7}$Leung Center for Cosmology and Particle Astrophysics, National Taiwan University, \\ Taipei 10617, Taiwan}}\\

\end{center}

\vspace{.7cm}

\hrule \vspace{0.3cm}

\begin{abstract}
Spectator fields which provide additional tensor degrees of freedom, on top of the standard metric tensor perturbations, can produce significant amounts of gravitational waves (GWs). Employing the effective field theory approach for spin-2 fields, we identify a characteristic prediction of this class of scenarios: whenever the spin-2 sector undergoes a localized non-adiabatic evolution during inflation, the linear mixing between the metric and extra tensor modes transmits this feature to the GW spectrum as oscillations in scale. The leading oscillation period is determined analytically by the time at which the localized feature occurs and by the propagation speed of the extra tensor mode outside the feature. The same coupling, when sufficiently strong, also drives an in-time oscillation of the superhorizon modes that is reminiscent of neutrino flavor oscillations. Moreover, parity-violating operators in the spin-2 EFT can imprint chiral signatures on the resulting GW background. Such chirality is an important possible signature for features in the non-minimal kinetic coupling, and becomes closely tied to large observable enhancement for features in the sound speed of the extra tensor mode. We consider a concrete scenario in which the spin-2 field generates observable chiral GWs with characteristic oscillatory patterns. These results identify robust signatures that can be probed with future GW detectors, while the exact amplitude, peak position, and parameter values remain model dependent.
\end{abstract}
\vspace{0.5cm}

\hrule
\def\thefootnote{\arabic{footnote}}
\setcounter{footnote}{0}

\thispagestyle{empty}

\newpage

\section{Introduction}\label{sec-introduction}
Inflation has become a key concept in modern theoretical cosmology, offering a widely accepted framework to address several challenges in cosmology such as the origin of large-scale structure of the Universe \cite{Mukhanov:2005sc,Weinberg:2008zzc}. The quantum fluctuations of the inflaton field serve as a source for the observed large-scale structure in the Universe. Inflation predicts an almost scale-invariant, Gaussian, and adiabatic spectrum for the Cosmic Microwave Background (CMB) anisotropies which is strongly supported by cosmological observations \cite{COBE:1992syq,WMAP:2012nax,Planck:2018jri}.

Despite its successes, the inflationary paradigm raises important questions about the fundamental physics of the early Universe. One key question involves the possible presence of additional fields beyond the minimal inflaton, which could have influenced the dynamics of inflation. These fields may include scalar, vector, or higher-spin fields, each introducing new effects and potentially observable signatures. Among them, extra spin-2 fields, on top of the standard massless graviton of General Relativity (GR), are especially compelling. Such spin-2 fields naturally arise in modified gravity theories like massive gravity \cite{deRham:2010kj,Hinterbichler:2011tt,deRham:2014zqa} and bi-gravity \cite{Hassan:2011zd,Hinterbichler:2012cn}; in metric-affine gravity \cite{Aoki:2019snr,Aoki:2020rae,Aoki:2020zqm,Aoki:2023sum}, where both torsion and nonmetricity can generate spin-2 modes; or in the matter sector as composite states induced by a non-trivial background, such as an $SU(2)$ non-Abelian gauge field with an isotropic vacuum expectation value in a cosmological setting \cite{Maleknejad:2011jw,Aoki:2025uwz}.

It is usually taken for granted that the contribution of spectator fields to the GWs is negligible since their contribution to the background energy density is always small. However, spectator fields which provide additional tensor modes can produce a significant amount of GWs which is comparable to the one provided by the curvature perturbations \cite{Gorji:2023ziy,Gorji:2023sil}.\footnote{See also Ref. \cite{Garcia:2025yit} for a recent study to compute GWs produced by a scalar spectator field which is detectable even if its contribution is smaller than the one provided by the curvature perturbations.}

Observationally, the modifications to the tensor sector can be probed through several channels. The CMB B-mode polarization is sensitive to the primordial GW spectrum and can potentially reveal signatures of additional spin-2 fields \cite{Maleknejad:2011jw,Maleknejad:2011sq,Adshead:2012kp,Maleknejad:2012fw,Adshead:2013nka,Namba:2013kia,Maleknejad:2014wsa,Maleknejad:2016qjz,Obata:2016tmo,Agrawal:2017awz,Agrawal:2018mrg,Dimastrogiovanni:2018uqy,Bordin:2018pca,Firouzjahi:2018wlp,Domcke:2018rvv,Watanabe:2020ctz,Gorji:2020vnh,Ishiwata:2021yne,LiteBIRD:2023zmo,Badger:2024ekb}. Recently, using the intrinsic alignment of galaxies to detect a GW background on intergalactic scales was proposed \cite{Okumura:2024xnd,Mikura:2025vaj}. Additionally, direct detections of high frequency GWs by ground-based interferometers like LIGO and Virgo, as well as future space-based observatories like LISA, Taiji and the Einstein Telescope, offer complementary probes of the tensor sector \cite{Bielefeld:2014nza,Bielefeld:2015daa,Caldwell:2016sut,Caldwell:2017chz,Thorne:2017jft,Iacconi:2019vgc,BeltranJimenez:2019xxx,LISACosmologyWorkingGroup:2019mwx,Iacconi:2020yxn,Fujita:2022jkc,Gorji:2023ziy,Gorji:2023sil}. There are many potential sources of high-frequency GWs, which can be generally divided into two main categories, early-universe and late-time sources. Late-time sources are typically astrophysical, such as mergers of supermassive black holes or neutron stars \cite{Sathyaprakash:2009xs,Regimbau:2011rp,Christensen:2018iqi}. In contrast, early-universe sources arise from high-energy processes in the very early times, such as inflation, primordial phase transitions, or other mechanisms occurring before big bang nucleosynthesis \cite{Maggiore:1999vm,Caprini:2018mtu}. Distinguishing between these two classes represents a central challenge in GW cosmology. Among the most powerful diagnostics are parity violation \cite{Bielefeld:2014nza,Bielefeld:2015daa,Caldwell:2016sut,Caldwell:2017chz,Thorne:2017jft,BeltranJimenez:2019xxx,LISACosmologyWorkingGroup:2019mwx,Fujita:2022jkc,Machado:2018nqk,Salehian:2020dsf,Namba:2020kij,Cai:2021uup,Cai:2022lec,Unal:2023srk,Bari:2023rcw} and anisotropy \cite{Namikawa:2015prh,Geller:2018mwu,Jenkins:2018kxc,Cusin:2019jhg,Cusin:2019jpv,Bartolo:2019oiq,Pitrou:2019rjz,Bartolo:2019yeu,ValbusaDallArmi:2020ifo,Liu:2020mru,Capurri:2021zli,Gardiner:2023zzr,Sah:2024oyg}. Since it is difficult to generate significant parity-violating signatures after big bang nucleosynthesis, the detection of chiral GWs would provide strong evidence for an early-universe origin. On the other hand, anisotropy is more often seen with astrophysical sources, though not always.

Even when we restrict our analysis to the case of spin-2 fields, a wide range of models must still be considered. Consequently, each early-universe model that predicts parity-violating GWs will lead to distinctive observational signatures. Therefore, the diversity of such models makes it difficult to draw general conclusions based on specific examples. To address this, we adopt the effective field theory (EFT) framework \cite{Arkani-Hamed:2003pdi,Cheung:2007st,Bordin:2018pca}, which allows us to identify universal, model-independent features associated with chiral GW production. In particular, we focus on scenarios where parity violation originates from interactions with additional spin-2 fields. The EFT approach enables a systematic study of the generic predictions of these setups and helps us isolate robust observational imprints that can be used to distinguish early-universe sources from astrophysical backgrounds.

Indeed, we find that, whenever the spin-2 sector undergoes a localized non-adiabatic evolution during inflation -- which in our concrete realizations corresponds to a Gaussian dip in either the non-minimal coupling $f$ or the propagation speed $c_t$ -- the linear mixing $\ncpl$ between the metric tensor perturbations $\gamma_{ij}$ and the extra tensor perturbations $t_{ij}$ transmits a sharp oscillatory pattern in $k$ to the GW spectrum. Both ingredients are independently necessary: without the feature the system is scale invariant and produces no $k$-fringes, while without the mixing the feature is confined to the $t$-sector and is invisible in $\gamma$. The mechanism that produces the fringes is model-independent. The leading fringe period is also model-independent, being determined only by the temporal location of the feature and by the propagation speed of $t^\lambda_{\bf k}$ outside it. The envelope shape in $k$ reflects the shape of the localized feature in time and is therefore common to our two phenomenological models, both of which use a Gaussian profile, while it would change for a different functional form. The peak amplitude of the fringes is the model-dependent piece and tracks the level of violation of the adiabaticity condition in the spin-2 sector. The analytical identification of the mechanism and the corresponding formulas are given in Sec.~\ref{sec-osc-mechanism}. Moreover, when the mixing is large, $|\ncpl|\geq 3/2$, the superhorizon dynamics features an additional, qualitatively different, in-time oscillation that is reminiscent of neutrino flavor oscillations.

The remainder of the paper is organized as follows. In Sec.~\ref{sec-EFT}, we formulate the EFT of a light spin-2 field during inflation and derive the quadratic action that characterizes the interaction between the metric tensor perturbations and the additional tensor perturbations provided by the spin-2 field. In Sec.~\ref{sec-inflation}, we clarify the roles of different EFT couplings and study the time evolution of the system during inflation. In Sec.~\ref{sec-power-spectra}, we present the quantization of the coupled system, compute the tensor power spectra, and identify the source of the in-$k$ oscillations analytically in Sec.~\ref{sec-osc-mechanism}. In Sec.~\ref{sec-SGWs}, we study the production of stochastic GWs within this framework. Sec.~\ref{sec-summary} is devoted to a summary and conclusions. Some technical parts are presented in the appendices.


\section{Parity-violating EFT of spin-2 field during inflation}\label{sec-EFT}

In the context of single-field inflation, the EFT framework provides a robust method for classifying all possible operators consistent with the symmetries of the inflationary background \cite{Arkani-Hamed:2003pdi,Cheung:2007st}. Extending this framework to include additional spin-2 fields allows for a systematic study of the possible interactions between the extra tensor modes of the spin-2 field and the standard tensor modes of the metric perturbations \cite{Bordin:2018pca}. In this section, we further generalize this approach by incorporating parity-violating operators. The readers who are only interested in the application of the setup to the GWs may wish to move directly to Sec.~\ref{sec-inflation}.

In single-field inflation, the inflaton field $\phi$ acquires a time-dependent vacuum expectation value, $\langle \phi(t) \rangle$, which spontaneously breaks time diffeomorphisms. The residual symmetry is the time-dependent spatial diffeomorphisms. We thus need to find all operators that are invariant under this residual symmetry. In order to do so, we work in the comoving slicing $t=t(\phi)$. Then, $\partial_\mu\phi=\delta_\mu^0$ and $g^{\alpha\beta}\partial_\alpha\phi\partial_\beta\phi=g^{00}$ are building blocks. Since $g^{00}$ does not contribute to the tensor perturbations of interest in this work, we neglect it in our analysis. The future-directed timelike unit vector normal to the constant-time hypersurfaces is then given by
\begin{align}\label{n-def}
n_{\mu} = - \frac{\delta_\mu^0}{\sqrt{-g^{00}}} \, ,
\qquad
n_\alpha n^\alpha = -1\,.
\end{align}
The corresponding induced metric on the spatial hypersurfaces is given by
\begin{align}\label{h}
h_{\mu\nu} = g_{\mu\nu} + n_\mu n_\nu \,.
\end{align}
It satisfies $h^\mu{}_{\alpha} n^\alpha = 0$. We can define time and spatial derivatives as usual,
\begin{align}\label{derivatives-def}
\pounds_{n} \, ,
\qquad
D_\mu \equiv h_{\mu}{}^\alpha\nabla_\alpha \, ,
\end{align}
where $\pounds_{n}$ is the Lie derivative along the direction of $n^{\mu}$.

The covariant derivative of $n_\mu$, $\nabla_\mu n_{\nu}$, is also a building block. The projection on the spatial hypersurface is characterized by the extrinsic curvature
\begin{align}\label{K-def}
K_{\mu\nu} = h_{(\mu|}{}^\alpha \nabla_\alpha n_{|\nu)} \,.
\end{align}
We do not need to include the projection of $\nabla_\mu n_{\nu}$ along $n^\mu$, which determines the acceleration, since $a_\mu = n^\alpha \nabla_\alpha n_\mu = - D_\mu \ln(-g^{00})/2$ and, as noted above, $g^{00}$ does not contribute to the tensor perturbations of interest in this work. Moreover, the three-dimensional spatial Riemann tensor $^{(3)}R_{\mu\nu\sigma\rho}$ is also a building block. One can explicitly show that these quantities are invariant under the residual symmetry of the EFT, namely time-dependent spatial diffeomorphisms.

For the gravitational sector, we take the free part to be the Einstein-Hilbert action ($\hbar=1=c$ and $\Mpl=1/\sqrt{8\pi{G}}$),
\begin{align}\label{EH-action}
S_{\rm E.H.} = \frac{\Mpl^2}{2} \int \D^4x \sqrt{-g} R \, ,
\end{align}
where $R$ is the four-dimensional Ricci scalar and $\Mpl$ denotes the reduced Planck mass. We have considered a minimal setup by considering the four-dimensional Ricci scalar. For example, we could include a term like $K_{\alpha\beta}K^{\alpha\beta}$ in the action which would modify the GW speed \cite{Cheung:2007st}. Moreover, we could consider non-minimal couplings involving $\Sigma^{\mu\nu}$, such as $\Sigma^{\alpha\beta} \Sigma_{\alpha\beta} R$, but we omit these to maintain simplicity.

We are interested in an effective description of additional traceless-transverse spatial tensor perturbations, on top of the metric tensor perturbations. It is thus reasonable to look for a symmetric, spatial, traceless rank-2 tensor
\begin{align}\label{Sigma-def}
\Sigma^{\mu\nu} = \Sigma^{\nu\mu}\,;
\qquad
n_\alpha\Sigma^{\alpha\mu} = 0 \, ,
\qquad
h_{\alpha\beta}\Sigma^{\alpha\beta} = 0 \, .
\end{align}
Conditions \eqref{Sigma-def} leave us with five degrees of freedom, as expected for a massive spin-2 field.

The spatial metric $h^{\mu\nu}$ and three-dimensional spatial Levi-Civita tensor $\epsilon^{\mu\nu\rho}\equiv{n}_\alpha\epsilon^{\alpha\mu\nu\rho}$, where $\epsilon^{\alpha\mu\nu\rho}$ is the usual four-dimensional Levi-Civita tensor, are invariant quantities on the spatial hypersurfaces. Thus, we can use them to write down the free action for $\Sigma^{\mu\nu}$. Restricting ourselves to terms that are quadratic in $\Sigma^{\mu\nu}$ and considering at most one time and one spatial derivative, we find
\begin{align}
S_{\Sigma} = \frac{1}{2} \int \D^4x \sqrt{-g}
&\Big[
C_K\, \dot{\Sigma}^{\alpha\beta} \dot{\Sigma}_{\alpha\beta}
- C_G \, D_\mu \Sigma^{\alpha\beta} D^\mu \Sigma_{\alpha\beta}
- C_T D_\mu \Sigma^{\mu\alpha} D^\nu \Sigma_{\nu\alpha}
- C_M \, \Sigma^{\alpha\beta} \Sigma_{\alpha\beta}
\nonumber
\\
& - 2 C_P \, \epsilon^{\mu\nu\alpha} h^{\beta\rho}\Sigma_{\mu\beta} D_\nu \Sigma_{\alpha\rho}
- 2 C_D\, \epsilon^{\mu\nu\alpha} h^{\beta\rho} \dot{\Sigma}_{\mu\beta} D_\nu \Sigma_{\alpha\rho} \Big] \, ,
\label{action-free-Sigma}
\end{align}
where a dot denotes the covariant derivative along the normal vector $n^\alpha \nabla_\alpha$, and $C_{K,G,T,M,P,D}$ are unspecified functions of time $t$. Note that all functions are dimensionless except $C_P$, which has dimension of mass. Applying this EFT to a specific system, one can parameterize $C_P$ in terms of the scale provided by the system under consideration. For example, a natural choice in FLRW background can be the Hubble parameter. Note also that $\pounds_{n}\Sigma^{\mu\nu}=n^\alpha\nabla_\alpha\Sigma^{\mu\nu}-\Sigma^{\alpha\nu}\nabla_\alpha{n}^\mu-\Sigma^{\mu\alpha}\nabla_\alpha{n}^\nu=n^\alpha\nabla_\alpha\Sigma^{\mu\nu}-\Sigma^{\alpha\nu}K_\alpha{}^\mu-\Sigma^{\mu\alpha}K_\alpha{}^\nu$ and since $K_{\mu\nu}$ is a building block of the EFT, we can work with either $\pounds_n$ or $n^\alpha\nabla_\alpha$ without loss of generality.

The remaining part is the interaction action between gravity and the spin-2 field $\Sigma^{\mu\nu}$. We are interested in metric tensor perturbations which are responsible for GWs. Thus, it is more convenient to work with the traceless part of the extrinsic curvature, the shear tensor,
\begin{align}\label{shear-def}
\sigma_{\mu\nu} = K_{\mu\nu} - \frac{1}{3} K h_{\mu\nu} \,;
\qquad
K=h^{\alpha\beta}K_{\alpha\beta} \, ,
\end{align}
where $K$ is the trace part of the extrinsic curvature. Looking at the interaction up to the quadratic order in $\sigma_{\mu\nu}$ and $\Sigma^{\mu\nu}$ which include at most one time and one spatial derivative, we end up with the following interaction action,
\begin{align}\label{action-int}
S_{\rm int} = \Mpl \int \D^4x \sqrt{-g} &\Big[
C_\rho \,\sigma_{\alpha\beta} \Sigma^{\alpha\beta}
+ C_\kappa \, \sigma_{\alpha\beta} \dot{\Sigma}^{\alpha\beta}
+ C_\eta\, \epsilon^{\mu\nu\alpha} {h}^{\beta\rho}\sigma_{\mu\beta} D_\nu \Sigma_{\alpha\rho}
\Big] \, ,
\end{align}
where coefficients $C_{\rho,\kappa,\eta}$ are unspecified functions of time $t$. Note that $C_{\kappa,\eta}$ are dimensionless while $C_\rho$ has dimension of mass. Additionally, the interactions labeled by $C_{\rho,\kappa}$ preserve parity, whereas the interaction labeled by $C_\eta$ does not. Since the interaction $\sigma_{\alpha\beta} \Sigma^{\alpha\beta}$ is first-order in derivatives, while the others are second-order, one might argue that the interactions labeled by $C_{\kappa,\eta}$ are suppressed relative to $\sigma_{\alpha\beta} \Sigma^{\alpha\beta}$ within the framework of derivative expansion in the EFT. However, as the time and spatial derivatives behave differently in FLRW background, for the modes deep inside the horizon, the extra derivatives in $\sigma_{\alpha\beta} \dot{\Sigma}^{\alpha\beta}$ and $\epsilon^{\mu\nu\alpha} {h}^{\beta\rho}\sigma_{\mu\beta} D_\nu \Sigma_{\alpha\rho}$ make them more relevant than $\sigma_{\alpha\beta} {\Sigma}^{\alpha\beta}$. We thus keep both of these terms.

\subsection{Tensor perturbations in cosmological background}\label{sec-EFT-cosmology}

Let us now implement our EFT setup to cosmology. The spatially flat FLRW metric including tensor perturbations is given by
\begin{align}\label{BG}
&{g}_{\mu\nu} {\D}x^\mu {\D}x^\nu
= - N(t)^2 {\D}t^2 + a(t)^2 \left( \delta_{ij} + h_{ij} \right) {\D}x^i {\D}x^j \, ,
\end{align}
where $N$ and $a$ are the lapse function and scale factor, respectively, while $h_{ij}$ characterizes the traceless $\delta^{ij}h_{ij}=0$ and transverse $\partial^mh_{mi}=0$ metric tensor perturbations. The shear tensor \eqref{shear-def} does not have a background value and starts at the level of perturbations, which makes it more relevant than the extrinsic curvature for our purpose. The spin-2 field $\Sigma^{\mu\nu}$ does not have a background value and the extra tensor modes encoded in it will show up only at the level of perturbations.

In the FLRW background \eqref{BG}, the normal vector and covariant derivative along it, defined in \eqref{n-def}, reduce to
\begin{align}
n_{\mu}=-N\delta^{0}_\mu \, ,
\qquad
n^\alpha\nabla_\alpha=\frac{1}{N}\frac{{\rm D}}{\D t} \, .
\end{align}
Normalizing metric tensor perturbations as $\gamma_{ij}\equiv\Mpl{h}_{ij}/2$, the Einstein-Hilbert action \eqref{EH-action} gives the usual quadratic action,
\begin{align} \label{action-free-gamma}
S_{\rm E.H.} &= \frac{1}{2} \int \D^3x\, \D{t}\, N a^3 \left[
\left( \dot{\gamma}_{ij} \right)^2
- \frac{1}{a^2}\left( \partial_i{\gamma}_{jk} \right)^2
\right] \, ,
\end{align}
where, here and in what follows, the dot denotes the partial derivative with respect to time, $\dot{~}=\partial/\partial t$.
The spin-2 field has five degrees of freedom. Performing the usual scalar-vector-tensor decomposition in cosmological background, we find that it includes one scalar, two vector, and two tensor modes/helicities \cite{Gorji:2023cmz}. Focusing only on the tensor modes, with which we are interested in this paper, at the linear level in perturbations we have
\begin{align}\label{Sigma-O(1)}
\Sigma^{\mu\nu} = a^{-2} \left(\begin{matrix}
0 & 0
\\
0 & t^{ij}
\end{matrix} \right) \, ,
\qquad
\Sigma_{\mu\nu} = a^{2} \left(\begin{matrix}
0 & 0
\\
0 & t_{ij}
\end{matrix} \right) \, ,
\end{align}
where $t_{ij}=\delta_{im}\delta_{jn}t^{mn}$ and $\partial_it^{ij}=0=\delta_{ij}t^{ij}$. The factor $a^{-2}$ in the first expression is chosen such that $t^{ij}$ is the canonical field. Substituting \eqref{BG} and \eqref{Sigma-O(1)} in action \eqref{action-free-Sigma}, we find the quadratic action for the tensor helicities of the spin-2 field,
\begin{align}\label{action-2nd-Sigma}
S_{\Sigma} &=
\frac{1}{2} \int \D^3x\, \D{t}\, N a^3 \left[ C_K
\left( \dot{t}_{ij} \right)^2
- \frac{1}{a^2} C_G \left( \partial_i{t}_{jk} \right)^2 - {\tilde C}_M t_{ij} t^{ij} + \frac{2}{a}  {\tilde C}_P\, \varepsilon_{ijk} t^{kn} \partial^i t_{n}{}^j \right]
\, ,
\end{align}
where $\varepsilon_{ijk}$ is the usual three-dimensional Euclidean Levi-Civita symbol with $\varepsilon_{123}=1$. In obtaining the above action, we have performed an integration by parts to put the time derivative in the interaction labeled by $C_D$ on the coefficient and we have defined
\begin{align}
{\tilde C}_M \equiv C_M - 2 C_G H^2 \, ,
\qquad
{\tilde C}_P \equiv C_P - 2 H C_D - \dot{C}_D \, ,
\end{align}
where $H=\dot{a}/a$ is the Hubble parameter. Note that the $C_T$ term does not contribute to the helicity-2 modes due to the transverse condition. Choosing the parametrization for the free coefficients as
\begin{align}\label{parametrization-CI}
C_K = f^2 \, ,
\qquad
C_G = f^2 c_t^2 \, ,
\qquad
{\tilde C}_M = f^2 m^2 \, ,
\qquad
{\tilde C}_P = f^2 H \pvc \, ,
\end{align}
where $c_t$, $m$, and $\pvc$ are functions of time, we find
\begin{align}
S_{\Sigma} &=
\frac{1}{2} \int \D^3x\, \D{t}\, N a^3 f^2 \left[
\left( \dot{t}_{ij} \right)^2
- \frac{c_t^2}{a^2} \left( \partial_i{t}_{jk} \right)^2
-m^2 {t}_{ij}^2
+ \frac{2}{a} \pvc {H} \varepsilon^{ijk} t_{im} \partial_j t_{k}{}^m
\right]
\, .
\end{align}
Note that the Hubble parameter is the only dynamical scale in the FLRW background and, in this regard, parametrization \eqref{parametrization-CI} is a reasonable choice.

Substituting \eqref{BG} in the definition of shear \eqref{shear-def}, we find
\begin{align}\label{shear-O(1)}
&\delta_1\sigma_{\mu\nu} = \frac{a^2}{2} \left(\begin{matrix}
0 & 0
\\
0 & \dot{h}_{ij}
\end{matrix} \right) \, ,
\end{align}
where $\delta_1$ denotes first order perturbations.
Using this result together with \eqref{BG} and \eqref{Sigma-O(1)} in the interaction action \eqref{action-int}, we find
\begin{align}\label{action-int-Sigma}
S_{\rm int} &=
\int \D^3x\, \D{t}\, N a^3 \left[
\ncpl {H} \, \dot{\gamma}_{ij} t^{ij}
+ \kappa \, \dot{\gamma}_{ij} \dot{t}^{ij}
+ \frac{\eta}{a} \varepsilon^{ijk} \dot{\gamma}_{im} \partial_j t_{k}{}^m
\right] \, ,
\end{align}
in which we have considered the parametrization,
\begin{align}\label{ncpl-def}
&C_\rho = \ncpl H \, ,
\end{align}
where $\ncpl$ is a dimensionless parameter. Working with conformal time $\tau$ defined as $N=a$, we find the expression for the total action $S=S_{\rm E.H.} + S_{\Sigma} + S_{\rm int}$,\footnote{It is worth mentioning that although we have assumed the Einstein-Hilbert action in the gravity sector, parity-violating operators built purely from the metric tensor perturbations arise at higher order than the parity-violating operator proportional to ${\tilde C}_P$ in the spin-2 sector in Eq.~\eqref{action-2nd-Sigma}. The reason is that the relevant tensor perturbation of the extrinsic curvature is its traceless part, the shear $\sigma_{\mu\nu}$, which has no background value. In particular, for tensor modes one finds $\delta_1\sigma_{ij}=a^2\dot h_{ij}/2$, see Eq.~\eqref{shear-O(1)}, so the terms proportional to $Hh_{ij}$ contained in $K_{ij}$ cancel after subtracting the trace. Therefore the lowest-order parity-violating gravity operator, $\epsilon^{\mu\nu\alpha}h^{\beta\rho}\sigma_{\mu\beta}D_\nu\sigma_{\alpha\rho}$, gives schematically $\epsilon^{ijk}\dot\gamma_{i\ell}\partial_j\dot\gamma_k{}^\ell$, and is third order in derivatives. By contrast, the parity-violating operator in the spin-2 sector gives $\epsilon^{ijk}t_{i\ell}\partial_j t_k{}^\ell$, which contains only one derivative.}
\begin{align}\nonumber
S &= \frac{1}{2} \int \D^3x\, \D\tau\, a^2 \left[
\left( {\gamma}'_{ij} \right)^2
- \left( \partial_i{\gamma}_{jk} \right)^2
\right]
\\ \nonumber
&+ \frac{1}{2} \int \D^3x\, \D\tau\, a^2 f^2 \left[
\left( {t}'_{ij} \right)^2
- c_t^2 \left( \partial_i{t}_{jk} \right)^2
- m^2 a^2 t_{ij}^2
+ 2 \pvc {\cal H}  \varepsilon^{ijk}  t_{im} \partial_j t_{k}{}^m
\right]
\\ \label{action-total}
&+
\int \D^3x\, \D\tau\, a^2 \gamma'_{ij} \Big(
\ncpl\, {\cal H} t^{ij}
+ \kappa \, t'^{ij}
+ \eta \, \varepsilon^{imn} \partial_m t_{n}{}^j
\Big) \, ,
\end{align}
where a prime denotes derivative with respect to the conformal time and ${\cal H}=aH=a'/a$ is the conformal Hubble parameter.

In a fully de Sitter invariant setup, light higher-spin fields are strongly constrained. In particular, demanding long-lived superhorizon fluctuations for a massive spin-${ s}$ field during inflation runs into a unitarity obstruction: for spinning fields the helicity-0 component becomes ghost-like unless the Higuchi bound is satisfied $m^2 > { s}({ s}-1)H^2$ \cite{Higuchi:1986py,Arkani-Hamed:2015bza,Bordin:2016ruc,Kehagias:2017cym}. For spin-2 this implies $m^2>2H^2$, which excludes the would-be ``light'' regime $m^2<2H^2$ needed for non-decaying superhorizon modes. As we consider a quasi-de Sitter expansion during inflation, one may wonder whether the Higuchi bound holds approximately. This expectation, however, relies on the approximate de Sitter isometries being realized in the spin-2 sector. Indeed, in the EFT it can be evaded: a sizable coupling to the inflaton/foliation breaks the relevant de Sitter isometries, notably boosts, so the assumptions underlying the Higuchi argument need not apply in the same way \cite{Bordin:2018pca}. This is analogous to the vector case, where an inflaton-dependent gauge kinetic function breaks conformal invariance and allows vector perturbations to persist on superhorizon scales \cite{Ratra:1991bn,Watanabe:2009ct}. Likewise, with sufficiently strong inflaton/foliation couplings one can relax the Higuchi bound for spin-2 and higher-spin fields, as explored in \cite{Kehagias:2017cym,Baumann:2017jvh,Goon:2018fyu,Bordin:2018pca}.\footnote{Strictly speaking, the spin-2 sector need not have exactly zero background energy density once it is coupled to the inflaton/foliation, so in that narrow sense it is not a ``spectator.'' Nevertheless, in our regime of interest its energy density is assumed to remain subdominant compared to the inflaton, so its backreaction on the background can be neglected and it is effectively a spectator throughout.}

The EFT used above to derive the quadratic action \eqref{action-total} can be viewed as a generalization of the framework developed in Ref.~\cite{Bordin:2018pca} to include parity-violating operators. Another interesting but conceptually different setup that can provide additional tensor modes is based on an $SU(2)$ non-Abelian gauge field with an isotropic vacuum expectation value in a cosmological background~\cite{Maleknejad:2011jw,Maleknejad:2011sq}. Very recently, an EFT description for these scenarios has been formulated in Ref.~\cite{Aoki:2025uwz}. Although our EFT differs from the models based on $SU(2)$ gauge fields~\cite{Adshead:2012kp,Maleknejad:2012fw,Adshead:2013nka,Namba:2013kia,Maleknejad:2014wsa,Obata:2016tmo,Maleknejad:2016qjz,Agrawal:2017awz,Agrawal:2018mrg,Domcke:2018rvv,Watanabe:2020ctz,Ishiwata:2021yne,Dimastrogiovanni:2025snj}, it is still useful to compare the symmetry structures of these frameworks. All possible interactions for the $SU(2)$ tensor EFT are classified in Ref.~\cite{Aoki:2025uwz}. An apparently allowed interaction in our EFT would be a term of the form $\varepsilon^{ijk} \gamma_{im} \partial_j t_k{}^{m}$ which directly couples $\gamma_{ij}$ and $t_{ij}$ in a parity-violating manner. However, such a term is actually forbidden in our setup because the tensor that is invariant under \textit{time-dependent spatial diffeomorphisms} is the extrinsic curvature, $K_{ij} \propto \dot{\gamma}_{ij}$. In contrast, in $SU(2)$ gauge field models the situation is different. The presence of a non-vanishing and isotropic spatial vacuum expectation value $A^a_{\mu} = A(t)\, \delta^a_{\mu}$ explicitly breaks spatial diffeomorphisms. The theory is instead invariant under a \textit{residual symmetry given by a diagonal combination of internal $SU(2)$ gauge transformations and spatial rotations}. This different symmetry-breaking pattern allows for operators such as $\varepsilon^{ija} \gamma_{im} \partial_j t_a{}^{m}$ (see Refs.~\cite{Agrawal:2017awz,Agrawal:2018mrg}), which are absent in our EFT. Therefore, the structure of allowed parity-violating interactions in our EFT is fundamentally different from those in scenarios based on $SU(2)$ gauge fields.

\section{Evolution during inflation}\label{sec-inflation}

In the previous section, we have found the quadratic action \eqref{action-total} that describes interactions between the metric tensor perturbations $\gamma_{ij}$ and additional tensor perturbations $t_{ij}$ which are provided by a spin-2 field during inflation. Interactions labeled by $\kappa$ and $\eta$ have one more time/spatial derivative compared with the interaction labeled by $\ncpl$. Thus, $\kappa$ and $\eta$ are dominant for the modes deep inside the horizon $k\gg{\cal H}$ while they are negligible in the superhorizon limit $k\ll{\cal H}$. On the other hand, $\ncpl$ dominates in the superhorizon limit $k\ll{\cal H}$ and it is negligible in the deep-subhorizon limit. In this regard, $\kappa$ and $\eta$ will affect the spectrum of GWs through the adiabatic initial condition which is imposed when the modes are deep inside the horizon. In Appendix \ref{app-deep-inside-horizon}, we show that $\kappa$ and $\eta$ change the propagation speeds of the diagonalized tensor modes so that the metric-like mode is no longer luminal at early times. Late-time observations of GWs from binary black hole and neutron star mergers have revealed that the speed of GWs should be equal to the speed of light with very high accuracy. There will be no conflict with observations if the speed of GWs differs significantly from the speed of light in vacuum at early times but approaches it at late time. In the following, we impose the conservative assumption that the metric-like tensor speed remains luminal at all times and we neglect the effects of $\kappa$ and $\eta$.

Going to Fourier space,
\begin{align}\label{Fourier-def}
    X_{ij}(\tau,{\bf x}) = \int \frac{\D^3{k}}{(2\pi)^3} e^{i{\bf k}\cdot{\bf x}} \sum_{\lambda=\pm} X^\lambda_{\bf k}(\tau) e^\lambda_{ij}({\bf k}) \,;
    \qquad
    X_{ij}=\left(\gamma_{ij},t_{ij}\right) \, ,
\end{align}
where the circular polarization tensors satisfy $\delta^{ij} e^\lambda_{ij}({\bf k}) = 0 = k^i e^\lambda_{ij}({\bf k})$, to respect the traceless-transverse conditions, together with the orthonormality condition
$e^\lambda_{ij} ({\bf k}) e^{\lambda'}_{ij}(-{\bf k}) = \delta^{\lambda\lambda'}$. Ignoring the interactions $\kappa$ and $\eta$, it is straightforward to show that the action \eqref{action-total} takes the form
\begin{align}
    \label{action-helicity}
    \begin{split}
        S &= \frac{1}{2} \sum_{\lambda} \int \frac{\D^3k}{(2\pi)^3}\, \D\tau\, a^2
        \left[
        \left| {\gamma}'^\lambda_{\bf k} \right|^2
        - k^2 \left| {\gamma}^\lambda_{\bf k} \right|^2 \right]
        \\
        &
        + \frac{1}{2} \sum_{\lambda} \int \frac{\D^3k}{(2\pi)^3}\, \D\tau\, a^2  f^2 \left[
        \left| {t}'^\lambda_{\bf k} \right|^2
        - \left( c_t^2 k^2 + 2 \lambda {\pvc} k {\cal H} + m^2 a^2 \right) \left| {t}^\lambda_{\bf k} \right|^2
        \right]
        \\
        &
        + \frac{1}{2} \sum_{\lambda} \int \frac{\D^3k}{(2\pi)^3} \D\tau\, a^2
        \ncpl {\cal H} \left( {\gamma}'^\lambda_{\bf k} {t}^{\ast\lambda}_{\bf k} + \mbox{c.c.} \right) \, ,
    \end{split}
\end{align}
where $\lambda=+1$ and $\lambda=-1$ denote the two circular helicities. In obtaining the above result, we have used the identity $i \delta_{im}\varepsilon^{mnl} k_n e^\lambda_{lj} ({\bf k}) = \lambda  k\, e^\lambda_{ij}({\bf k})$.

During inflation, the scale factor is $a=-1/(H_{\rm inf}\tau)$ with $-\infty<\tau<0$, where we ignore slow-roll corrections so that the Hubble parameter during inflation, $H_{\rm inf}$, is treated as constant. Thus, for $m\ll H_{\rm inf}$, we can ignore the mass term during inflation, while the spectrum of $t_{ij}$ is Boltzmann suppressed in the case $m\gg H_{\rm inf}$. From now on, we ignore the mass term during inflation. This is of course our choice, and one may also study the case $m\sim H_{\rm inf}$ to possibly find a new feature.\footnote{The mass may dominate after inflation and, interestingly, $t_{ij}$ can play the role of spin-2 dark matter \cite{Gorji:2023cmz}.}

Varying the action \eqref{action-helicity} with respect to the complex Fourier amplitudes, treating $\gamma^\lambda_{\bf k}$ and $t^\lambda_{\bf k}$ as independent variables from their complex conjugates, we find
\begin{align}\label{EoM-gamma-inf}
    \gamma''^\lambda_{\bf k} + 2 \frac{a'}{a} \gamma'^\lambda_{\bf k} + k^2 \gamma^\lambda_{\bf k}
    &=
    - \ncpl \frac{a'}{a}
    \left[ t'^\lambda_{\bf k} + \frac{\left( a a' \right)'}{ a a'} t^\lambda_{\bf k}
    \right] - \ncpl' \frac{a'}{a} t^\lambda_{\bf k} \, ,
    \\
    \label{EoM-t-inf}
    t''^\lambda_{\bf k} + 2 \frac{(af)'}{af} t'^\lambda_{\bf k} + \left( c_t^2k^2 + 2 \lambda \pvc k \frac{a'}{a} \right) t^\lambda_{\bf k}
    &= \frac{ \ncpl}{f^2} \frac{a'}{a} \gamma'^\lambda_{\bf k} \, .
\end{align}
Since the system is linear, the quantum mode functions introduced in the next section obey the same equations as the corresponding classical Fourier amplitudes. In this section, we use $\gamma^\lambda_{\bf k}$ and $t^\lambda_{\bf k}$ as generic classical solutions of the coupled system in order to study their time evolution. The quantization of the system and the computation of the power spectra are presented in Sec.~\ref{sec-power-spectra}. Thus, for a given form of $f(\tau)$, $c_t(\tau)$, $\pvc(\tau)$, and $\ncpl(\tau)$, we can, in principle, solve Eqs.~\eqref{EoM-gamma-inf} and \eqref{EoM-t-inf} to find the time evolution of $\gamma^\lambda_{\bf k}$ and $t^\lambda_{\bf k}$.

To qualitatively understand the roles of the EFT couplings $f(\tau)$, $c_t(\tau)$, $\pvc(\tau)$, and $\ncpl(\tau)$, let us first consider the case in which they do not change significantly in time, so that they can be treated as constants. We can then easily see that:

\begin{itemize}
    \item For the modes deep inside the horizon, $k\gg{\cal H}$, assuming $c_t={\cal O}(1)$, the effects of the couplings $\ncpl$ and $\pvc$ are negligible. In this regime the decoupled positive-frequency solutions are
    \begin{align}\label{gamma-t-DIH}
        \gamma^\lambda_{\bf k}
        = \frac{1}{a} \frac{e^{-i k\tau}}{\sqrt{2k}}\,,
        \qquad
        t^\lambda_{\bf k} = \frac{1}{af} \frac{e^{-i c_t k\tau}}{\sqrt{2c_tk}} \,.
    \end{align}
    Here Eq.~\eqref{gamma-t-DIH} should be understood as the normalization of the two decoupled positive-frequency basis solutions deep inside the horizon. In the quantum treatment, these two basis solutions correspond to the two independent mode columns introduced in Sec.~\ref{sec-power-spectra}. If $c_t$ varies in time, the WKB phase $c_t k\tau$ should be replaced by $\int^\tau c_t(\tilde \tau) k\,\D\tilde \tau$.

    \item Around the time of horizon crossing, $k={\cal H}$, or equivalently $-k\tau=1$, both $\ncpl$ and $\pvc$ become efficient.

    \item For $\pvc>0$ and $\lambda=-1$, the modes satisfying $c_tk/{\cal H}<2\pvc/c_t$ in Eq.~\eqref{EoM-t-inf} become tachyonic and are enhanced. For $\pvc<0$ the same statement applies to the opposite helicity. For $2|\pvc|/c_t<1$, this enhancement happens only for the superhorizon modes $c_tk< {\cal H}$, while for $2|\pvc|/c_t>1$ subhorizon modes with $1<c_t k/{\cal H}<2|\pvc|/c_t$ can also be enhanced. This tachyonic-like growth happens for a wide range of modes and leads to parity violation at all scales. Thus, very large values of $|\pvc|$ will make the system unstable.

    \item For $\pvc=0$, $f={\cal O}(1)$, and $c_t={\cal O}(1)$, the amplitudes $\gamma^\lambda_{\bf k}$ and $t^\lambda_{\bf k}$ interact with each other through $\ncpl$. The usual superhorizon freezing may then fail for both $\gamma^\lambda_{\bf k}$ and $t^\lambda_{\bf k}$.

    \item The amplitude of $t^\lambda_{\bf k}$ is proportional to $f^{-1}$. Moreover, from the WKB solution \eqref{gamma-t-DIH}, we see that, at least for the modes deep inside the horizon, the amplitude of $t^\lambda_{\bf k}$ is also proportional to $c_t^{-1/2}$.
\end{itemize}

Before turning to time-dependent couplings, it is useful to note that if $f$, $c_t$, $\ncpl$, and $\pvc$ are all constant, then Eqs.~\eqref{EoM-gamma-inf}-\eqref{EoM-t-inf}, written in terms of the dimensionless variable $x\equiv k\tau$ and the rescaled fields $a\gamma^\lambda_{\bf k}$ and $af\,t^\lambda_{\bf k}$, contain no explicit $k$ dependence. The full evolution from $x\to-\infty$ to $x\to 0^-$ is therefore $k$-independent, and the resulting tensor power spectrum is exactly scale invariant. In order to produce any $k$-dependent feature in ${\cal P}_h$, in particular oscillations in $k$, one must explicitly break this scaling, i.e. introduce a localized time dependence in at least one of the couplings.

\medskip
Based on the above observations, and for the sake of simplicity, we assume that $\ncpl$ and $\pvc$ are constant. We allow $\ncpl$ to be large enough to produce the superhorizon mixing effect discussed below, while $\pvc$ is kept sufficiently small to avoid a broad tachyonic instability. This already provides an ${\cal O}(1)$ modification since the usual scenario for GWs corresponds to $\ncpl=0$. On the other hand, if we assume $f,c_t={\cal O}(1)$ with no localized time dependence, we will not get any significant scale-dependent effects from the couplings $f$ and $c_t$. Considering $f,c_t\ll1$ at all times is also not a good choice since the modes are then enhanced at all scales, which is not what we are interested in. For example, we do not want a significant enhancement at CMB scales since there is an upper bound on the amplitude of primordial GWs. Therefore, in order to have a consistent setup, we need to assume a time dependence for $f$ and $c_t$ which turns on only for a finite period of time. As we are working within the EFT framework and do not have a specific underlying model, we look for general functional forms for $f$ and $c_t$ which capture robust features of this type of scenario. In the standard case, $f=1$ and $c_t={\cal O}(1)$. We therefore assume that $f=1$ and $c_t=c_{t,0}$ hold at the beginning and at the end of inflation, while for a finite period of time, for example over $2$ to $5$ e-folds at sub-CMB scales, they become time dependent. In this way, only the modes which leave the horizon during this period are modified, leading to a scale-dependent spectrum. This time dependence may arise from a coupling between the spin-2 field and the inflaton. The precise time dependence determines the shape of the spectrum. The simplest possibility is a power spectrum with a peak. As mentioned above, we expect that smaller values of $f$ and $c_t$ enhance the amplitude of $t^\lambda_{\bf k}$ and therefore also the amplitude of $\gamma^\lambda_{\bf k}$, since the two interact through $\ncpl$. To obtain such a peaked power spectrum, we consider the form
\begin{align}\label{f}
    F_i(\tau,\tau_\ast) = c_i - {\cal A} \exp\left[-\frac{\left(\ln\left[\tau/\tau_\ast\right]\right)^2}{2\sigma^2H_{\rm inf}^2}\right] \,;
    \qquad
    F_i = \{f,c_t\} \, ,
    \quad
    c_i = \{1,c_{t,0}\} \, ,
\end{align}
where $i=1,2$ and $\tau_\ast$, ${\cal A}$, $\sigma$, and $c_{t,0}$ are free parameters, with $0<{\cal A}<c_i$ in the corresponding case, so that the profile remains positive. We also use the shorthand $\tilde\sigma\equiv\sigma H_{\rm inf}$ below.
\begin{figure}[htbp]
    \centering
    \includegraphics[width=0.8\textwidth]{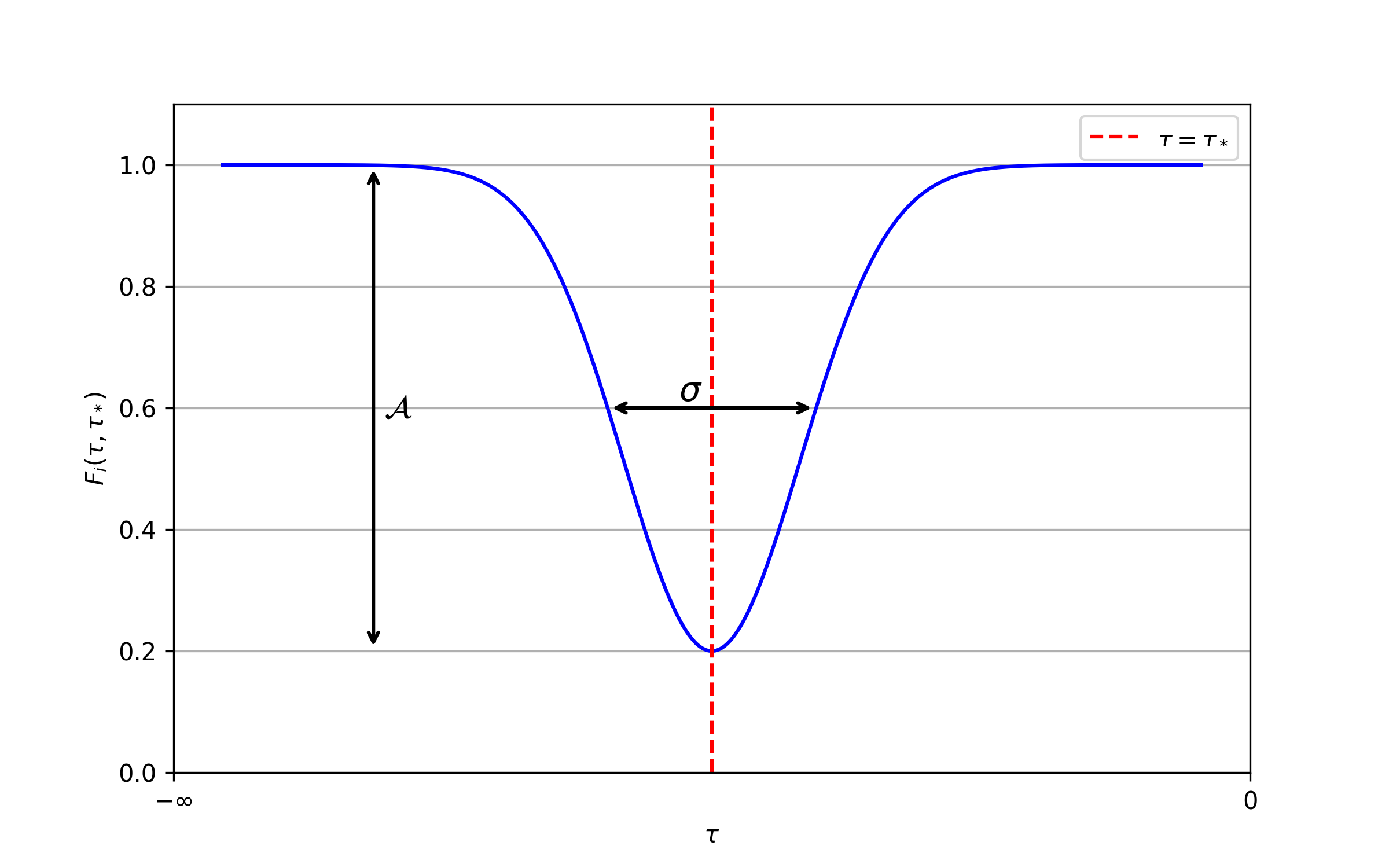}
    \caption{Function $F_i(\tau,\tau_\ast)$ defined in \eqref{f} is plotted for $c_i=1$, $\sigma H_{\rm inf}=1$, and $\tau_\ast=-100$. The Gaussian dip with height ${\cal A}$ and width $\sigma$ at the time $\tau_\ast$ leads to a peak in the power spectrum around $k\sim{k}_\ast=-1/\tau_\ast$.}
    \label{fig:f}
\end{figure}
The ansatz \eqref{f} has a clear meaning in terms of cosmic time $t=\int a(\tau)\D\tau$: it represents a function with a Gaussian dip of width $\sigma$, located at the time $t_\ast=-\ln\left(-H_{\rm inf}\tau_\ast\right)/H_{\rm inf}$, and approaching the constant value $c_i$ for $t\gg t_\ast$ and $t\ll t_\ast$. As shown in Fig.~\ref{fig:f}, the amplitude is characterized by ${\cal A}$ and $c_{t,0}$, while the scale dependence is controlled by $\tau_\ast$ and $\sigma$.

Based on the above discussion, in the following we focus on the two cases
\begin{align}\label{Cases}
    \begin{cases}
        \ncpl={\rm const.},\quad \pvc\lesssim{\cal O}(1),
        \quad
        c_t=c_{t,0},
        \quad \mbox{and} \quad &f(\tau,\tau_\ast)=F_1(\tau,\tau_\ast) , \qquad \mbox{Case I}
        \\
        \ncpl={\rm const.},\quad \pvc\lesssim{\cal O}(1),
        \quad
        f=1,
        \quad \mbox{and} \quad
        &c_t(\tau,\tau_\ast)=F_2(\tau,\tau_\ast) . \qquad \mbox{Case II}
    \end{cases}
\end{align}
We will numerically solve Eqs.~\eqref{EoM-gamma-inf} and \eqref{EoM-t-inf} for Cases I and II for a mode with comoving momentum $k$ which starts deep inside the horizon, $-k\tau\gg1$, until the time it becomes superhorizon, $-k\tau\ll1$. In this section, the numerical solutions are used to illustrate the time evolution of representative classical Fourier amplitudes. The physical power spectra are computed in the next section by evolving the two independent quantum mode functions and summing over them. Before doing so, it is instructive to gain some analytical understanding of the setup in useful limits. Below, we first study the effects of the couplings $f$ and $c_t$ in order to better understand the roles of the parameters in \eqref{f} for Cases I and II. Then we consider the dynamics of the modes inside and far outside the horizon separately. The origin of the in-$k$ oscillations is discussed in Sec.~\ref{sec-osc-mechanism}, after the numerical spectra are displayed.

\subsection{Time varying non-minimal coupling $f$ and sound speed $c_t$}\label{sec-time-varying}

We have assumed that $\ncpl$ and $\pvc$ are both constant while we have considered the nontrivial functional form \eqref{f} for $f$ and $c_t$. In this subsection, in order to get analytical understanding of the choice \eqref{f} for the non-minimal coupling $f$ and the sound speed $c_t$, we look at the limit $\ncpl=0$. In this limit, $\gamma^\lambda_{\bf k}$ and $t^\lambda_{\bf k}$ decouple. Since $\gamma^\lambda_{\bf k}$ is then completely decoupled, the metric tensor mode gives the standard result and satisfies the adiabatic condition inside the horizon. The nontrivial part is the evolution of $t^\lambda_{\bf k}$ through the dip.

Working with the normalized field $af t^\lambda_{\bf k}$ in \eqref{EoM-t-inf}, the homogeneous $t$-mode has the effective frequency
\begin{align}\label{omega-t-full-inf}
	\omega_t^2(\tau)
	=
	c_t^2(\tau) k^2
	+
	2\lambda \pvc k \frac{a'}{a}
	-
	\frac{(af)''}{af} \, .
\end{align}
For constant $\pvc$, a necessary WKB condition is
\begin{align}\label{wkb-full-condition}
2\pi\left|\frac{\omega_t'}{\omega_t^2}\right| \ll 1 \, .
\end{align}
This condition applies when $\omega_t^2$ is positive. If $\omega_t^2$ becomes negative for one helicity, the mode is no longer in an oscillatory WKB regime and instead undergoes tachyonic growth. During inflation $a'/a\simeq-1/\tau>0$ as $\tau<0$. Therefore, for $\pvc>0$, the parity-violating term lowers $\omega_t^2$ for the helicity $\lambda=-1$ and raises it for $\lambda=+1$.

It is useful to first isolate the violation of adiabaticity caused by the localized time dependence of $f$ or $c_t$. Setting $\pvc=0$, Eq.~\eqref{omega-t-full-inf} reduces to
\begin{align}\label{omega-t-pvc-zero}
	\omega_{t}^2(\tau) = c_t^2 k^2 - \frac{(af)''}{af} \, .
\end{align}
We now evaluate the WKB condition in this limit for the two Cases.

For Case I, $c_t=c_{t,0}$ and the dip is in $f$. Evaluating \eqref{wkb-full-condition} with $\pvc=0$ at the center of the dip, $\tau=\tau_\ast$, and assuming that the mode remains subhorizon at that time, gives
\begin{align}\label{adiabatic-condition-case-I}
	\frac{\pi}{|c_{t,0}k\tau_\ast|^3}
	\left(4+\frac{5 {\cal A}}{|1- {\cal A}|\,\tilde\sigma^2}\right) \ll 1
	\qquad \mbox{Case I}\, .
\end{align}
For $|1- {\cal A}|\ll1$, this condition reduces parametrically to
\begin{align}\label{adiabatic-conditions-I-simple}
	\tilde\sigma\sqrt{|1- {\cal A}|} \gg
	\frac{\sqrt{5\pi}}{|c_{t,0}k\tau_\ast|^{3/2}}\,
	\qquad \mbox{Case I}\, .
\end{align}
Thus a sharper and deeper dip in $f$ makes the $t$-mode more non-adiabatic. In this case the term $-(af)''/(af)$ can become large and negative during the dip. Therefore, even for $\pvc=0$, Case I can produce strong enhancement through the non-adiabatic time dependence of $f$.

For Case II, $f=1$ and the dip is in $c_t$. We denote the minimum value of the sound speed by
\begin{align}
	c_{t,\min}=c_{t,0}-{\cal A}.
\end{align}
For $\pvc=0$, the dip in $c_t$ lowers the positive gradient term $c_t^2 k^2$ but does not make it negative. A necessary condition for applying a subhorizon WKB expansion through the bottom of the dip is therefore
\begin{align}\label{adiabatic-condition-case-II-bottom}
	|c_{t,\min} k\tau_\ast|\gg1 \, .
\end{align}
A separate adiabaticity condition on the slopes of the Gaussian profile follows from $|c_t'|/(c_t^2 k)\ll1$ over the region where the profile varies. Parametrically this requires
\begin{align}\label{adiabatic-condition-case-II-slope}
	\frac{{\cal A}}{\tilde\sigma\,c_{t,\min}^2 |k\tau_\ast|}\ll1
	\qquad \mbox{Case II}\, ,
\end{align}
up to an order-one numerical factor determined by where on the Gaussian slope the maximum is evaluated. Therefore decreasing $c_{t,\min}$ makes the evolution increasingly non-adiabatic and eventually invalidates a perturbative WKB treatment near the bottom of the dip. However, unlike Case I, the sound-speed dip by itself does not make $\omega_t^2$ negative when $\pvc=0$.

For $\pvc\neq0$, the helicity-dependent term in \eqref{omega-t-full-inf} can make one helicity tachyonic. In Case II, near the bottom of the dip, $c_t\simeq c_{t,\min}$. We also use $a'/a|_{\tau_\ast}=k_\ast$, which follows from $\tau_\ast=-1/k_\ast$ in de Sitter. For modes for which the $-a''/a$ contribution is subleading compared with the gradient and parity-violating terms, Eq.~\eqref{omega-t-full-inf} can be written as
\begin{align}\label{omega-caseII-bottom-ktach}
	\omega_t^2(\tau_\ast)
	\simeq
	c_{t,\min}^2 k
	\left[
	k+\lambda\,{\rm sgn}(\pvc)\,k_{\rm tach}
	\right],
\end{align}
where the positive momentum scale $k_{\rm tach}$ is defined by
\begin{align}\label{ktach-inflation}
	k_{\rm tach}
	\equiv
	\frac{2|\pvc|}{c_{t,\min}^2}\,k_\ast .
\end{align}
Therefore the helicity $\lambda=-{\rm sgn}(\pvc)$ becomes tachyonic for
\begin{align}\label{tachyonic-band-inflation}
	k\lesssim k_{\rm tach}.
\end{align}
For $\pvc>0$ this corresponds to $\lambda=-1$, while for $\pvc<0$ it corresponds to $\lambda=+1$. The estimate \eqref{tachyonic-band-inflation} receives order-one corrections from the time dependence of the profile and from the precise time at which the mode passes through the dip. Thus in Case II a small sound speed has two different effects. It can make the evolution non-adiabatic, as measured by \eqref{adiabatic-condition-case-II-bottom} and \eqref{adiabatic-condition-case-II-slope}, and it can also enlarge the tachyonic momentum range through \eqref{ktach-inflation}. The latter effect is helicity dependent and is the origin of the strong chirality in the large Case II signals.

\subsection{Modes inside the horizon}

To study the modes inside the horizon, it is more convenient to work with the variables
\begin{align}
    H^\lambda_{\bf k} \equiv a \gamma^\lambda_{\bf k} \, ,
    \qquad
    T^\lambda_{\bf k} \equiv a f t^\lambda_{\bf k} \, ,
\end{align}
in terms of which Eqs.~\eqref{EoM-gamma-inf} and \eqref{EoM-t-inf}, for constant $\ncpl$, take the forms
\begin{align}\label{EoM-H-inf}
    {H^{\lambda}_{\bf k}}'' + \left( k^2-\frac{a''}{a} \right) H^\lambda_{\bf k}
    &=
    - \frac{\ncpl}{f}
    \left(
    \frac{a'}{a} {T^\lambda_{\bf k}}'
    - \frac{a' f'}{a f} T^\lambda_{\bf k}
    + \frac{a''}{a} T^\lambda_{\bf k}
    \right) \, ,
    \\ \label{EoM-T-inf}
    {T^\lambda_{\bf k}}'' + \left( c_t^2 k^2 + 2 \pvc \lambda k \frac{a'}{a}-\frac{(af)''}{af}\right) T^\lambda_{\bf k}
    &=
    \frac{\ncpl}{f} \frac{a'}{a}
    \left(
    {H^\lambda_{\bf k}}' - \frac{a'}{a} H^\lambda_{\bf k}
    \right) \, .
\end{align}
For the modes deep inside the horizon, $|k\tau|\to\infty$, the effects of both $\ncpl$ and $\pvc$ are negligible and we recover the WKB solutions \eqref{gamma-t-DIH} as expected. Going to the next order and keeping the first corrections in the limit $k\gg{\cal H}$, and away from the localized non-adiabatic part of the dip where derivatives of $f$ or $c_t$ dominate, we have
\begin{align}\label{EL-H}
    &{H^{\lambda}_{\bf k}}'' + \wh^2 H^\lambda_{\bf k} = - \ncpl {\cal H} {T^\lambda_{\bf k}}' \,;
    &&\wh \equiv k \, ,
    \\
    \label{EL-T}
    &{T^\lambda_{\bf k}}'' + \wt^2 T^\lambda_{\bf k}
    = \ncpl {\cal H} {H^\lambda_{\bf k}}' \,;
    &&\wt \equiv \sqrt{c_t^2k^2+ 2 \lambda \pvc k {\cal H}} \, .
\end{align}
Substituting
\begin{align}\label{H-T}
    H^\lambda_{\bf k}
    &=
    \frac{-i}{\wh}
    \left(
    {\mathsf c}_{+,\lambda} \omega_{+,\lambda} X^\lambda_{\bf k}
    +
    {\mathsf c}_{-,\lambda} \omega_{-,\lambda} Y^\lambda_{\bf k}
    \right) \, ,
    \\
    T^\lambda_{\bf k}
    &=
    -{\mathsf c}_{-,\lambda} X^\lambda_{\bf k}
    +
    {\mathsf c}_{+,\lambda} Y^\lambda_{\bf k} \, ,
\end{align}
in \eqref{EL-H} and \eqref{EL-T}, and setting ${\mathsf c}_{+,\lambda}^2+ {\mathsf c}_{-,\lambda}^2=1$, we obtain
\begin{align}\label{Hamiltons-EoM}
    X''_{\bf k} + \omega_{+,\lambda}^2 X_{\bf k} =0 \, ,
    \qquad
    Y''_{\bf k} + \omega_{-,\lambda}^2 Y_{\bf k} =0 \, ,
\end{align}
where
\begin{align}\label{omega-pm}
    \omega_{\pm,\lambda}^2 &=
    \frac{1}{2} \left[\wh^2+\wt^2 + \ncpl^2 {\cal H}^2 \pm \sqrt{\Delta_\lambda} \right] \, ,\\
    \label{alpha-def}
    {\mathsf c}_{\pm,\lambda} &\equiv \frac{1}{\sqrt{2}}
    \left[
    1\pm\frac{\wh^2-\wt^2- \ncpl^2 {\cal H}^2}{\sqrt{\Delta_\lambda}}
    \right]^{1/2} \, ,
\end{align}
and
\begin{align}
    \Delta_\lambda \equiv
    \left[\wh^2-\wt^2-\ncpl^2 {\cal H}^2\right]^2 + 4 \ncpl^2 {\cal H}^2 \wh^2 \, .
\end{align}
For the modes inside the horizon, the positive-frequency WKB solutions of \eqref{Hamiltons-EoM} are
\begin{align}\label{X-Y}
    X^\lambda_{\bf k}
    &=
    i \frac{\wh}{\omega_{+,\lambda}}
    \frac{e^{-i\int^\tau \omega_{+,\lambda}(\tilde \tau)\, d\tilde \tau}}{\sqrt{2\omega_{+,\lambda}}}\,,
    \\
    Y^\lambda_{\bf k}
    &=
    \frac{e^{-i\int^\tau \omega_{-,\lambda}(\tilde \tau)\, d\tilde \tau}}{\sqrt{2\omega_{-,\lambda}}} \, ,
\end{align}
where the constants of integration are chosen such that we recover \eqref{gamma-t-DIH} in the limit $\ncpl=0=\pvc$. More concretely, for $\ncpl=0=\pvc$ and $c_t<1$, we find $\wh=k$ and $\wt=c_tk$, which, after substituting into \eqref{omega-pm} and \eqref{alpha-def}, give $\omega_{+,\lambda}=k$, $\omega_{-,\lambda}=c_tk$, ${\mathsf c}_{+,\lambda}=1$, and ${\mathsf c}_{-,\lambda}=0$. Substituting these results into \eqref{H-T}, we find $X^\lambda_{\bf k} = i H^\lambda_{\bf k}$ and $Y^\lambda_{\bf k} = T^\lambda_{\bf k}$, which shows how \eqref{X-Y} reduces to \eqref{gamma-t-DIH} in the deep-inside-the-horizon limit where the effects of $\ncpl$ and $\pvc$ can be ignored.

\begin{figure}[ht]
    \begin{subfigure}[b]{0.5\linewidth}
        \centering
        \includegraphics[width=0.9\linewidth]{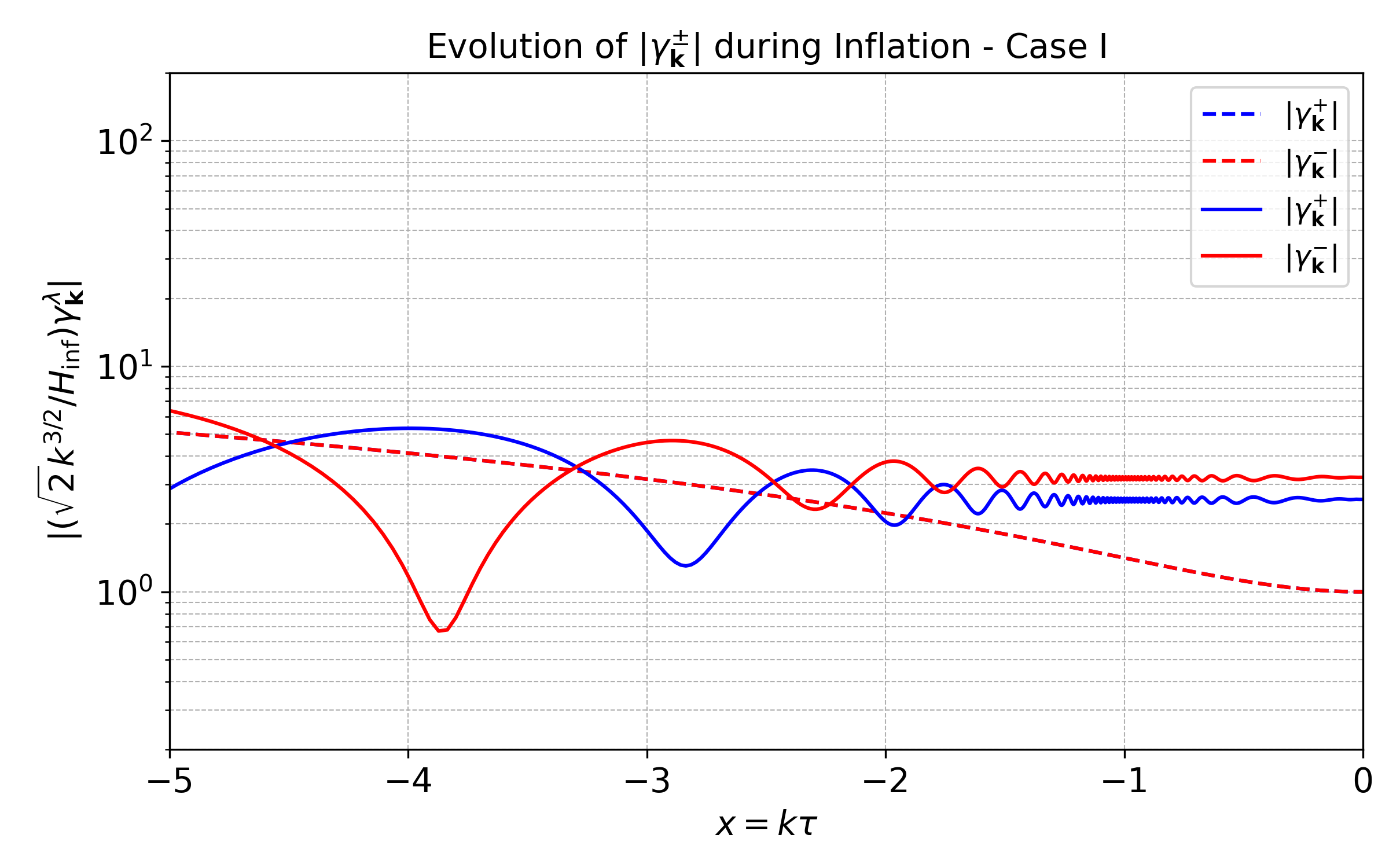}
    \end{subfigure}
    \begin{subfigure}[b]{0.5\linewidth}
        \centering
        \includegraphics[width=0.9\linewidth]{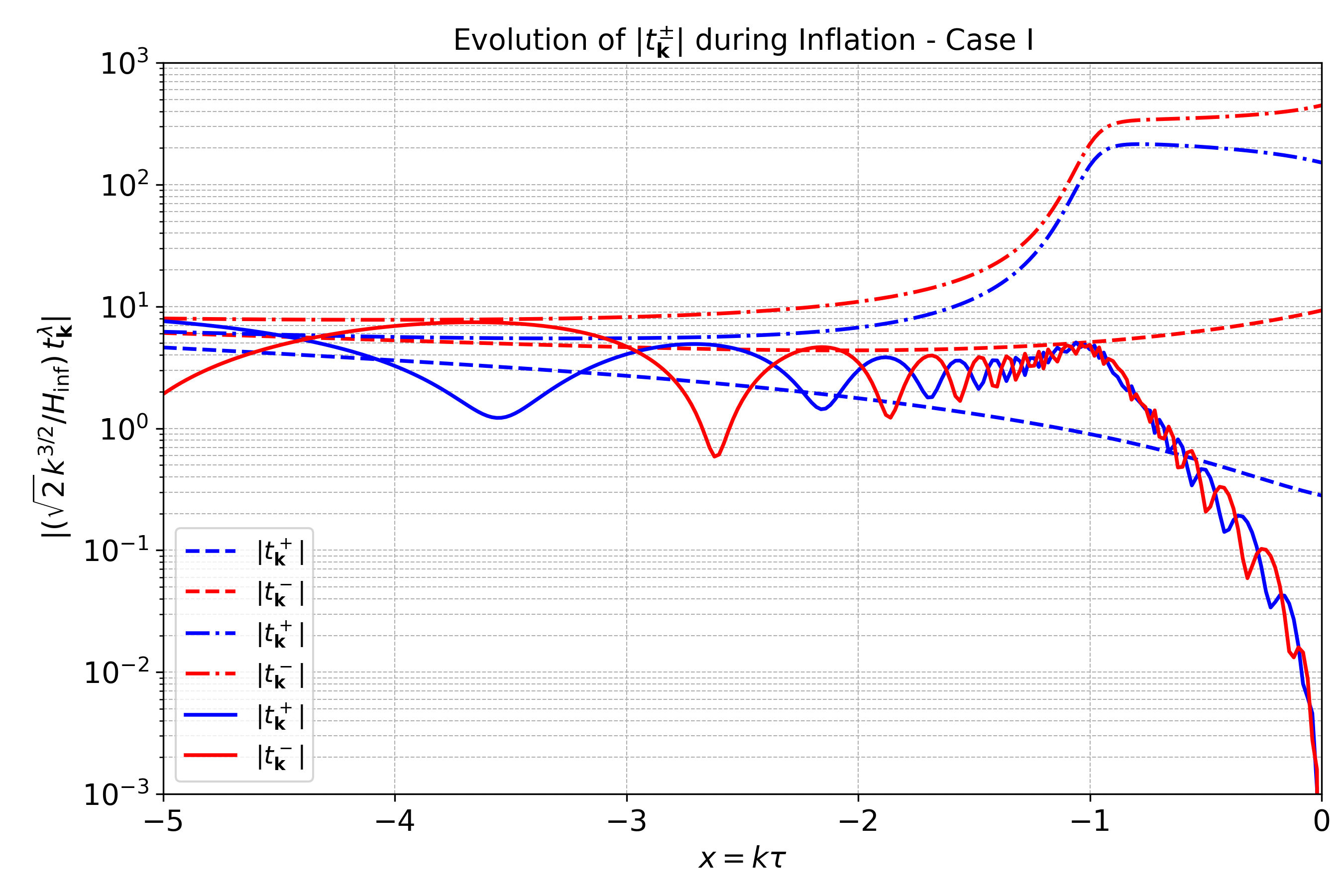}
    \end{subfigure}
    \begin{subfigure}[b]{0.5\linewidth}
        \centering
        \includegraphics[width=0.9\linewidth]{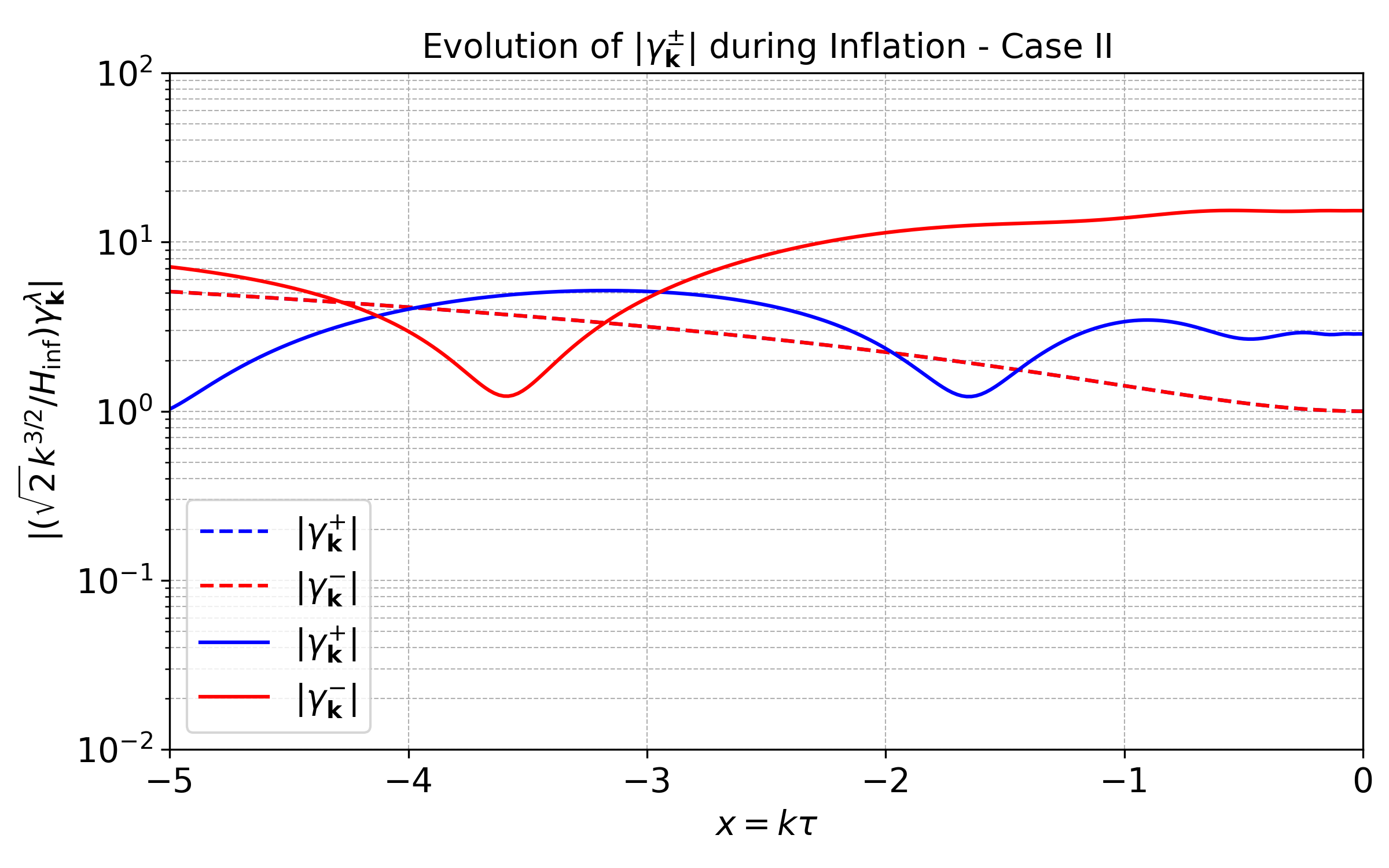}
    \end{subfigure}
    \begin{subfigure}[b]{0.5\linewidth}
        \centering
        \includegraphics[width=0.9\linewidth]{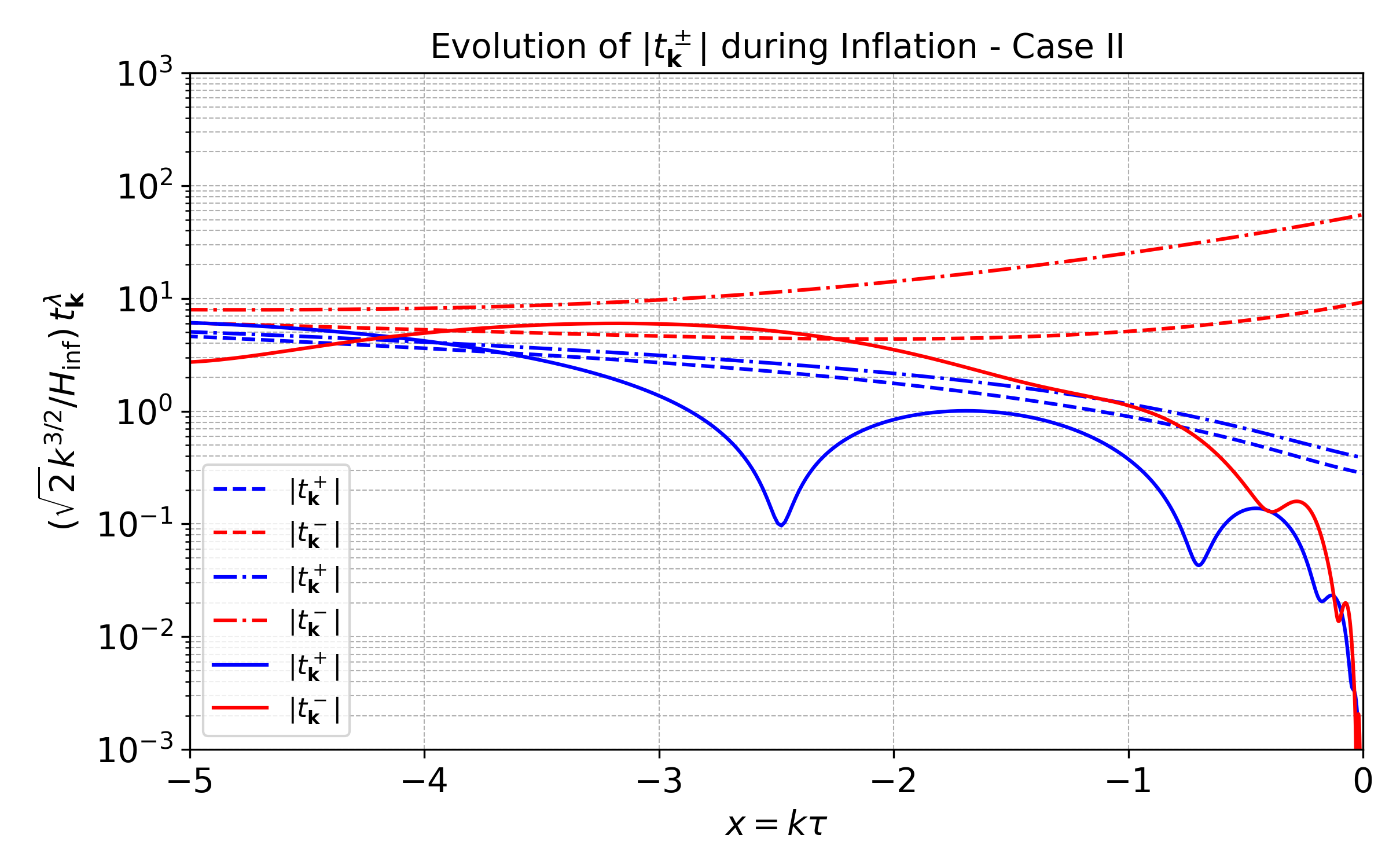}
    \end{subfigure}
    \caption{Time evolution of representative classical Fourier amplitudes $|\tilde\gamma^\lambda_{\bf k}|$ and $|\tilde t^\lambda_{\bf k}|$ during inflation for the mode $k=k_\ast=-1/\tau_\ast$. The blue/red curves show the $+/-$ helicities. The plotted solutions are meant to illustrate the mixing dynamics of the coupled system; the physical power spectra are computed in Sec.~\ref{sec-power-spectra} by evolving the independent quantum mode functions and summing their contributions. The dashed curves correspond to the standard case $\ncpl=0$, $f=1$ and $c_t=c_{t,0}$ while the dotted-dashed curves for ${t}^{\pm}_{\bf k}$ illustrate the case with $\ncpl=0$ but $f/c_t$ given by Case I/II. The figures are plotted for $\ncpl=5$, $\pvc=1$, $\sigma=H_{\rm inf}^{-1}$, $c_{t,0}=0.9$, and $H_{\rm inf}/\Mpl=10^{-5}$. We have considered $|1-{\cal A}|=10^{-2}$ and $|c_{t,0}-{\cal A}|=0.1$ for Case I and Case II, respectively.}
    \label{fig:gamma-t-time-f}
\end{figure}

\subsection{Modes far outside of horizon}\label{sec-modes-far-outside-horizon}

Let us now look at the modes far outside of the horizon, $-k\tau\to0$. We assume that $\tau_\ast\ll\tau_{\rm end}$ so that, for $\tau_\ast\ll\tau<\tau_{\rm end}$, we can safely ignore the effects of the time dependence of $f$ or $c_t$ in Case I and Case II. Thus, Case I and Case II have similar behavior at superhorizon scales, $-k\tau\to0$, where Eqs.~\eqref{EoM-gamma-inf} and \eqref{EoM-t-inf} simplify to
\begin{align}\label{EoM-gamma-SH}
    \gamma''^\lambda_{\bf k} + 2 {\cal H} \gamma'^\lambda_{\bf k}
    &=
    - \ncpl {\cal H}
    \left(
    t'^{\lambda}_{\bf k} + 3 {\cal H} t^{\lambda}_{\bf k}
    \right) \, ,
    \\
    \label{EoM-t-SH}
    t''^\lambda_{\bf k} + 2 {\cal H} t'^\lambda_{\bf k}
    &=
    \ncpl {\cal H} \gamma'^{\lambda}_{\bf k} \, ,
\end{align}
where we have used the fact that ${\cal H}'={\cal H}^2$ during inflation. Integrating once Eq.~\eqref{EoM-gamma-SH}, we find
\begin{align}\label{gamma-prime}
    \gamma'^\lambda_{\bf k} = \frac{C_4}{a^2} - \ncpl {\cal H} t^\lambda_{\bf k} \, .
\end{align}
Substituting the above result into Eq.~\eqref{EoM-t-SH}, we obtain
\begin{align}
    \label{EoM-t-SH-reduced}
    t''^\lambda_{\bf k} + 2 {\cal H} t'^\lambda_{\bf k}
    + \ncpl^2 {\cal H}^2 t^{\lambda}_{\bf k}
    &=
    \ncpl {\cal H} \frac{C_4}{a^2} \, ,
\end{align}
which is solved as
\begin{align}\label{t-sol-a}
    t^\lambda_{\bf k}
    =
    \frac{C_4}{\ncpl H_{\rm inf} a^3}
    +
    \frac{1}{a^{\frac{3}{2}}}
    \Big(
    C_2\, a^{\frac{3}{2}\nu} + C_3\, a^{-\frac{3}{2}\nu}
    \Big) \,;
    \qquad
    \nu \equiv \sqrt{1-\frac{4}{9}\ncpl^2} \, .
\end{align}
Substituting \eqref{t-sol-a} into \eqref{gamma-prime}, we find
\begin{align}\label{gamma-prime-sol-a}
    \gamma'^\lambda_{\bf k}
    =
    - \frac{\ncpl {\cal H}}{a^{\frac{3}{2}}}
    \Big(
    C_2\, a^{\frac{3}{2}\nu} + C_3\, a^{-\frac{3}{2}\nu}
    \Big) \, ,
\end{align}
which after integration gives
\begin{align}\label{gamma-sol-a}
    \gamma_{\bf k}^\lambda
    &=
    C_1 + \frac{2\ncpl}{3 a^{\frac{3}{2}}}
    \left[
    \frac{C_2}{1-\nu} a^{\frac{3}{2}\nu}
    +
    \frac{C_3}{1+\nu} a^{-\frac{3}{2}\nu}
    \right] \, .
\end{align}

For $\ncpl\to0$ ($\nu\to1$), Eqs.~\eqref{EoM-gamma-SH} and \eqref{EoM-t-SH} have the solutions $\gamma_{\bf k}^\lambda\to\text{const.}$ and $t_{\bf k}^\lambda\to\text{const.}$ for $\tau\to0$. Taking carefully the limit $\tau\to0$ and $\ncpl\to0$ (or $\nu\to1$) in \eqref{gamma-sol-a} and \eqref{t-sol-a}, we find $\gamma_{\bf k}^\lambda\to C_1$ and $t_{\bf k}^\lambda\to C_2$ as expected. Thus, there is no enhancement in $\gamma^\lambda_{\bf k}$ even if $t^\lambda_{\bf k}$ has a large amplitude. In this case, the adiabatic normalization \eqref{gamma-t-DIH} gives $\gamma_{\bf k}^\lambda\to C_1\approx H_{\rm inf}/\sqrt{2k^3}$, and we approximately recover the usual amplitude of vacuum primordial GWs.

For $-3/2 < \ncpl < 3/2$ ($0< \nu \leq 1$), while relaxing $\ncpl\ll1$, we approximately have $\gamma_{\bf k}^\lambda\to C_1$ and $t_{\bf k}^\lambda\to0$. However, there is enhancement in $\gamma^\lambda_{\bf k}$ since the large amplitude of $t^\lambda_{\bf k}$ is transferred into $\gamma^\lambda_{\bf k}$ through the mixing. Note that the value of $\gamma_{\bf k}^\lambda\to C_1$ is different from $H_{\rm inf}/\sqrt{2k^3}$ and it should be determined by matching the superhorizon solution to the subhorizon one, for example around horizon crossing, $k\tau=-1$.

More interestingly, if $\ncpl\geq3/2$ or $\ncpl\leq-3/2$, then $\nu$ becomes purely imaginary, $\nu=i|\nu|$, where $|\nu|=\sqrt{4\ncpl^2/9-1}$. The $C_{2,3}$ terms in \eqref{gamma-sol-a} and \eqref{t-sol-a} oscillate with frequencies $\pm |\nu|H_{\rm inf}$ in cosmic time $t=\int a(\tau)\, d\tau$ and with damping amplitudes $a^{-3/2}$. Therefore, we have a constant solution $C_1$ with damped oscillations on top of it. These behaviors are shown for $\ncpl=5$ and some relevant values of the other couplings in Fig.~\ref{fig:gamma-t-time-f}, where we solved the system numerically. This in-time oscillation at superhorizon scales is intrinsic to the linear mixing alone: it is the consequence of the coupled superhorizon system being diagonalized into modes whose exponents become complex when $4\ncpl^2/9>1$, so that the propagation eigenstates evolve as $e^{\pm i|\nu|H_{\rm inf}t}$ with overall damping $a^{-3/2}$ in cosmic time, while GW observations are made in the $\gamma$-basis rather than in the diagonalized basis. In this precise sense, the in-time superhorizon oscillation is reminiscent of neutrino flavor oscillations~\cite{Caldwell:2016sut}, with $\gamma_{ij}$ and $t_{ij}$ playing the role of ``flavor'' eigenstates and the diagonalized modes the role of ``propagation'' eigenstates. We emphasize, however, that this analogy applies only to the superhorizon in-time oscillations identified here. The oscillations in $k$ that will be visible in the GW spectrum have a different origin, identified analytically in Sec.~\ref{sec-osc-mechanism}, and the neutrino analogy does not apply to them.

Let us also comment on the effects of the extra tensor modes on the CMB spectrum. The metric tensor modes that leave the horizon at CMB scales are also affected by the source provided by $t^\lambda_{\bf k}$ as long as $\ncpl\neq0$. However, we have assumed that $\tau_\ast$, shown in Fig.~\ref{fig:f}, occurs after the last CMB mode leaves the horizon during inflation, so that the amplitude of $t^\lambda_{\bf k}$ is small at CMB scales. Therefore, the CMB spectrum is not significantly affected by the extra tensor mode $t^\lambda_{\bf k}$ in our scenario. Of course, if one assumes that $\tau_\ast$ occurs during the period when CMB modes leave the horizon, then there can be an observable effect in the CMB B-mode polarization. We leave this possibility for future study.

\section{Power spectra}\label{sec-power-spectra}

We now present the quantization of the coupled tensor system and the corresponding formula for the power spectra. Since the metric tensor mode $\gamma^\lambda_{\bf k}$ and the extra tensor mode $t^\lambda_{\bf k}$ are linearly mixed during inflation, the system should be quantized as a coupled two-oscillator system. We therefore expand the field operators as \cite{Achucarro:2010da,Dimastrogiovanni:2023oid}
\begin{align}
	\hat{\gamma}_{\bf k}^\lambda(\tau)
	&=
	\sum_{I=1}^2
	\left(
	U_{\gamma I}^\lambda(\tau)\,\hat a_I^\lambda({\bf k})
	+
	U_{\gamma I}^{\lambda *}(\tau)\,\hat a_I^{\lambda\dagger}(-{\bf k})
	\right),
	\label{eq:gamma-operator-body}
	\\
	\hat{t}_{\bf k}^\lambda(\tau)
	&=
	\sum_{I=1}^2
	\left(
	U_{t I}^\lambda(\tau)\,\hat a_I^\lambda({\bf k})
	+
	U_{t I}^{\lambda *}(\tau)\,\hat a_I^{\lambda\dagger}(-{\bf k})
	\right),
	\label{eq:t-operator-body}
\end{align}
where the creation and annihilation operators satisfy
\begin{equation}\label{eq-a-ad}
	[\hat a_I^\lambda({\bf k}),\hat a_J^{\lambda'\dagger}({\bf q})]
	=
	(2\pi)^3 \delta_{IJ}\delta^{\lambda\lambda'}\delta^{(3)}({\bf k}-{\bf q}),
\end{equation}
with all other commutators vanishing.

The two pairs $(U_{\gamma 1}^\lambda,U_{t1}^\lambda)$ and $(U_{\gamma 2}^\lambda,U_{t2}^\lambda)$ are the two independent mode columns of the coupled system. Substituting \eqref{eq:gamma-operator-body} and \eqref{eq:t-operator-body} into the operator equations of motion, one finds that each column satisfies the same coupled classical equations,
\begin{align}
	U_{\gamma I}^{\lambda\prime\prime}
	+2\frac{a'}{a}U_{\gamma I}^{\lambda\prime}
	+k^2 U_{\gamma I}^{\lambda}
	&=
	-\ncpl \frac{a'}{a}
	\left[
	U_{tI}^{\lambda\prime}
	+
	\frac{(aa')'}{aa'}\,U_{tI}^{\lambda}
	\right]
	-\ncpl' \frac{a'}{a} U_{tI}^{\lambda},
	\label{eq:UgammaI-body}
	\\
	U_{tI}^{\lambda\prime\prime}
	+2\frac{(af)'}{af}U_{tI}^{\lambda\prime}
	+\left(c_t^2k^2+2\lambda \pvc k \frac{a'}{a}\right)U_{tI}^{\lambda}
	&=
	\frac{\ncpl}{f^2}\frac{a'}{a}U_{\gamma I}^{\lambda\prime},
	\label{eq:UtI-body}
\end{align}
for $I=1,2$.

\subsection{Inflationary era}

The initial conditions are imposed deep inside the horizon, where the mixing is negligible and the system approaches the adiabatic vacuum. In that regime, the two oscillator sectors can be chosen such that the first column starts in the $\gamma$-like positive-frequency mode
\begin{align}
	\begin{aligned}
		U_{\gamma 1}^\lambda(\tau_{\rm in})
		&=
		\frac{1}{a(\tau_{\rm in})} \frac{e^{-ik\tau_{\rm in}}}{\sqrt{2k}},
		&
		U_{\gamma 1}^{\lambda\prime}(\tau_{\rm in})
		&= -
		\left(ik+\frac{a'}{a}\right)_{\tau_{\rm in}}
		U_{\gamma 1}^\lambda(\tau_{\rm in}),
		\\
		U_{t1}^\lambda(\tau_{\rm in})&=0,
		&
		U_{t1}^{\lambda\prime}(\tau_{\rm in})&=0,
	\end{aligned}
\end{align}
and the second column starts in the $t$-like positive-frequency mode
\begin{align}
	\begin{aligned}
		U_{\gamma 2}^\lambda(\tau_{\rm in})&=0,
		&
		U_{\gamma 2}^{\lambda\prime}(\tau_{\rm in})&=0,
		\\
		U_{t2}^\lambda(\tau_{\rm in})
		&=
		\frac{1}{a(\tau_{\rm in})f(\tau_{\rm in})} \frac{e^{-i\int^{\tau_{\rm in}} c_t(\tilde\tau)k\,d\tilde\tau}}{\sqrt{2c_t(\tau_{\rm in})k}},
		&
		U_{t2}^{\lambda\prime}(\tau_{\rm in})
		&=
		-\left[
		i c_tk+\frac{(af)'}{af}+\frac{c_t'}{2c_t}
		\right]_{\tau_{\rm in}}
		U_{t2}^\lambda(\tau_{\rm in}),
	\end{aligned}
\end{align}
where the last line is the WKB initial condition at leading adiabatic order. In our setup, $f\to1$ and $c_t\to c_{t,0}$ at early time.

Once the two columns are evolved during inflation, the tensor power spectrum for a given helicity is obtained by summing the contribution of the two oscillator sectors,
\begin{equation}
	{\cal P}_{h}^{\lambda}(k,\tau)
	=
	\frac{2k^3}{\pi^2 \Mpl^2}
	\sum_{I=1}^2
	\left|U_{\gamma I}^{\lambda}(\tau)\right|^2.
	\label{eq:Ph-inflation-body}
\end{equation}
The total tensor power spectrum is then
\begin{equation}
	{\cal P}_{h}(k,\tau)
	=
	\sum_{\lambda}{\cal P}_{h}^{\lambda}(k,\tau).
	\label{eq:Ph-total-body}
\end{equation}
This is the correct generalization of the usual single-field formula to the coupled case: one must sum over the two independent mode columns. Equivalently, the variance of the metric tensor mode is
$\langle 0|\hat\gamma^\lambda_{\bf k}(\tau)\hat\gamma^{\lambda\dagger}_{\bf k}(\tau)|0\rangle= \sum_{I=1}^2\left|U_{\gamma I}^{\lambda}(\tau)\right|^2$.

In the figures below we plot the dimensionless normalized spectrum
\begin{equation}
	\widehat{\cal P}_{h}^{\lambda}(k,\tau)
	\equiv
	\left(\frac{\pi\Mpl}{H_{\rm inf}}\right)^2 {\cal P}_{h}^{\lambda}(k,\tau)
	=
	\sum_{I=1}^2\left|\tilde U_{\gamma I}^{\lambda}(x)\right|^2,
	\qquad x\equiv k\tau,
\end{equation}
where $\tilde U_{\gamma I}^{\lambda}\equiv(\sqrt{2}k^{3/2}/H_{\rm inf})U_{\gamma I}^{\lambda}$. Thus the plotted quantity differs from ${\cal P}_h^\lambda$ only by the constant factor $(H_{\rm inf}/\pi\Mpl)^2$.

\begin{figure}[!ht]
	\begin{subfigure}[b]{0.5\linewidth}
		\centering
		\includegraphics[width=1\linewidth]{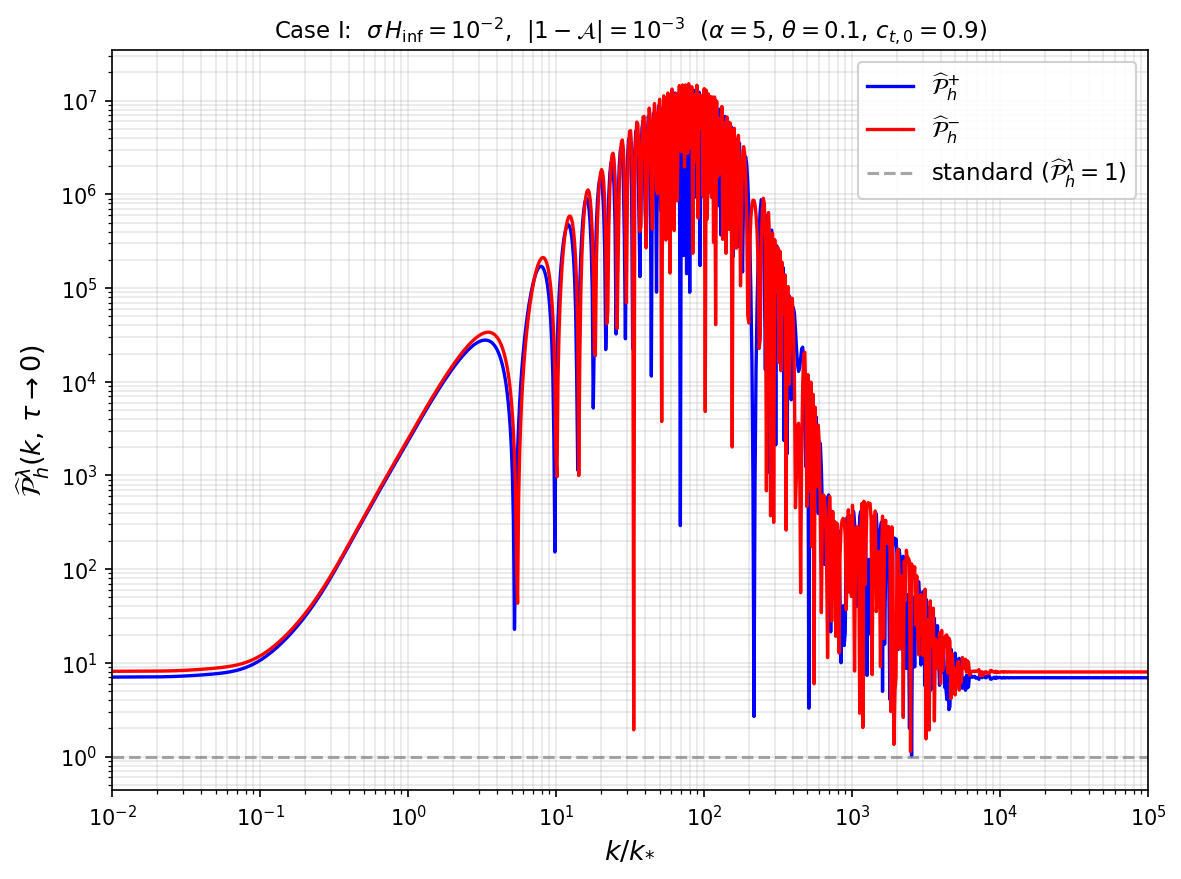}
	\end{subfigure}
	\begin{subfigure}[b]{0.5\linewidth}
		\centering
		\includegraphics[width=1\linewidth]{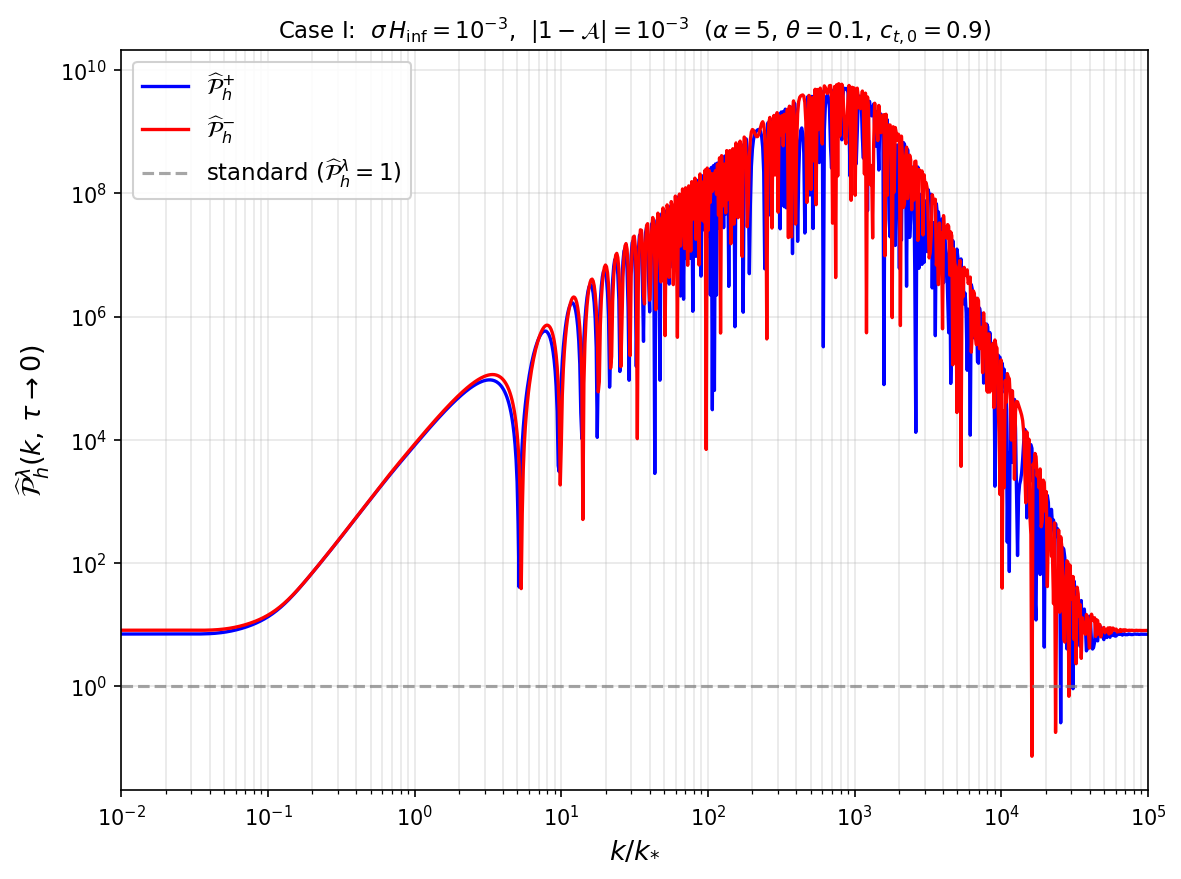}
	\end{subfigure}
	\caption{Normalized power spectra of GWs, $\widehat{\cal P}_h^\lambda=\sum_I|\tilde U_{\gamma I}^\lambda|^2$, as a function of scale at the end of \textit{inflation} $\tau\to0$ for Case I.}
	\label{fig:gamma-PS-f-inf}
\end{figure}

\begin{figure}[!ht]
	\begin{subfigure}[b]{0.5\linewidth}
		\centering
		\includegraphics[width=1\linewidth]{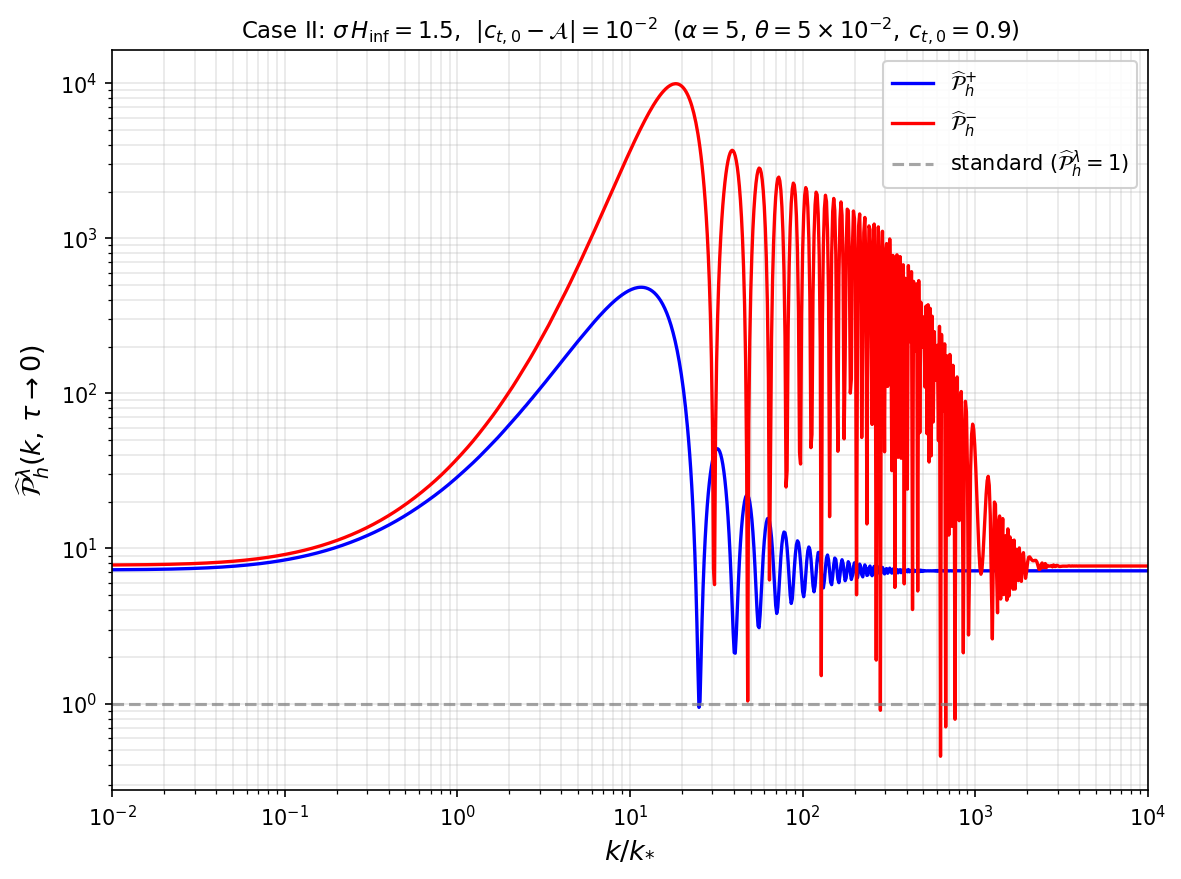}
	\end{subfigure}
	\begin{subfigure}[b]{0.5\linewidth}
		\centering
		\includegraphics[width=1\linewidth]{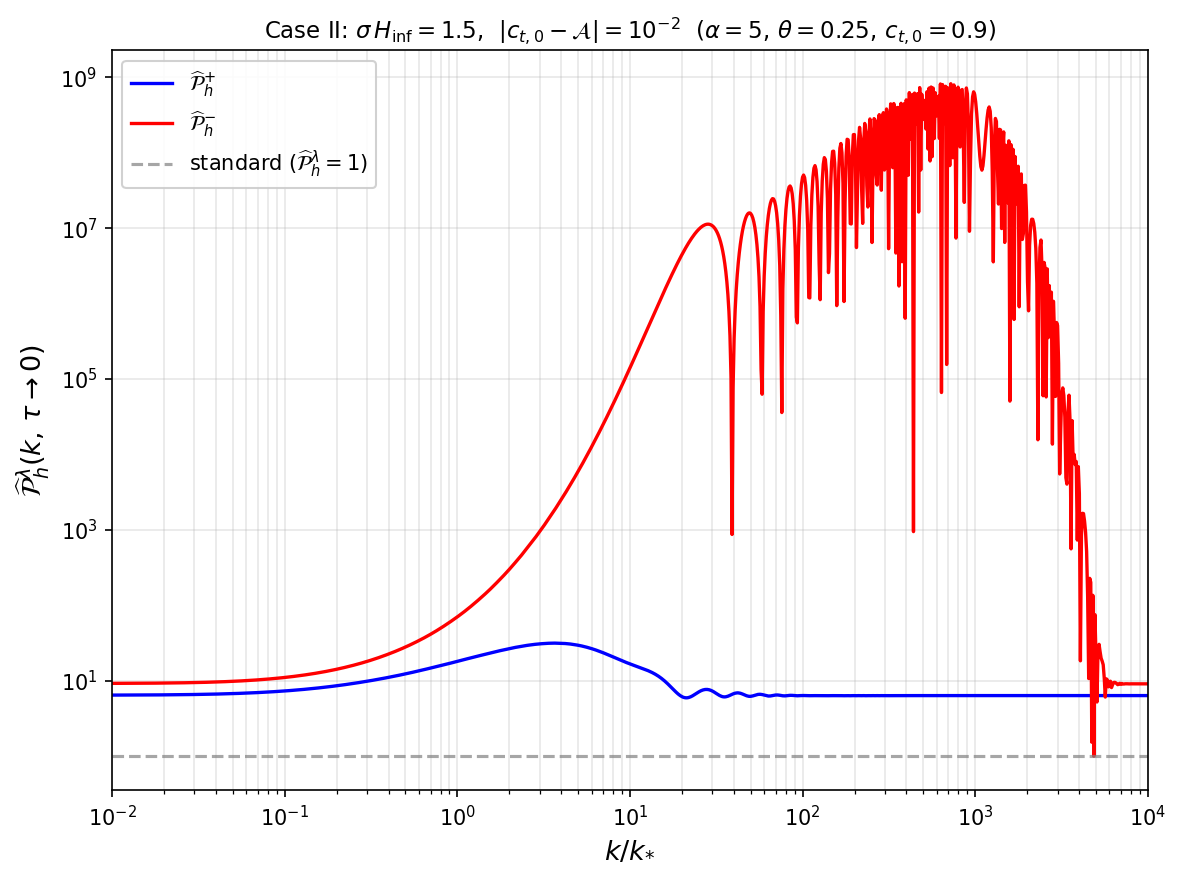}
	\end{subfigure}
	\caption{Normalized power spectra of GWs, $\widehat{\cal P}_h^\lambda=\sum_I|\tilde U_{\gamma I}^\lambda|^2$, as a function of scale at the end of \textit{inflation} $\tau\to0$ for Case II.}
	\label{fig:gamma-PS-ct-inf}
\end{figure}

In Figs.~\ref{fig:gamma-PS-f-inf} and \ref{fig:gamma-PS-ct-inf} we show the helicity-resolved power spectra of $\gamma^\lambda_{\bf k}$, computed from \eqref{eq:Ph-inflation-body}, evaluated at the end of inflation $\tau\to 0^-$, i.e. while the linear mixing $\ncpl$ is still on. Both Cases already display pronounced oscillations in $k$ for $k\gtrsim k_\ast$. The difference between the two Cases can be understood from the effective frequency \eqref{omega-t-full-inf} of the canonically normalized $t$-mode.

In Case~I the dip is in $f$. The term $-(af)''/(af)$ can become negative enough during the dip to violate adiabaticity, and even to make $\omega_t^2$ negative for part of the evolution. Thus a large enhancement of the GW spectrum in Case~I does not require a large parity-violating parameter $\pvc$: the non-adiabatic $f$-term itself can provide the enhancement. Parity violation remains an important possible signature, but it is not required in order to obtain a large amplitude. This is why, for the values used in Fig.~\ref{fig:gamma-PS-f-inf}, $\pvc=10^{-1}$ and $c_{t,0}=0.9$, the two helicities are close to each other even though the spectrum is strongly enhanced.

In Case~II the dip is in $c_t$. The term $c_t^2k^2$ is always positive, so a dip in $c_t$ by itself does not make $\omega_t^2$ negative. Also, a small $c_t$ should not be interpreted as giving a huge enhancement only from the WKB normalization $|t|\propto c_t^{-1/2}$; the actual enhancement depends on adiabaticity and on Bogoliubov production. Numerically, in the examples we studied, Case~II without appreciable parity violation can still give a moderate enhancement, up to roughly four orders of magnitude in the power spectra. A much larger observable enhancement in Case~II instead requires appreciable $\pvc$, because the helicity-dependent term in \eqref{omega-t-full-inf} can drive one helicity tachyonic when the sound speed is small.

The approximate tachyonic momentum range near the bottom of the sound-speed dip is given by \eqref{ktach-inflation}, with $c_{t,\min}$ the minimum sound speed. With the values used in Fig.~\ref{fig:gamma-PS-ct-inf}, $c_{t,\min}=10^{-2}$ and $\pvc=5\times10^{-2}$ or $0.25$, the scale $k_{\rm tach}$ is much larger than $k_\ast$. The resulting $t^\lambda$ field is therefore strongly chiral, and the chirality is transferred to $\gamma^\lambda$ through the source on the right-hand side of \eqref{eq:UgammaI-body}. The scale $k_{\rm tach}$ characterizes the tachyonic part of the enhancement; it does not by itself determine the oscillatory structure of the spectrum. The latter is controlled by the localized time dependence of the feature and is discussed next. Within this class of features, where the large Case~II signal is produced by a dip in $c_t$ together with parity violation, the need for appreciable parity violation is robust, although the exact amplitude, peak position, and extent of the oscillatory band remain model dependent.

\subsection{Origin of the in-$k$ oscillations}\label{sec-osc-mechanism}

The end-of-inflation power spectra of $\gamma^\lambda_{\bf k}$ shown in Figs.~\ref{fig:gamma-PS-f-inf} and \ref{fig:gamma-PS-ct-inf} exhibit pronounced oscillations as a function of $k$ for $k\sim k_\ast$. In this subsection we identify the underlying mechanism analytically and quote the resulting fringe period and envelope. The complete derivation, including the explicit form of the relevant quantities in both Cases and the identification of the phases, is presented in Appendix~\ref{app-osc-mechanism}.

Two ingredients are independently necessary for the in-$k$ fringes to appear: a localized feature in $f$ or $c_t$ during inflation, and a non-zero linear mixing $\ncpl$ between $\gamma^\lambda_{\bf k}$ and $t^\lambda_{\bf k}$. The first is needed because, as noted above, with all EFT couplings constant the inflationary system is exactly scale invariant under $k\tau\to x$, so that the Gaussian dip~\eqref{f} is the source of any $k$-dependence. The second is needed because, in the absence of mixing, the dip affects only $t^\lambda_{\bf k}$, which is not directly observed. Neither ingredient alone is sufficient.

For modes that are sub-horizon during the dip, the latter acts as a non-adiabatic event for the canonically normalized field $T^\lambda_{\bf k}\equiv af\,t^\lambda_{\bf k}$. It induces a deviation $\delta\omega_t^2(\tau)$ in the squared WKB frequency of $T^\lambda_{\bf k}$, centered around $\tau\sim\tau_\ast$ and computed explicitly in both Cases in Appendix~\ref{app-osc-mechanism}. This deviation excites a small admixture of the negative-frequency WKB branch of $T^\lambda_{\bf k}$, parametrized by a Bogoliubov coefficient $\beta_k$. In the perturbative regime this coefficient is given by the integral of $\delta\omega_t^2(\tau)$ against $e^{-2ic_{t,0}k\tau}$, see Eq.~\eqref{app-beta-formula}. After the dip, $|T^\lambda_{\bf k}|^2$ contains an interference cross-term that oscillates as $|\beta_k|\cos(2c_{t,0}k/k_\ast+\varphi_0)$, where $\varphi_0$ is a $k$-independent phase at leading WKB order. The mixing $\ncpl$ then transmits this oscillating amplitude to the metric channel via Eq.~\eqref{EoM-gamma-inf}, so that the helicity-resolved GW power spectrum acquires the leading template
\begin{equation}\label{Ph-analytical-body}
	\mathcal{P}_h^\lambda(k)\,\simeq\,\mathcal{P}_{h,{\rm smooth}}^\lambda(k)\,\left[\,1\,+\,|\beta_k|^2\,+\,2|\beta_k|\cos\!\left(\frac{2c_{t,0}k}{k_\ast}\,+\,\varphi_0\right)\,\right] \, .
\end{equation}
Here $\mathcal{P}_{h,{\rm smooth}}^\lambda(k)$ is the smooth peaked envelope that one would obtain in the strictly adiabatic limit $\beta_k\to 0$ with the dip in $f$ or $c_t$ still in place. It is not the spectrum obtained by removing the dip entirely, which would be flat in $k$ by scale invariance. In practice, $\mathcal{P}_{h,{\rm smooth}}^\lambda(k)$ can be obtained by averaging the full $\mathcal{P}_h^\lambda(k)$ over one fringe period, which removes the cosine cross-term; a precise operational definition suited to the deep-dip regime is given in Eq.~\eqref{Psmooth-def-app} of Appendix~\ref{app-osc-mechanism}. At leading WKB order the bracketed fringe factor in Eq.~\eqref{Ph-analytical-body} is helicity independent: the parity-violating coupling $\pvc$ is taken constant in time and so does not enter the localized non-adiabatic kick $\delta\omega_t^2$ that produces $\beta_k$. The dominant helicity dependence is then contained in the smooth envelope $\mathcal{P}_{h,{\rm smooth}}^\lambda(k)$, while subleading WKB corrections can shift the phase of the fringes in a helicity-dependent way. In the strongly chiral Case~II examples, the same smoothed envelope remains a useful way to display the numerical spectra, but the simple bracketed expression should be regarded as an analytic guide to the feature-induced fringes rather than a complete description of the tachyonic regime.
\begin{figure}[!ht]
	\centering
	\includegraphics[width=0.95\linewidth]{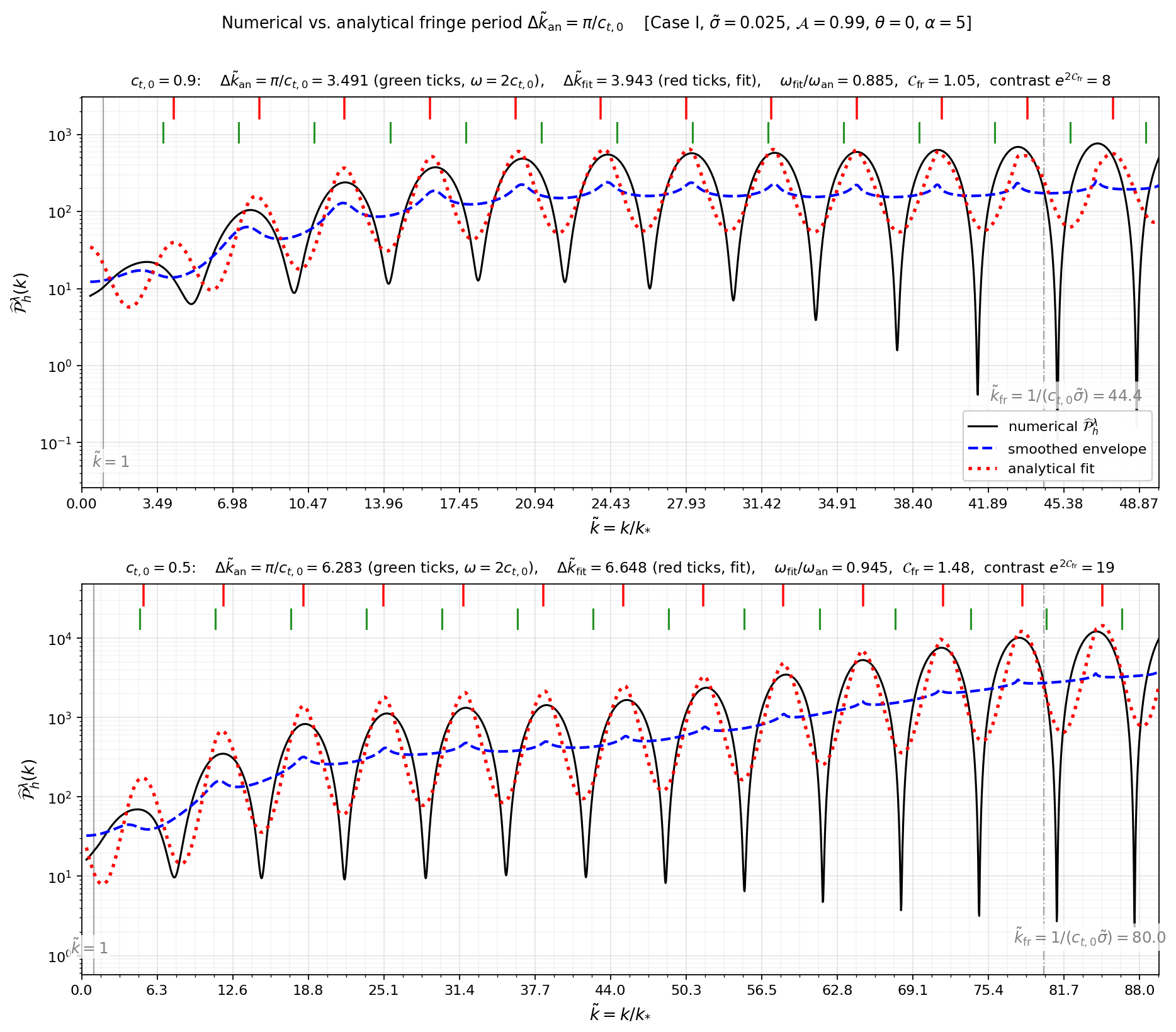}
	\caption{Comparison of the full numerical normalized spectrum $\widehat{\mathcal P}_h^\lambda(k)=\sum_I|\tilde U_{\gamma I}^\lambda|^2$ (black solid) with the log-space fitted template~\eqref{Ph-fit-log-body} (red dotted) and the smoothed envelope $\widehat{\mathcal P}_{h,{\rm smooth}}^\lambda(k)$ defined in~\eqref{Psmooth-def-app} (blue dashed), at the end of \textit{inflation} $\tau\to 0$ for Case I. We considered two values of $c_{t,0}=0.9$ (top) and $c_{t,0}=0.5$ (bottom). The red tick marks at the top of each panel are peak positions of the fitted residual cosine with frequency $\omega_{\rm fit}$; the green tick marks use the leading-WKB analytical frequency $\omega_{\rm an}=2c_{t,0}$, equivalently the spacing $\Delta k=\pi k_\ast/c_{t,0}$ from~\eqref{Delta-k}, with the same fitted phase. The vertical gray lines mark the lower and upper edges $k_\ast$ and $k_{\rm fr}$ of the excited window~\eqref{window-body}. The symbol $\mathcal C_{\rm fr}$ printed in the panel titles denotes the fitted log-space contrast defined in Eq.~\eqref{Ph-fit-log-body}.}
	\label{fig:fringe-comparison}
\end{figure}

Reading off the period of the cosine in $k$, we obtain the universal leading oscillation period in $k$-space (fringe spacing)
\begin{equation}\label{Delta-k}
	\Delta k\simeq\frac{\pi\,k_\ast}{c_{t,0}}\,,\qquad k_\ast\equiv-\frac{1}{\tau_\ast} \qquad\mbox{Case I and Case II}\,.
\end{equation}
This follows from the leading WKB phase $\Theta(\tau)\simeq c_{t,0}k\tau$ evaluated at the feature time $\tau=\tau_\ast$. The constant $\varphi_0$ only shifts where the fringes sit on the $k$-axis without changing their spacing.

The Bogoliubov coefficient $\beta_k$ in Eq.~\eqref{Ph-analytical-body} is given by the integral~\eqref{app-beta-formula} or equivalently \eqref{beta-u-integral-app}, which cannot be performed analytically in general. We therefore treat the narrow-dip regime $\tilde\sigma\ll 1$ and the broad-dip regime $\tilde\sigma\gtrsim 1$ separately in Appendix~\ref{app-osc-mechanism}. Below we present the results and explain the physics in each case.

For the narrow-dip regime $\tilde\sigma\ll1$, \eqref{beta-u-integral-app} reduces to the Fourier integral of the Gaussian profile chosen in Eq.~\eqref{f} which can be evaluated through a saddle-point which gives the leading envelope (for the detail of the derivation, see Appendix~\ref{app-osc-mechanism})
\begin{equation}\label{beta-envelope}
	\frac{|\beta_k|}{|\beta_{k_\ast}|}\simeq \exp\!\left[-\tfrac{1}{2}\bigl(2c_{t,0}\,\tilde\sigma\bigr)^2\!\left((k/k_\ast)^2-1\right)\right] \, ,
\end{equation}
up to slowly varying prefactors in $k$. A sharper dip (smaller $\tilde\sigma$) therefore yields a broader fringe envelope in $k$, in agreement with the numerical spectra. The Gaussian form of~\eqref{beta-envelope} is a direct consequence of the Gaussian ansatz~\eqref{f}: a different functional form for the feature would give a different envelope shape. The reason the same leading formula applies in both Cases~I and~II is that the same Gaussian ansatz is used in both, with the same temporal-width parameter $\tilde\sigma$. The fringes therefore live in a finite window of comoving momenta,
\begin{equation}\label{window-body}
	k_\ast\,\lesssim\,k\,\lesssim\,k_{\rm fr}\equiv\frac{k_\ast}{c_{t,0}\,\tilde\sigma} \qquad (\tilde\sigma\ll 1)\, ,
\end{equation}
whose two edges have distinct physical origins. The lower edge is set by the sound horizon at the time of the dip: modes with $c_{t,0}k|\tau_\ast|\lesssim 1$ have already exited the horizon when the dip switches on, are frozen on superhorizon scales, and are not captured by the subhorizon Bogoliubov picture; they retain the nearly scale-invariant tensor amplitude characteristic of the unmixed vacuum. The upper edge $k_{\rm fr}$ is set by the temporal width of the feature: the relevant comparison is between the temporal period $\sim 1/(c_{t,0}k)$ of the mode and the duration $|\Delta\tau_{\rm dip}|\sim \tilde\sigma|\tau_\ast|$ of the dip. Slow modes $k\lesssim k_{\rm fr}$ oscillate less than one cycle through the dip and cannot average out the perturbation, so they are non-adiabatic and acquire an $\mathcal{O}(1)$ $\beta_k$. Fast modes $k\gg k_{\rm fr}$ oscillate many times during the dip, adiabatically track the instantaneous $\omega_t(\tau)$, and exit with exponentially suppressed $|\beta_k|$. Note that this is the opposite of the more familiar resolution argument for spatial features: for a temporal feature it is the long-period modes that fail to average out the perturbation. The width of the window $k_{\rm fr}/k_\ast=1/(c_{t,0}\tilde\sigma)$ controls the number of visible fringes, in agreement with the spectra where narrower $\tilde\sigma$ yields a broader excited window with more oscillations.\footnote{In the narrow-dip and perturbative WKB regime, where the localized change of the squared frequency can be treated as a small scattering potential around $c_{t,0}k$, Appendix~\ref{app-osc-mechanism} gives the parametric estimates
\begin{equation}\label{betak-summary}
|\beta_{k_\ast}|_{\rm I}^{\rm pert}\sim
\frac{{\cal A}}{c_{t,0}\,(1-{\cal A})\,\tilde\sigma}\,e^{-2(c_{t,0}\tilde\sigma)^2},
\qquad
|\beta_{k_\ast}|_{\rm II}^{\rm pert}\sim
\frac{{\cal A}\,(2c_{t,0}-{\cal A})}{c_{t,0}}\,\tilde\sigma\,e^{-2(c_{t,0}\tilde\sigma)^2},
\end{equation}
up to order-one factors. They break down in the deep-dip regime where $f_{\min}=1-{\cal A}$ or $c_{t,\min}=c_{t,0}-{\cal A}$ becomes very small, in which case the numerical integration of the full coupled system is required. They make explicit, however, that Case~I is especially sensitive to the sharpness of the dip through derivatives of $f$, while Case~II is especially sensitive to the small sound speed through the violation of the sound-horizon WKB conditions~\eqref{adiabatic-condition-case-II-bottom}-\eqref{adiabatic-condition-case-II-slope} and through the smooth WKB amplitude $|t_{\bf k}|\propto c_t^{-1/2}$.}

For the broad-dip regime $\tilde\sigma\gtrsim 1$, the dip is no longer localized in conformal time near $\tau_\ast$ but covers many $e$-folds, extending in $\tau$ from $\tau_\ast e^{+\tilde\sigma}$ to $\tau_\ast e^{-\tilde\sigma}$ (recall $\tau_\ast<0$), corresponding to a duration $\Delta{\cal N}=2\tilde\sigma$. The narrow-dip approximation used to derive~\eqref{beta-envelope}, which treats the temporal period of the mode as constant across the Gaussian profile of the dip, then fails. A simple estimate for the oscillatory support in this case is obtained by asking whether the mode is still inside the sound horizon while the dip is active. This gives
\begin{equation}\label{window-broad-body}
	k_\ast\,\lesssim\,k\,\lesssim\,k_{\rm fr}\equiv\frac{k_\ast\,e^{+\tilde\sigma}}{c_{t,0}}
	\qquad (\tilde\sigma\gtrsim 1)\, ,
\end{equation}
with width factor $\sim e^{+\tilde\sigma}/c_{t,0}$ rather than $\sim 1/(c_{t,0}\tilde\sigma)$. The two formulas~\eqref{window-body} and~\eqref{window-broad-body} share the lower edge $k_\ast$ and cross over around $\tilde\sigma\sim 1$ at the level of an order-of-magnitude estimate. For Case I, we use $\tilde\sigma\sim 10^{-2}$, which is in the narrow-dip regime~\eqref{window-body}, and the analytical envelope~\eqref{beta-envelope} is accurate. For Case~II, we use $\tilde\sigma\sim 1$--$2$, which is in or near the broad-dip regime; there the oscillatory window is described only parametrically by~\eqref{window-broad-body} when the tachyonic amplification is not the dominant effect.

The windows~\eqref{window-body} and \eqref{window-broad-body} characterize the range over which the cosine modulation in Eq.~\eqref{Ph-analytical-body} has appreciable support. The smooth amplitude enhancement encoded in $\mathcal{P}_{h,\rm smooth}^\lambda(k)$ is a separate effect, controlled by the depth of the dip and by the transfer from the spin--2 sector to the metric tensor. Thus the enhancement scale and the fringe envelope need not coincide. In the strongly chiral Case~II spectra there is also the tachyonic scale $k_{\rm tach}$ introduced in \eqref{ktach-inflation}. For the parameters of Fig.~\ref{fig:gamma-PS-ct-inf}, one finds $k_{\rm tach}\gg k_{\rm fr}$. A purely analytic comparison of these scales would not determine the full range of visible oscillations, because the derivation of \eqref{Ph-analytical-body} assumes a small perturbation $\delta\omega_t^2$ and does not apply once one helicity becomes significantly tachyonic during the dip. The numerical solution shows that the oscillatory pattern persists across the full enhanced band in this case. This is therefore a genuine feature of the coupled evolution, rather than a direct consequence of the perturbative window estimates \eqref{window-body} and \eqref{window-broad-body}.

In summary, the in-$k$ oscillations of the GW spectrum are Bogoliubov-interference fringes seeded by a non-adiabatic event in the $t$-sector around $\tau\sim\tau_\ast$ and transmitted to the metric channel by $\ncpl$. The leading period is set by the temporal location of the feature. The envelope is set by the temporal width and shape of the feature. The peak amplitude is the model-dependent piece, and tracks the violation of the relevant adiabaticity conditions.

To make the agreement with these analytical formulas quantitative at the fringe-by-fringe level, in Fig.~\ref{fig:fringe-comparison} we compare the full numerical normalized spectrum $\widehat{\mathcal P}_h^\lambda(k)$ with a log-space implementation of the fringe template and with the smoothed envelope defined precisely in Eq.~\eqref{Psmooth-def-app} of Appendix~\ref{app-osc-mechanism}. The parity-violating coupling is switched off, $\pvc=0$, so that both helicities give the same spectrum and any in-$k$ oscillation comes purely from the Bogoliubov interference described above. Because the plotted examples are in the deep-dip regime and have very deep nulls, the actual fitting curve used for the figure is
\begin{equation}\label{Ph-fit-log-body}
	\widehat{\mathcal P}_{h,{\rm fit}}^\lambda(k)
	=
	\widehat{\mathcal P}_{h,{\rm smooth}}^\lambda(k)
	\exp\!\left[\mathcal C_{\rm fr}
	\cos\!\left(\omega_{\rm fit}\frac{k}{k_\ast}+\varphi_{\rm fit}\right)\right] ,
\end{equation}
where $\mathcal C_{\rm fr}$ is a fitted log-space fringe contrast. This phenomenological log-space fit tests the fringe period and phase of Eq.~\eqref{Ph-analytical-body}; it should not be interpreted as an independent analytic prediction for the deep-dip amplitude. The analytical leading value corresponds to $\omega_{\rm an}=2c_{t,0}$, or $\Delta k=\pi k_\ast/c_{t,0}$. The log-space template tracks the numerical curve fringe-by-fringe across the excited window~\eqref{window-body}, including the deep nulls between successive peaks. The ratio of fitted fringe spacings between the two panels, $\Delta k(c_{t,0}=0.5)/\Delta k(c_{t,0}=0.9)\simeq 1.7$, agrees with the analytical scaling $\Delta k\propto 1/c_{t,0}$ from~\eqref{Delta-k} (predicting $0.9/0.5=1.8$) up to a small correction. This residual mismatch is consistent with subleading WKB phase corrections, and it decreases as the temporal width of the feature is reduced.

\subsection{Radiation era}

As we have mentioned above, the couplings $f, \ncpl, \pvc$ and the sound speed $c_t$ are functions of time through their dependence on the inflaton field which decays to the Standard Model particles at the end of inflation. For the sake of simplicity, we assume that the interaction terms vanish at this time such that $f(\tau_{\rm end})=1$, $c_t(\tau_{\rm end})=c_{t,0}$, $\ncpl(\tau_{\rm end})=0$, and $\pvc(\tau_{\rm end})=0$. Thus,
\begin{align}
	\ncpl = 0 \, , \qquad
	\mbox{during radiation domination} \, ,
\end{align}
and the metric tensor mode decouples from the extra tensor mode and satisfies the standard equation
\begin{equation}
	U_{\gamma I}^{\lambda\prime\prime}
	+2\frac{a'}{a}U_{\gamma I}^{\lambda\prime}
	+k^2 U_{\gamma I}^{\lambda}
	=0,
	\qquad
	\mbox{during radiation domination}.
	\label{eq:UgammaI-RD-body}
\end{equation}
The two independent columns are still needed, but now they evolve independently with the same standard tensor equation.

To evolve the system into radiation domination, one uses the values of $U_{\gamma I}^{\lambda}$ and $U_{\gamma I}^{\lambda\prime}$ obtained at the end of inflation as the initial conditions for the radiation era. More precisely, if $\tau=\tau_{\rm end}$ denotes the matching time, then for each $I=1,2$ one imposes continuity of the mode and its first derivative,
\begin{align}
	U_{\gamma I}^{\lambda,{\rm RD}}(\tau_{\rm end})
	&=
	U_{\gamma I}^{\lambda,{\rm inf}}(\tau_{\rm end}),
	\\
	U_{\gamma I}^{\lambda,{\rm RD}\prime}(\tau_{\rm end})
	&=
	U_{\gamma I}^{\lambda,{\rm inf}\prime}(\tau_{\rm end}).
	\label{eq:matching-Ugamma-body}
\end{align}
In this way, the superhorizon solution during radiation domination is fully determined by the inflationary evolution.

Therefore, the helicity-resolved power spectrum during radiation domination is
\begin{equation}
	{\cal P}_{h}^{\lambda,{\rm RD}}(k,\tau)
	=
	\frac{2k^3}{\pi^2 \Mpl^2}
	\sum_{I=1}^2
	\left|U_{\gamma I}^{\lambda,{\rm RD}}(\tau)\right|^2.
	\label{eq:Ph-RD-body}
\end{equation}
In particular, for superhorizon modes during radiation domination, one evaluates \eqref{eq:Ph-RD-body} at a time such that $k\tau\ll1$. This is the quantity that we use as the initial tensor power spectrum before horizon re-entry. After that, one may continue the standard radiation-era evolution up to horizon crossing and later times.

In summary, the computation of the tensor power spectrum proceeds as follows: one imposes adiabatic initial conditions deep inside the horizon during inflation for the two mode columns, evolves the coupled system \eqref{eq:UgammaI-body}-\eqref{eq:UtI-body}, computes the inflationary spectrum from \eqref{eq:Ph-inflation-body}, then matches the $\gamma$-modes to the decoupled radiation-era equation \eqref{eq:UgammaI-RD-body}, and finally computes the radiation-era spectrum from \eqref{eq:Ph-RD-body}. The same procedure is implemented numerically in our analysis. The numerical method is explained in more detail in Appendix~\ref{app-numerical-setup}.

\begin{figure}[!ht]
	\begin{subfigure}[b]{0.5\linewidth}
		\centering
		\includegraphics[width=1\linewidth]{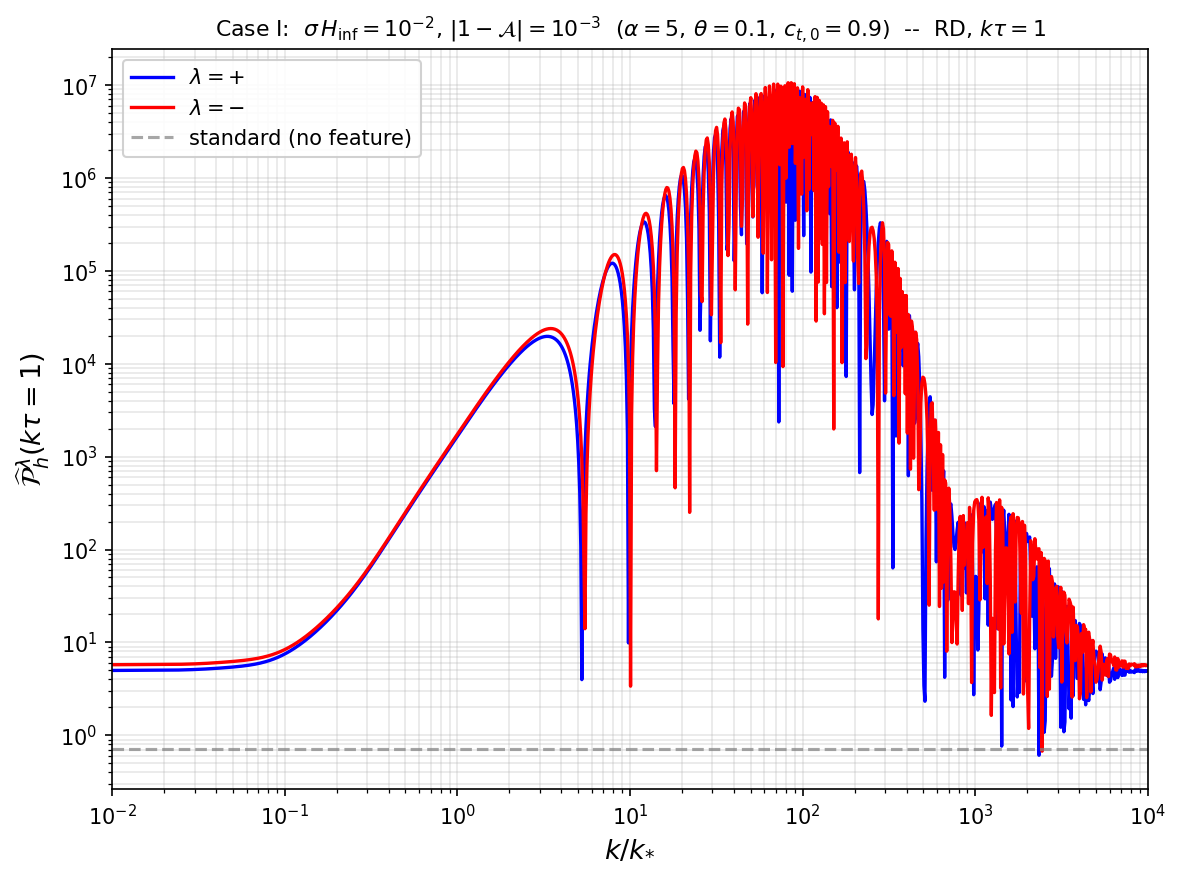}
	\end{subfigure}
	\begin{subfigure}[b]{0.5\linewidth}
		\centering
		\includegraphics[width=1\linewidth]{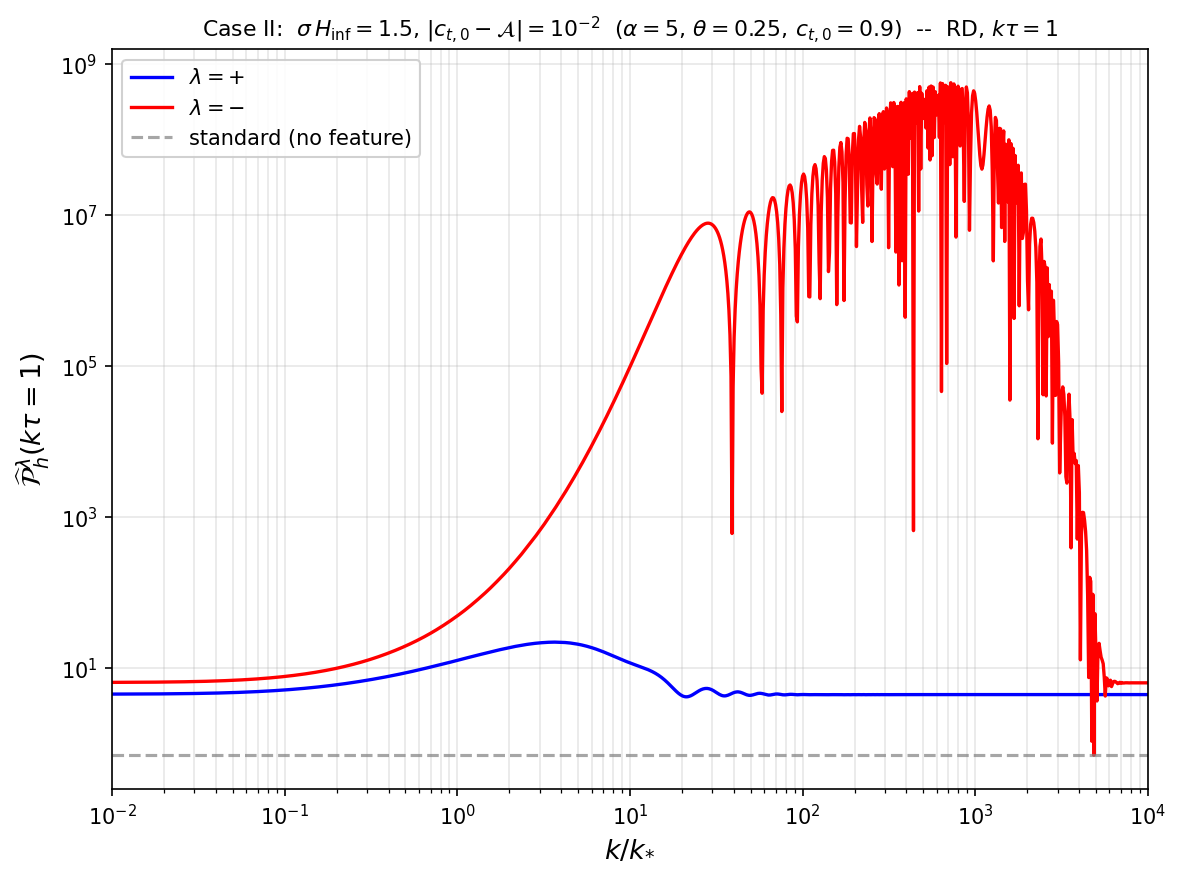}
	\end{subfigure}
	\caption{Normalized power spectra of GWs as a function of scale at the time of horizon re-entry $k\tau=1$ during the \textit{radiation era} for Case I and Case II.}
	\label{fig:gamma-PS-f}
\end{figure}

In Fig.~\ref{fig:gamma-PS-f} we show the same helicity-resolved power spectra, evaluated this time at the time of horizon re-entry $k\tau=1$ during the radiation era, for Case I and Case II respectively. We have used the values of $U_{\gamma I}^{\lambda}(\tau)$ at the late inflationary time $\tau\to 0^-$ as initial conditions to solve the decoupled radiation-era equation~\eqref{eq:UgammaI-RD-body} during the radiation era (see Appendix~\ref{app-numerical-setup} for the details). In both Cases the power spectra have peaks around $k_\ast$ and exhibit clear oscillations in $k$ for $k\gtrsim k_\ast$, with the same fringe period as at the end of inflation. Moreover, depending on the value of $\pvc$, different polarizations behave differently.

Comparing Fig.~\ref{fig:gamma-PS-f} with the corresponding end-of-inflation spectra in Figs.~\ref{fig:gamma-PS-f-inf} and \ref{fig:gamma-PS-ct-inf} confirms two important points. First, the fringe pattern is essentially unchanged across the transition $\ncpl\neq 0\to\ncpl=0$ at $\tau=\tau_{\rm end}$. The transition is continuous in the mode functions and their first derivatives, see Eq.~\eqref{eq:matching-Ugamma-body}, and a constant superhorizon mode remains constant across it; the same in-$k$ structure that was present at the end of inflation is therefore preserved in the radiation era. The switching-off of $\ncpl$ neither produces nor erases the oscillations; it merely decouples the $\gamma$- and $t$-sectors after the imprint has already been transferred. Second, the overall normalization of the radiation-era spectra in Fig.~\ref{fig:gamma-PS-f} relative to those at the end of inflation reflects the standard radiation-era tensor evolution from $\tau\to 0^-$ to horizon re-entry, with the spectral shape essentially preserved.

The origin of the in-$k$ oscillations is the Bogoliubov-interference mechanism described in Sec.~\ref{sec-osc-mechanism}. For the values used in our figures, $c_{t,0}=0.9$, the leading formula predicts $\Delta k/k_\ast\simeq\pi/0.9\approx3.5$, in agreement with what is seen numerically in both the inflationary and radiation-era spectra. We stress that this fringe pattern is already present in the $\gamma$-spectrum at the end of inflation, while $\ncpl\neq0$, as Figs.~\ref{fig:gamma-PS-f-inf} and \ref{fig:gamma-PS-ct-inf} make explicit. The in-$k$ oscillations are also a different effect from the in-time superhorizon oscillation of Sec.~\ref{sec-modes-far-outside-horizon}, to which the neutrino-oscillation analogy applies; the analogy does not apply to the in-$k$ fringes.

Moreover, we have numerically confirmed that the amplitude of the peak of the power spectrum is very sensitive to $\sigma$ and $|c_{t,0}-{\cal A}|$ for Case I and Case II, respectively: the smaller these quantities, the larger the enhancement. In Case I this is tied to the derivatives of the dip in $f$ and to the violation of the WKB condition~\eqref{adiabatic-condition-case-I}. In Case II, the sensitivity to $|c_{t,0}-{\cal A}|$ arises because the minimum sound speed $c_{t,\min}=c_{t,0}-{\cal A}$ controls both the WKB amplitude of the $t$-mode, $|t|\propto c_t^{-1/2}$ in \eqref{gamma-t-DIH}, and the validity of the sound-horizon WKB conditions~\eqref{adiabatic-condition-case-II-bottom}-\eqref{adiabatic-condition-case-II-slope}. For the large Case~II signal shown in Fig.~\ref{fig:gamma-PS-f}, however, the dominant effect is the parity-violating tachyonic growth of one helicity during the dip, not the normalization factor alone. The oscillations remain visible over the full enhanced Case~II band in the numerical solution, while Sec.~\ref{sec-osc-mechanism} explains the robust origin of their period and gives the corresponding feature-width estimate in the regime where the WKB scattering picture is applicable.

The radiation-era spectra shown in this subsection are also the direct input for the stochastic-background calculation below: the same two $\gamma$-mode columns that determine ${\cal P}_h^\lambda$ at horizon re-entry enter the energy-density spectrum $\Omega_{\rm GW}^\lambda$.

\section{Stochastic gravitational wave background}\label{sec-SGWs}

Having found the power spectra at superhorizon scales and at horizon re-entry, let us now look at the spectral dimensionless energy density of GWs, defined as
\begin{align}\label{Omega-def-total}
    \Omega_{\rm GW}(k,\tau) &= \sum_\lambda \Omega^\lambda_{\rm GW}(k,\tau) \, .
\end{align}
In the coupled system, one must sum over the two oscillator sectors. Thus, the helicity-resolved GW energy density is
\begin{align}\label{Omega-def}
    \Omega^\lambda_{\rm GW}(k,\tau)
    &=
    \frac{k^3}{12\pi^2\Mpl^2a^2H^2}
    \sum_{I=1}^2
    \left[
    \big| U_{\gamma I}^{\lambda\,\prime}(k,\tau) \big|^2
    +
    k^2 \big| U_{\gamma I}^{\lambda}(k,\tau) \big|^2
    \right] ,
\end{align}
where the prime denotes derivative with respect to conformal time. This is the direct analogue of the single-field expression, with the important difference that the metric tensor perturbation receives contributions from both independent oscillator sectors.

For modes deep inside the horizon during radiation domination, one has approximately $U_{\gamma I}^{\lambda\,\prime}\simeq i k U_{\gamma I}^{\lambda}$, and Eq.~\eqref{Omega-def} reduces to
\begin{align}\label{Omega-subhorizon}
    \Omega_{\rm GW}(k,\tau)
    \simeq
    \frac{1}{12}
    \left(\frac{k}{aH}\right)^2
    {\cal P}_{h}(k,\tau) \, ,
\end{align}
where ${\cal P}_{h}=\sum_\lambda{\cal P}_{h}^{\lambda}$.

For a primordial superhorizon spectrum propagated through radiation domination, the oscillation-averaged present-day abundance is more accurately written as
\begin{align}\label{eq:omegaGW}
\Omega_{{\rm GW},0}(k) h^2
\simeq
\frac{\Omega_{r,0} h^2}{24}
\left(\frac{g_*(T_{\rm hc})}{g_{*,0}}\right)
\left(\frac{g_{*s}(T_{\rm hc})}{g_{*s,0}}\right)^{-4/3}
{\cal P}_h(k) \, .
\end{align}
Here $T_{\rm hc}$ is the temperature at horizon re-entry, $g_*$ and $g_{*s}$ are the effective relativistic degrees of freedom for the energy and entropy densities, respectively, and the subscript zero denotes their present values. The factor $1/24$ includes the standard oscillation average of the radiation-era tensor transfer function. If instead one uses an instantaneous estimate directly at horizon crossing, the numerical coefficient differs by an order-one factor. A more precise late-time evolution can be obtained by solving the tensor transfer equation and matching to the WKB solution, as in Refs.~\cite{Watanabe:2006qe,Bernal:2019lpc}.

As explained in the previous section, for $\ncpl\ll1$ the superhorizon solution during inflation approaches the usual vacuum result, so that $\sum_{I=1}^2 |U_{\gamma I}^\lambda|^2 \approx H_{\rm inf}^2/(2k^3)$. Matching this to the constant superhorizon solution during radiation domination, one recovers the usual scale-invariant tensor power spectrum at horizon re-entry,
\begin{align}
    {\cal P}_{h}(k)\approx \frac{2H_{\rm inf}^2}{\pi^2\Mpl^2}\, .
\end{align}
Using the CMB upper bound on the tensor amplitude, the corresponding present abundance is very small. Thus, for $\ncpl\ll1$, the standard vacuum prediction for primordial GWs is approximately recovered.

\begin{figure}[!ht]
    \begin{subfigure}[b]{0.5\linewidth}
        \centering
        \includegraphics[width=1\linewidth]{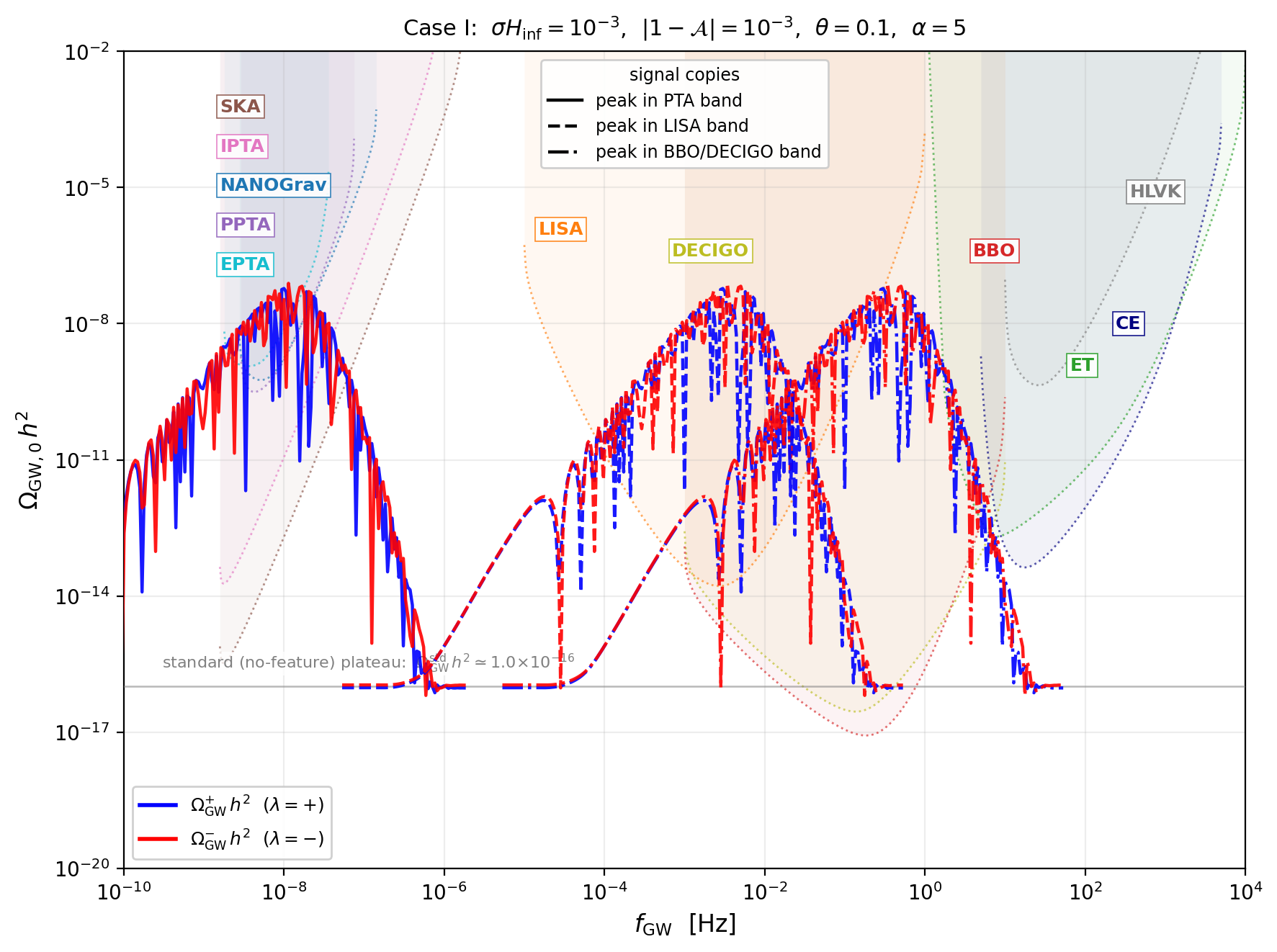}
    \end{subfigure}
    \begin{subfigure}[b]{0.5\linewidth}
        \centering
        \includegraphics[width=1\linewidth]{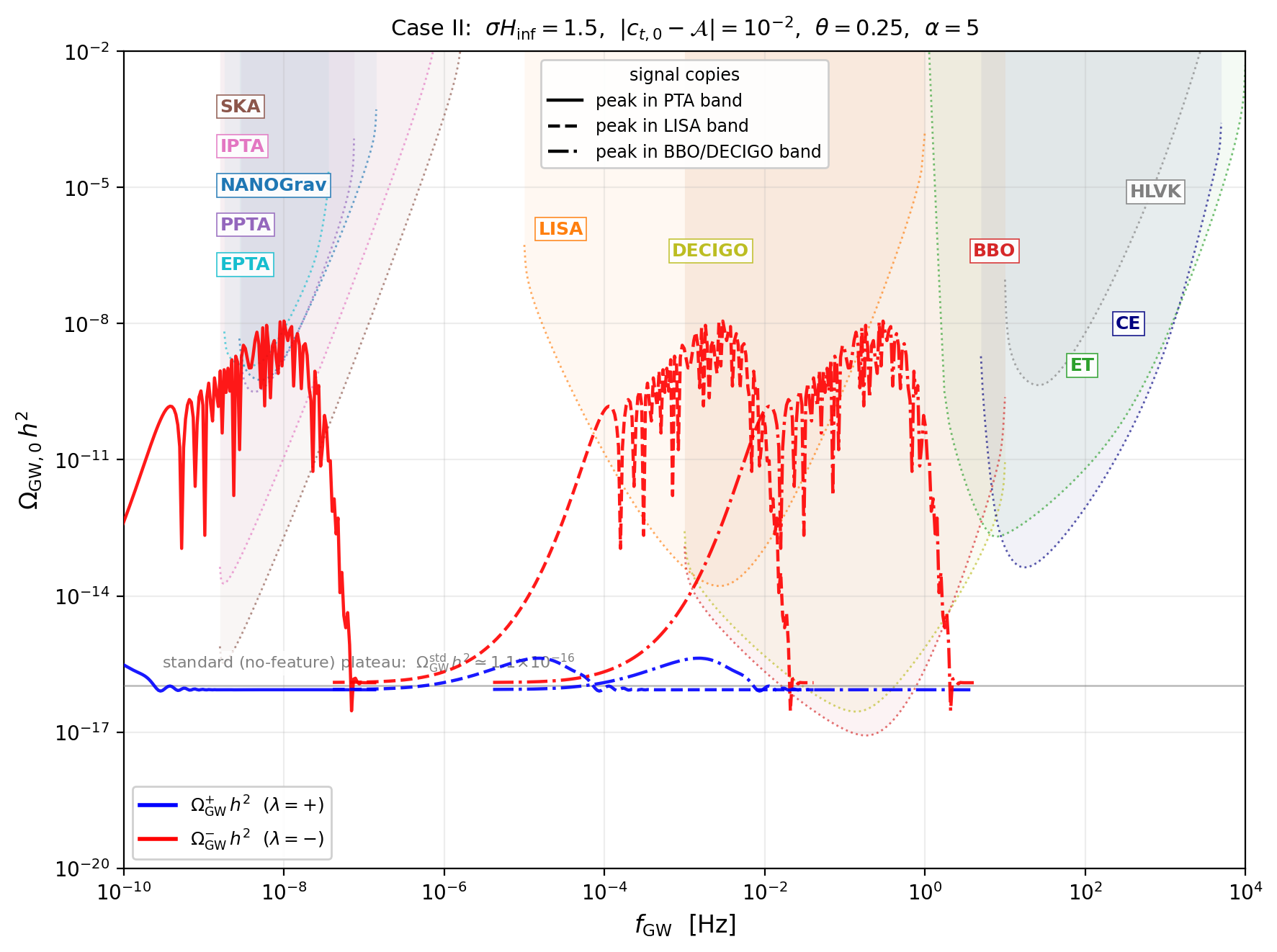}
    \end{subfigure}
    \caption{Present-day chiral GW abundance $\Omega_{\rm GW,0}^{\lambda}h^2$ for Case I (left) and Case II (right), overlaid on the power-law-integrated sensitivities of current and planned GW detectors. We used $c_{t,0}=0.9$ and $H_{\rm inf}/\Mpl=10^{-5}$. In each panel the same spectrum is shown three times, with its peak placed at the centre of the PTA (solid), LISA (dashed), and BBO/DECIGO (dot-dashed) bands through the corresponding choice of $k_\ast$; $\lambda=+$ is blue, $\lambda=-$ is red. The gray horizontal line marks the no-feature plateau set by the $\ncpl$-enhanced Bunch-Davies vacuum. Although $\pvc$ is at the same order in both Case I and Case II, the dip in $c_t$ in Case II locally suppresses the gradient $c_t^2 k^2$ relative to the parity term, making Case II markedly more chiral than Case I.}
    \label{fig:GW-exp-f-ct}
\end{figure}

On the other hand, for $|\ncpl|\geq3/2$, $\gamma_{\bf k}^\lambda$ is enhanced due to the enhancement in $t_{\bf k}^\lambda$ caused by the dip in $f/c_t$ in Case I/II. We expect the largest enhancement to occur for modes with momenta close to $k_\ast$ when $\pvc=0$. In Fig.~\ref{fig:GW-exp-f-ct}, we plot the present GW spectrum $\Omega_{{\rm GW},0}$ for different polarizations in both Case I (left panel) and Case II (right panel) for $\ncpl=5$. As can be seen, there are two characteristic features: i) the GW spectrum has oscillatory features in $k$ with the universal leading period~\eqref{Delta-k}, and ii) different polarizations behave differently, leading to chiral GWs. These are the features of the initial power spectra at the time of horizon re-entry, shown in Fig.~\ref{fig:gamma-PS-f}, that are directly imprinted in the GW spectrum. The leading period is the same one derived in Sec.~\ref{sec-osc-mechanism}: a localized non-adiabatic event in the $t$-sector at a time $\tau_\ast$, transmitted to the metric channel by the linear mixing $\ncpl\neq0$, gives $\Delta k\simeq\pi k_\ast/c_{t,0}$ at leading WKB order. The detailed shape of the envelope and the peak amplitude depend on the profile of the feature and are different in the two phenomenological models considered here. In the strongly chiral Case~II spectrum in the right panel, the oscillatory pattern persists over the full numerically enhanced interval, reflecting the combined effect of the localized feature, the mixing, and the tachyonic growth of one helicity. The same Bogoliubov-interference template, with the scalar sound speed $c_s$ in place of $c_{t,0}$, underlies the oscillatory GW signatures of sharp-feature inflationary scenarios in which the excited state lives in the curvature sector and is transferred to the GW spectrum at second order in scalars~\cite{Fumagalli:2020nvq,Fumagalli:2021dtd,Fumagalli:2021mpc}; the distinctive aspect of the present construction is that the linear mixing $\ncpl$ transmits the template to the metric channel at first order, preserving potentially large modulations and allowing for chiral asymmetry through the parity-violating EFT operators in the spin-2 sector. Given the robustness of the period, it is interesting to examine whether such oscillations can be observable in the GW spectrum today \cite{Caldwell:2018giq,Smith:2019wny,Zhou:2020kkf,Fumagalli:2020nvq,Fumagalli:2021dtd}.

The physical origin of the chirality differs in the two Cases. In Case~I the large enhancement can already be generated by the non-adiabatic term $-(af)''/(af)$, so a large $\pvc$ is not required in order to obtain a large $\Omega_{\rm GW}$. The parity-violating operators then provide an additional chiral imprint whose size depends on $\pvc$. In Case~II, by contrast, the dip lowers the positive gradient term $c_t^2k^2$ but does not by itself make $\omega_t^2$ negative. Without appreciable parity violation, the examples we studied give only a moderate enhancement, up to about $\Omega_{\rm GW}\sim10^{-12}$. The much larger Case~II signal in Fig.~\ref{fig:GW-exp-f-ct} is therefore tied to the parity-violating tachyonic growth of one helicity and is consequently strongly chiral. For the parameters of the right panel, the tachyonic scale $k_{\rm tach}$ is much larger than the oscillatory scale estimated from the feature width. The persistence of oscillations throughout the numerically enhanced band is then a result of the full coupled evolution. This conclusion should be understood within the class of features described here by a dip in $c_t$; the exact signal size, peak position, and extent of the oscillatory band remain model dependent.

We have already found the power spectrum for GWs at the time of horizon re-entry, as shown in Fig.~\ref{fig:gamma-PS-f} for Case I and Case II, respectively. There, we have used $H_{\rm inf}/\Mpl=10^{-5}$. The value of $H_{\rm inf}/\Mpl$ should be consistent with the bound coming from CMB B-mode polarization at very low frequencies. More precisely 
\begin{align}
\frac{H_{\rm inf}}{\Mpl}=\pi\sqrt{\frac{{\cal P}_h}{2}}=\pi\sqrt{\frac{r{\cal P}_S}{2}} \,,
\end{align}
where ${\cal P}_S$ is the power spectrum of the scalar curvature perturbation and $r$ is the tensor-to-scalar ratio. Using the Planck result $\mathcal{P}_S \simeq 2.1\times 10^{-9}$ and $r< 0.056$ \cite{Planck:2013jfk,Planck:2018jri}, we find $H_{\rm inf}/\Mpl<2.4\times10^{-5}$. Therefore, any value of $H_{\rm inf}/\Mpl< {\cal O}(10^{-5})$ is consistent with this bound. Using $H_{\rm inf}/\Mpl=10^{-5}$, we compute the GW spectrum, given by Eq.~\eqref{eq:omegaGW}, as a function of the physical frequency $f_{\rm GW}=k/(2\pi a_0)$. The results are shown in Fig.~\ref{fig:GW-exp-f-ct}. In plotting the sensitivity curves, we used the data from Ref.~\cite{Schmitz:2020syl}.

\section{Summary and conclusions}\label{sec-summary}

Spectator fields that provide additional tensor perturbations, on top of the usual metric tensor perturbations, can produce a significant amount of GWs due to possible linear mixing with the metric tensor modes \cite{Gorji:2023ziy}. To understand the general, model-independent features of such scenarios, we used the EFT approach for spin-2 fields, which includes all possible operators including parity-violating ones. With the EFT setup in hand, we studied the general role of the free EFT parameters.

A characteristic prediction of this class of scenarios is the appearance of oscillatory features in the GW sector, of two qualitatively different kinds.
\begin{itemize}
\item An in-time oscillation of the superhorizon modes, which appears whenever the linear mixing is strong enough, $|\ncpl|\geq3/2$. This effect is intrinsic to the mixing alone, does not require any feature in the spin-2 sector, and is reminiscent of neutrino flavor oscillations, with $\gamma_{ij}$ and $t_{ij}$ playing the role of ``flavor'' eigenstates and the diagonalized modes the role of ``propagation'' eigenstates.
\item Oscillations in $k$ of the GW power spectrum, which require both $\ncpl\neq0$ and a localized non-adiabatic event in the spin-2 sector during inflation. Given any such localized feature at a conformal time $\tau_\ast$, the linear mixing transmits Bogoliubov-interference fringes from the $t$-sector to the metric channel, producing a robust oscillatory pattern. We have shown analytically that the leading fringe period $\Delta k\simeq\pi k_\ast/c_{t,0}$ is set only by the location $\tau_\ast$ of the feature and by the propagation speed $c_{t,0}$ outside it, and is therefore the same in both of the phenomenological models we considered, as confirmed by the numerical spectra in both the perturbative and the chiral examples. The shape of the envelope is Gaussian in $k/k_\ast$ in our setup as a direct consequence of the Gaussian profile~\eqref{f} used in both Cases; a different functional form for the feature would yield a different envelope shape. The peak amplitude is the model-dependent piece and tracks the violation of the relevant adiabaticity conditions.
\end{itemize}
Moreover, the parity-violating EFT operators give rise to chiral GWs.

Unlike scalar-induced GW scenarios, where scalar perturbations non-linearly source metric tensor perturbations, our setup is linear. This leads to a direct correspondence between the extra tensor modes and the metric tensor modes, causing their power spectra to have very similar shapes. Exploiting this fact, we studied two phenomenological models for the spin-2 sector: one with non-minimal coupling (Case I) and another with a varying sound speed (Case II). In both cases the resulting GW spectrum exhibits in-$k$ oscillations with the leading period $\Delta k\simeq\pi k_\ast/c_{t,0}$, while the amplitude of the peak tracks the level of violation of the adiabatic conditions: Case~I is especially sensitive to the sharpness of the dip in $f$, and Case~II to the minimum sound speed $c_{t,\min}=c_{t,0}-{\cal A}$. The role of chirality is, however, different in the two Cases. In Case~I, the non-adiabatic term $-(af)''/(af)$ can itself drive a large enhancement, so appreciable chirality is not required for enhancement, although it remains an important possible signature. In Case~II, the dip in $c_t$ suppresses the positive gradient term $c_t^2k^2$ but does not by itself make the frequency tachyonic; obtaining a large observable enhancement in this class of models is therefore tied to the helicity-dependent tachyonic growth produced by the parity-violating term, and hence to a strongly chiral GW background. The analytical discussion explains the origin of the fringe period and gives the expected oscillatory support in the regimes where the feature can be treated as a localized WKB scattering event. In the strongly chiral Case~II examples, the full numerical solution shows that the oscillations persist throughout the larger tachyonically enhanced band. The precise size of the signal, the position of the peak, the extent of the oscillatory band, and the exact parameter values are model dependent. We demonstrated that both models can produce a significant amount of chiral GWs featuring oscillatory patterns that could be detectable by future GW observatories. Our results provide a useful framework to understand the general characteristics of this class of scenarios.

\vspace{0.7cm}

{\bf Acknowledgments:} JG is supported by PID2022-136224NB-C22, funded by MCIN/AEI
/10.13039/501100011033/FEDER, UE, and by 2021-SGR00872 funded by AGAUR. MAG thanks Teruaki Suyama for useful discussions. The work of MAG was supported by IBS under the project code IBS-R018-D3.  F.H. is supported by Homer Dodge postdoctoral fellowship. F.H.  thanks the Mitchell Institute  at  Texas A \& M University for their hospitality and support during the final stages of this project.
He is also grateful to the organizers of workshop of Center for Theoretical Underground Physics and Related Areas (CETUP* - 2024 and 2025), The Institute for Underground Science at Sanford
Underground Research Facility (SURF), Lead, South Dakota
for their hospitality and financial support.
This work was supported in part by JSPS KAKENHI No.~JP24K00624.\\

OpenAI ChatGPT and Anthropic Claude were used to assist with language editing and the discussion of some analytical and numerical calculations.
All results were independently verified by the authors.
\vspace{0.2cm}

\appendix

\section{Role of the couplings $\kappa$ and $\eta$}\label{app-deep-inside-horizon}

Going to Fourier space and decomposing tensor modes in terms of polarization tensors as in Eq.~\eqref{Fourier-def}, and then using the properties of the polarization tensors, the quadratic action \eqref{action-total} takes the form
\begin{align}
    \label{action-total-helicity-ibp}
    \begin{split}
        S &= \frac{1}{2} \sum_{\lambda} \int \frac{\D^3k}{(2\pi)^3}\, \D\tau\, a^2
        \left[
        \left| {\gamma}'^\lambda_{\bf k} \right|^2
        - k^2 \left| {\gamma}^\lambda_{\bf k} \right|^2 \right]
        \\
        &
        + \frac{1}{2} \sum_{\lambda} \int \frac{\D^3k}{(2\pi)^3}\, \D\tau\, a^2  f^2 \left[
        \left| {t}'^\lambda_{\bf k} \right|^2
        - \left( c_t^2 k^2 + 2 \lambda {\pvc} k {\cal H} + m^2 a^2 \right) \left| {t}^\lambda_{\bf k} \right|^2
        \right]
        \\
        &
        + \frac{1}{2} \sum_{\lambda} \int \frac{\D^3k}{(2\pi)^3}\, \D\tau\, a^2
        \left[ {\gamma}'^\lambda_{\bf k}
        \left(
        {\tilde \ncpl}_k \, {t}^{\ast\lambda}_{\bf k}
        + \kappa \, {t}'^{\ast\lambda}_{\bf k}
        \right) + \mbox{c.c.} \right] \, ,
    \end{split}
\end{align}
where, for simplicity, we defined
\begin{align}
    {\tilde \ncpl}_k \equiv \ncpl {\cal H} - \lambda \eta k \, .
\end{align}

Variation of \eqref{action-total-helicity-ibp} gives
\begin{align}\label{EoM-h-gen}
    \gamma''^\lambda_{\bf k} + 2 \frac{a'}{a} \gamma'^\lambda_{\bf k} + k^2 {\gamma}^\lambda_{\bf k}
    &=
    - \frac{1}{a^2} \frac{\D}{\D\tau}
    \left[
    a^2 \left( {\tilde \ncpl}_k t^{\lambda}_{\bf k} + \kappa t'^{\lambda}_{\bf k} \right)
    \right] \, ,
    \\
    \label{EoM-t-gen}
    t''^\lambda_{\bf k} + 2 \frac{(af)'}{af} t'^\lambda_{\bf k} + \left( c_t^2k^2 + 2 \lambda {\pvc} k \frac{a'}{a} + m^2 a^2 \right) {t}^\lambda_{\bf k}
    &=
    - \frac{1}{a^2f^2} \frac{\D}{\D\tau} \left(
    a^2 \kappa {\gamma}'^{\lambda}_{\bf k}
    \right)
    + \frac{{\tilde \ncpl}_k}{f^2} \gamma'^{\lambda}_{\bf k}
    \, .
\end{align}

We consider $f=1$ at early time $\tau\to-\infty$ during inflation so that the modes deep inside the horizon are not affected by $f$. For the modes deep inside the horizon, $k\gg{\cal H}$, we have ${\tilde \ncpl}_k\approx-\lambda\eta k$, while the parity-violating term proportional to $\pvc$ is subdominant. In this regime, the couplings $\kappa$ and $\eta$ dominate. Substituting the WKB ansatz
\begin{align}\label{WKB-ansatz}
    { \gamma}^{\lambda}_{\bf k} = A e^{-i\int\D\tau'\omega(\tau')} \, ,
    \qquad
    { t}^{\lambda}_{\bf k} = B e^{-i\int\D\tau'\omega(\tau')} \, ,
\end{align}
into Eqs.~\eqref{EoM-h-gen} and \eqref{EoM-t-gen}, one finds the following quartic equation for the frequency:
\begin{align}
    \left(1-\kappa^2\right) \omega^4
    - \left(1+c_t^2 + \eta^2\right) k^2 \omega^2 + c_t^2 k^4 = 0 \, .
\end{align}

Two comments are in order. First, since $\lambda^2=1$, the interaction $\eta$ does not generate chirality in this deep-subhorizon limit. Second, the effect of both $\eta$ and $\kappa$ is to modify the propagation speeds of the diagonalized combinations of ${\gamma}^{\lambda}_{\bf k}$ and ${t}^{\lambda}_{\bf k}$. It is then convenient to define
\begin{align}
    {\tilde c}_t \equiv \frac{c_t}{\sqrt{1-\kappa^2}} \, ,
    \qquad
    {\tilde \eta} \equiv \sqrt{\frac{\eta^2+\kappa^2}{1-\kappa^2}} \, ,
\end{align}
so that the frequency equation becomes
\begin{align}
    \omega^4
    - \left(1+{\tilde c}_t^2 + {\tilde \eta}^2\right) k^2\omega^2 + {\tilde c}_t^2 k^4 = 0 \, .
\end{align}
Solving this equation gives
\begin{align}\label{speeds-kappa-eta}
    c^2_{\pm} = \frac{1}{2} \left(
    1+\tilde{c}_t^2+\tilde{\eta}^2
    \pm\sqrt{\left(1-\tilde{c}_t^2\right){}^2+2\tilde{\eta}^2\left(1+\tilde{c}_t^2\right)+\tilde{\eta }^4}
    \right)\,,
\end{align}
where $\omega_\pm^2=c_\pm^2 k^2$. For $\eta=0=\kappa$, the two roots reduce to $1$ and $c_t^2$; their labels simply follow the ordering of the two speeds. For the parameter range used in the main text, $c_t<1$, the $+$ branch is continuously connected to the metric graviton and the $-$ branch to the extra tensor. Turning on $\eta$ or $\kappa$ shifts the metric-like speed away from unity, which motivates setting $\kappa=\eta=0$ in the main analysis under the conservative assumption that the metric-like GW speed is luminal.

\section{Derivation of the in-$k$ oscillation formulas}\label{app-osc-mechanism}

In this appendix we derive the analytical formulas for the leading fringe period~\eqref{Delta-k} and the envelope~\eqref{beta-envelope} quoted in Sec.~\ref{sec-osc-mechanism}. We focus on the regime in which the feature is localized, the relevant modes are well inside the sound horizon at the time of the feature, $c_{t,0}k|\tau_\ast|\gg1$, and the localized change of the squared WKB frequency can be treated perturbatively. This is the regime directly tested by the fringe comparison in Fig.~\ref{fig:fringe-comparison}.

\subsection{Setup and Bogoliubov system}\label{subapp-setup}

The equation of motion for the canonically normalized $t$-mode is Eq.~\eqref{EoM-T-inf}, which we rewrite as
\begin{equation}\label{T-eom-app}
	{T^\lambda_{\bf k}}''\,+\,\omega_t^2(\tau)\,T^\lambda_{\bf k}\,=\,\frac{\ncpl}{f}{\cal H}\left({H^\lambda_{\bf k}}'-{\cal H}H^\lambda_{\bf k}\right),
\end{equation}
with
\begin{equation}\label{omega-t-app}
	\omega_t^2(\tau)\,=\,c_t^2(\tau)\,k^2 + 2\lambda\pvc\,k\,\frac{a'}{a} - \frac{(af)''}{af}\, .
\end{equation}
To isolate the origin of the fringes we first set the slowly varying source on the right-hand side of \eqref{T-eom-app} aside and solve the homogeneous scattering problem for $T^\lambda_{\bf k}$. The mixing $\ncpl$ is then reinstated as the linear transfer from $T^\lambda_{\bf k}$ to $H^\lambda_{\bf k}$.

We split the frequency into a smooth background and a localized perturbation,
\begin{equation}\label{omega-split-app}
	\omega_t^2(\tau)=\omega_{t,0}^2(\tau)+\delta\omega_t^2(\tau),
\end{equation}
where the smooth background is obtained by setting $f=1$ and $c_t=c_{t,0}$:
\begin{equation}\label{omega-0-app}
	\omega_{t,0}^2(\tau)=c_{t,0}^2k^2-\frac{2\lambda\pvc k}{\tau}-\frac{2}{\tau^2}.
\end{equation}
Deep inside the horizon the leading piece is $c_{t,0}^2k^2$, so $\omega_{t,0}\simeq c_{t,0}k$.

In Case I, $c_t=c_{t,0}$ and $f=F_1$, so
\begin{equation}\label{delta-omega-I-general-app}
	\delta\omega_t^2\big|_{\rm I}(\tau)=-\frac{(aF_1)''}{aF_1}+\frac{a''}{a}
	=-2\frac{a'}{a}\frac{F_1'}{F_1}-\frac{F_1''}{F_1}.
\end{equation}
Introducing $u=\ln(\tau/\tau_\ast)$ and $\tilde\sigma=\sigma H_{\rm inf}$, so that $F_i(u)=c_i-\mathcal{A}e^{-u^2/(2\tilde\sigma^2)}$, we have
\begin{equation}\label{tau2-delta-I}
	\tau^2\delta\omega_t^2\big|_{\rm I}=\frac{3 F_{1,u}-F_{1,uu}}{F_1},
\end{equation}
where we have used $a\propto-1/\tau$ during inflation. At the center of the dip, $u=0$, this gives
\begin{equation}\label{delta-omega-I-peak-app}
	\delta\omega_t^2\big|_{\rm I}(\tau_\ast)=
	-\frac{\mathcal{A}}{1-\mathcal{A}}\frac{k_\ast^2}{\tilde\sigma^2}.
\end{equation}
In Case II, $f=1$ and $c_t=F_2$, so
\begin{equation}\label{delta-omega-II-general-app}
	\tau^2\delta\omega_t^2\big|_{\rm II}=\left[F_2^2-c_{t,0}^2\right]k^2\tau^2,
\end{equation}
and at the center of the dip
\begin{equation}\label{delta-omega-II-peak-app}
	\delta\omega_t^2\big|_{\rm II}(\tau_\ast)=
	-\mathcal{A}(2c_{t,0}-\mathcal{A})k^2.
\end{equation}

We now solve $T''+(\omega_{t,0}^2+\delta\omega_t^2)T=0$ perturbatively using the instantaneous Bogoliubov decomposition~\cite{Shtanov:1994ce,Kofman:1997yn,Salehian:2020dsf}. Writing
\begin{equation}\label{T-decomp-app}
	T^\lambda_{\bf k}(\tau)=\alpha_k(\tau)\frac{e^{-i\Theta(\tau)}}{\sqrt{2c_{t,0}k}}
	+\beta_k(\tau)\frac{e^{+i\Theta(\tau)}}{\sqrt{2c_{t,0}k}},
	\qquad
	\Theta(\tau)=\int^\tau\omega_{t,0}(\tilde\tau)\,\D\tilde\tau\simeq c_{t,0}k\tau,
\end{equation}
and imposing the usual variation-of-constants auxiliary condition $\alpha_k'\,e^{-i\Theta}+\beta_k'\,e^{+i\Theta}=0$ that fixes the redundancy of the two-function parametrization, one obtains the exact first-order system
\begin{equation}\label{alpha-beta-evol-app}
	\alpha_k'=-i\,\frac{\delta\omega_t^2}{2c_{t,0}k}\,\beta_k\,e^{+2i\Theta},
	\qquad
	\beta_k'=+i\,\frac{\delta\omega_t^2}{2c_{t,0}k}\,\alpha_k\,e^{-2i\Theta}.
\end{equation}
In writing Eq.~\eqref{alpha-beta-evol-app}, we have already absorbed into $\Theta$ the diagonal contributions of the exact two-function system. This is justified because they amount to a unitary phase rotation and do not affect the particle-production calculation that follows.\footnote{Explicitly, the exact system before absorption contains, in addition to the off-diagonal terms displayed in Eq.~\eqref{alpha-beta-evol-app}, diagonal pieces of the form $-i\,[\delta\omega_t^2/(2c_{t,0}k)]\,\alpha_k$ in $\alpha_k'$ and $+i\,[\delta\omega_t^2/(2c_{t,0}k)]\,\beta_k$ in $\beta_k'$. Integrating these alone gives $\alpha_k\to\alpha_k e^{-i\Phi}$ and $\beta_k\to\beta_k e^{+i\Phi}$ with the same real $\Phi(\tau)\equiv\int^\tau\delta\omega_t^2/(2c_{t,0}k)\,\D\tilde\tau$, so the rotation is unitary, leaves $|\alpha_k|$ and $|\beta_k|$ exactly invariant, and produces no particles by itself. The opposite signs of the phases of $\alpha_k$ and $\beta_k$ are the diagnostic that the rotation is consistent with the Wronskian conservation $|\alpha_k|^2-|\beta_k|^2=1$. The rotation is equivalently the next-order WKB correction $\Theta\to\Theta+\Phi\simeq\int^\tau\!\sqrt{\omega_{t,0}^2+\delta\omega_t^2}\,\D\tilde\tau$ to the unperturbed phase, sub-leading to the dominant $c_{t,0}k\tau$ in the regime $|\delta\omega_t^2|\ll\omega_{t,0}^2$ in which the leading-order Bogoliubov calculation is self-consistent.}

With adiabatic-vacuum boundary conditions $\alpha_k\to1$ and $\beta_k\to0$ before the dip, the first-order result after the dip is
\begin{equation}\label{app-beta-formula}
	\beta_k\simeq\frac{i}{2c_{t,0}k}\int_{-\infty}^{0}\D\tau\,\delta\omega_t^2(\tau)\,e^{-2ic_{t,0}k\tau}.
\end{equation}
The Wronskian condition $|\alpha_k|^2-|\beta_k|^2=1$ is conserved by Eq.~\eqref{alpha-beta-evol-app}.

Near the feature, $u=\ln(\tau/\tau_\ast)$ is small and $\tau=\tau_\ast e^u=-e^u/k_\ast$. Since $\D\tau=-e^u \D{u}/k_\ast$ and the orientation of the integral is reversed when converting from $\tau\in(-\infty,0)$ to $u\in(\infty,-\infty)$, Eq.~\eqref{app-beta-formula} becomes
\begin{equation}\label{beta-u-integral-app}
	\beta_k=\frac{i k_\ast}{2c_{t,0}k}\int_{-\infty}^{+\infty}\D{u}\,e^{-u}\,\tau^2\delta\omega_t^2(\tau_\ast e^u)\,
	\exp\left[2ic_{t,0}\frac{k}{k_\ast}e^u\right].
\end{equation}

We now evaluate this integral in two complementary regimes: narrow dip ($\tilde\sigma\ll 1$) and broad dip ($\tilde\sigma\gtrsim 1$).

\subsection{Narrow-dip regime $\tilde\sigma\ll 1$}\label{subapp-narrow}

The Gaussian factor $e^{-u^2/(2\tilde\sigma^2)}$ inherited from the dip~\eqref{f} has width $\tilde\sigma\ll 1$ and localizes the integrand at $|u|\lesssim\tilde\sigma$, so all slowly varying multipliers may be evaluated at the Gaussian center $u=0$. The factor $e^{-u}$ is slowly varying across a narrow feature and can be set to unity at leading order. Expanding $e^u\simeq 1+u$ in the phase and using the Gaussian Fourier transform gives
\begin{equation}\label{beta-explicit-app}
	\beta_k\simeq
	-\frac{i\,\mathcal{D}(k)\sqrt{2\pi}\tilde\sigma}{2c_{t,0}(k/k_\ast)}
	\exp\left[-2(c_{t,0}\tilde\sigma)^2(k/k_\ast)^2\right]
	\exp\left[2ic_{t,0}k/k_\ast\right],
\end{equation}
where $\mathcal{D}(k)$ is the dimensionless peak value of $-\tau^2\delta\omega_t^2$:
\begin{equation}\label{D-def-app}
	\mathcal{D}_{\rm I}(k_\ast)=\frac{\mathcal{A}}{(1-\mathcal{A})\tilde\sigma^2},
	\qquad
	\mathcal{D}_{\rm II}(k)=\mathcal{A}(2c_{t,0}-\mathcal{A})\left(\frac{k}{k_\ast}\right)^2.
\end{equation}
Equations \eqref{beta-explicit-app}-\eqref{D-def-app} give the perturbative estimates in Eq.~\eqref{betak-summary}.

At a time $\tau_0$ after the feature, the squared mode amplitude is
\begin{equation}\label{T-squared-app}
	|T^\lambda_{\bf k}(\tau_0)|^2=\frac{1}{2c_{t,0}k}
	\left[1+|\beta_k|^2+2\mathrm{Re}\left(\alpha_k\beta_k^*e^{-2i\Theta(\tau_0)}\right)\right].
\end{equation}
At leading order $\alpha_k\simeq1$. The phase of $\beta_k$ from \eqref{beta-explicit-app} is
\begin{equation}
	\arg\beta_k\simeq -\frac{\pi}{2}+\frac{2c_{t,0}k}{k_\ast},
\end{equation}
where we have taken into account the fact that ${\cal D}_{\rm I,II}$ are positive. So the interference term can be written as
\begin{equation}\label{interference-app}
	2\mathrm{Re}\left(\beta_k^*e^{-2i\Theta(\tau_0)}\right)
	\simeq2|\beta_k|\cos\left[2c_{t,0}k(\tau_0-\tau_\ast)+\varphi_{\rm c}\right],
\end{equation}
where $\varphi_{\rm c}$ collects $k$-independent pieces such as $-\pi/2$. Evaluating the late-time amplitude at $\tau_0\simeq0^-$ gives
\begin{equation}\label{app-interference-final}
	2\mathrm{Re}\left(\beta_k^*e^{-2i\Theta(\tau_0)}\right)
	\simeq2|\beta_k|\cos\left(\frac{2c_{t,0}k}{k_\ast}+\varphi_0\right).
\end{equation}
This is the origin of the bracketed factor in Eq.~\eqref{Ph-analytical-body}. A smooth transfer function from $T^\lambda_{\bf k}$ to $H^\lambda_{\bf k}$ through the linear mixing $\ncpl$ can shift $\varphi_0$ and the smooth envelope, but it does not change the leading spacing of the fringes.

The oscillation period in $k$-space (fringe spacing) follows immediately:
\begin{equation}\label{Deltak-derivation-app}
	\Delta k\,\frac{2c_{t,0}}{k_\ast}=2\pi
	\qquad\Longrightarrow\qquad
	\Delta k=\frac{\pi k_\ast}{c_{t,0}}.
\end{equation}
While this derivation is perturbative, the period \eqref{Deltak-derivation-app} only requires that the asymptotic WKB modes outside the dip be the same on both sides and that the matching occur near $\tau_\ast$; both conditions hold in the tachyonic regime as well. It is the magnitude of $\beta_k$, and hence the envelope and the visible window, that becomes non-perturbative in that case, while the cosine period is unchanged. This is consistent with the persistence of oscillations across the full enhanced band in the strongly chiral Case II numerics.

The envelope follows by normalizing the modulus of \eqref{beta-explicit-app} to its value at $k=k_\ast$ and ignoring slowly varying powers of $k$:
\begin{equation}\label{envelope-derivation-app}
	\frac{|\beta_k|}{|\beta_{k_\ast}|}\simeq
	\exp\left[-\frac12(2c_{t,0}\tilde\sigma)^2\left((k/k_\ast)^2-1\right)\right].
\end{equation}
The power-law prefactors in \eqref{beta-explicit-app} and \eqref{D-def-app} can change the relative height of successive peaks, especially in Case II, but they do not change the leading Gaussian cutoff or the period. The envelope~\eqref{envelope-derivation-app} sets the upper edge of the excited window, $k\lesssim k_{\rm fr}=k_\ast/(c_{t,0}\tilde\sigma)$, beyond which fast-oscillating modes adiabatically track $\omega_t(\tau)$ through the dip. The lower edge is set independently by the sound horizon at $\tau_\ast$: modes with $c_{t,0}k|\tau_\ast|\lesssim 1$ are already superhorizon when the dip switches on and lie outside the WKB regime in which the present perturbative calculation applies. We thus find the following window
\begin{equation}\label{window-app}
	k_\ast\,\lesssim\,k\,\lesssim\,k_{\rm fr}\equiv\frac{k_\ast}{c_{t,0}\,\tilde\sigma} \qquad (\tilde\sigma\ll 1)\,.
\end{equation}

\subsection{Broad-dip regime $\tilde\sigma\gtrsim 1$}\label{subapp-broad}

The reduction leading to Eq.~\eqref{beta-explicit-app} relied on linearizing $e^u\simeq 1+u$ in the phase of Eq.~\eqref{beta-u-integral-app}. This step is justified only when the Gaussian support $|u|\lesssim\tilde\sigma$ stays inside the small-$u$ regime, i.e.~when $\tilde\sigma\ll 1$. For $\tilde\sigma\gtrsim 1$ the phase $2c_{t,0}(k/k_\ast)e^u$ varies non-linearly across the Gaussian support, the Fourier-transform argument no longer applies, and the upper-edge formula $k_{\rm fr}=k_\ast/(c_{t,0}\tilde\sigma)$ in Eq.~\eqref{window-app} is no longer the right parametric guide.

In this regime the dip is no longer a localized event in conformal time. Its Gaussian support in $u$ of width $\tilde\sigma$ covers the conformal-time range $\tau_\ast e^{+\tilde\sigma}\lesssim\tau\lesssim\tau_\ast e^{-\tilde\sigma}$ (recall $\tau_\ast<0$), corresponding to a duration $\Delta{\cal N}=2\tilde\sigma$ in $e$-folds. A simple estimate of the oscillatory window is then obtained from two sound-horizon conditions, applied at the center and at the late edge of this support.

The first condition is sub-horizon at the dip center, where the perturbation $\delta\omega_t^2(\tau)$ peaks. A mode with $c_{t,0}k|\tau_\ast|\lesssim 1$ is already super-horizon at $\tau_\ast$, hence it is frozen and unable to respond to the perturbation at its peak. The Bogoliubov picture of mode excitation therefore requires
\begin{equation}\label{lower-edge-broad-app}
	c_{t,0}\,k\,|\tau_\ast|\,\gtrsim\,1 \qquad \Longleftrightarrow \qquad k\,\gtrsim\,k_\ast/c_{t,0}\sim k_\ast \, .
\end{equation}
This is the same lower edge as in the narrow-dip regime, up to order-one factors.

The second condition is that the mode be sub-horizon long enough to feel the dip while it is still active. Requiring the mode to still be sub-horizon at the late edge of the dip, $c_{t,0}k|\tau_\ast e^{-\tilde\sigma}|\gtrsim 1$, gives the upper-edge estimate
\begin{equation}\label{upper-edge-broad-app}
	k\,\lesssim\,\frac{k_\ast\,e^{+\tilde\sigma}}{c_{t,0}} \, .
\end{equation}
Combining~\eqref{lower-edge-broad-app} and~\eqref{upper-edge-broad-app} gives the broad-dip estimate
\begin{equation}\label{window-broad-app}
	k_\ast\,\lesssim\,k\,\lesssim\,k_{\rm fr}\equiv\frac{k_\ast\,e^{+\tilde\sigma}}{c_{t,0}}\qquad (\tilde\sigma\gtrsim 1) \, ,
\end{equation}
quoted in Eq.~\eqref{window-broad-body}, with width factor $\sim e^{+\tilde\sigma}/c_{t,0}$ instead of $\sim 1/(c_{t,0}\tilde\sigma)$. The two formulas~\eqref{window-app} and~\eqref{window-broad-app} share the lower edge $k_\ast$ and cross over around $\tilde\sigma\sim 1$ at the level of an order-of-magnitude estimate. A combined estimate of the upper edge is therefore
\begin{equation}\label{window-combined-app}
	\frac{k_{\rm fr}}{k_\ast}\sim\frac{1}{c_{t,0}}\,\max\!\left(\frac{1}{\tilde\sigma},\,e^{\tilde\sigma}\right).
\end{equation}
This formula should be read as a leading-WKB guide to the oscillatory window associated with the feature. It is not a sharp boundary of the full spectrum. In the narrow Case~I examples, where the dip is sharp, the numerical oscillatory window is consistent with the scaling $k_{\rm fr}/k_\ast\sim1/(c_{t,0}\tilde\sigma)$, up to the order-one uncertainty expected from a leading-WKB estimate. In the broad Case~II examples, Eq.~\eqref{window-combined-app} remains a useful scale for the feature-induced fringe estimate, while the full extent of the oscillatory band is determined by the numerical solution when the chiral tachyonic growth is large.

\subsection{Helicity dependence at leading order}\label{subapp-helicity}

Because $\pvc$ is constant, the parity-violating term is part of the smooth background frequency \eqref{omega-0-app}, not of the localized kick $\delta\omega_t^2$. As long as the evolution remains non-tachyonic and the WKB expansion is valid, the leading $\beta_k$ generated by the localized Gaussian feature is therefore helicity independent. Keeping the $\pvc$ term in the WKB phase gives
\begin{equation}
	\Theta_\lambda(\tau)=c_{t,0}k\tau-\frac{\lambda\pvc}{c_{t,0}}\ln|\tau|+\cdots,
\end{equation}
which can induce a subleading helicity-dependent phase shift. It does not alter the leading period \eqref{Deltak-derivation-app} in the perturbative WKB treatment. If the parity-violating term is large enough to make $\omega_t^2$ negative for one helicity, the helicity dependence and the full oscillatory range must be determined from the numerical solution.

\subsection{Numerical verification of the fringe formula}\label{subapp-fringe-numerical}

The leading-order template~\eqref{Ph-analytical-body} can be compared directly with the full numerical result in the regime illustrated in Fig.~\ref{fig:fringe-comparison}. For this comparison, we first need to define the smooth envelope $\mathcal{P}_{h,{\rm smooth}}^\lambda(k)$ in a clear way. In the cases where the template is applied, the oscillations can have very deep dips because $|\beta_k|$ can become close to one. For this reason, an ordinary average of $\mathcal{P}_h^\lambda(k)$ over one oscillation period is not the best choice: it is strongly affected by the high peaks and does not represent the middle of the oscillation well on a logarithmic plot. Since the spectrum is shown on a log scale, it is more natural to average the logarithm of the spectrum instead. We therefore define the smooth envelope by
\begin{equation}\label{Psmooth-def-app}
	\ln\mathcal{P}_{h,{\rm smooth}}^\lambda(k)
	\equiv
	\frac{1}{\Delta k}\int_{k-\Delta k/2}^{k+\Delta k/2}\D{k'}\,\ln\mathcal{P}_h^\lambda(k') \,,
	\qquad
	\Delta k=\frac{\pi k_\ast}{c_{t,0}}.
\end{equation}
Equivalently, $\mathcal{P}_{h,{\rm smooth}}^\lambda(k)$ is the geometric mean of $\mathcal{P}_h^\lambda(k)$ over one fringe period. This removes the fast oscillation and keeps the slowly varying envelope. In particular, the averaging window has the same width as one predicted period, so the oscillatory cosine term averages to zero at leading order:
$\int_0^{\Delta k}\cos(2c_{t,0}k'/k_\ast+\varphi_0)\D{k'}=0$. In log space the comparison template is implemented as
\begin{equation}\label{template-log-app}
	\ln\mathcal{P}_h^\lambda(k)
	\;\simeq\;
	\ln\mathcal{P}_{h,{\rm smooth}}^\lambda(k)
	\,+\,
	\mathcal C_{\rm fr}\cos\!\left(\omega_{\rm fit}\frac{k}{k_\ast}+\varphi_{\rm fit}\right),
\end{equation}
where $\mathcal C_{\rm fr}$ is a phenomenological fit parameter used in the numerical comparison and represents the common oscillation amplitude in log space. More explicitly, the fit assumes $\ln(\mathcal P_h^\lambda/\mathcal P_{h,\rm smooth}^\lambda)=\mathcal C_{\rm fr}\cos(\cdots)$ over the fitted range. Therefore, for each fitted fringe, the peak satisfies $\mathcal P_h^\lambda/\mathcal P_{h,\rm smooth}^\lambda=e^{+\mathcal C_{\rm fr}}$, while the corresponding null or dip satisfies $\mathcal P_h^\lambda/\mathcal P_{h,\rm smooth}^\lambda=e^{-\mathcal C_{\rm fr}}$. Thus the same peak-to-null contrast is assigned to each fitted oscillation, namely $\left.\mathcal P_h^\lambda\right|_{\rm peak}/\left.\mathcal P_h^\lambda\right|_{\rm null}=e^{2\mathcal C_{\rm fr}}$. The unknown contrast, phase, and fitted frequency are obtained by least-squares fitting~\eqref{template-log-app} to the numerical residual $\ln\mathcal{P}_h^\lambda(k)-\ln\mathcal{P}_{h,{\rm smooth}}^\lambda(k)$ in the excited window~\eqref{window-body}. The analytical leading prediction is recovered by setting $\omega_{\rm fit}=\omega_{\rm an}=2c_{t,0}$.

Figure~\ref{fig:fringe-comparison} shows the result for Case I, $\pvc=0$ (so that $\mathcal{P}_h^+=\mathcal{P}_h^-$ and any in-$k$ oscillation comes purely from the Bogoliubov interference of this appendix), $|1-\mathcal{A}|=10^{-2}$, $\tilde\sigma=2.5\times 10^{-2}$, and two values of $c_{t,0}$. The analytical log-space template tracks the numerical curve fringe-by-fringe across the entire excited window~\eqref{window-body}, including the deep nulls between consecutive peaks. The two sets of vertical ticks at the top of each panel separate the leading-order analytical period from the fitted numerical period: red ticks use the frequency and phase extracted from~\eqref{template-log-app}, while green ticks use the leading-WKB frequency $\omega_{\rm an}=2c_{t,0}$ with the same fitted phase. The ratio of fitted fringe spacings between the two panels is $\simeq 1.69$, consistent with the analytical scaling $c_{t,0}^{(0.9)}/c_{t,0}^{(0.5)}=1.80$ up to a small offset. This residual mismatch reflects subleading WKB phase corrections: the numerical period systematically exceeds the leading-order $\pi k_\ast/c_{t,0}$ by an amount that decreases as $\tilde\sigma$ is reduced. The analytical formulas of this appendix therefore reproduce the simulation quantitatively at the level expected from a leading-order WKB calculation. Together with the spectra in the main text, this supports the narrow-dip scaling $k_{\rm fr}/k_\ast\sim1/(c_{t,0}\tilde\sigma)$ for Case~I.

\section{Setup for numerical solution}\label{app-numerical-setup}

We want to find the time evolution of the tensor modes for a fixed comoving momentum $k$ and helicity $\lambda$. The mode starts deep inside the horizon during inflation, exits the horizon, and then later re-enters during radiation domination. Since the background and couplings are different in the two eras, we solve the system separately during inflation and during radiation domination, and then match the solutions at the transition.

For practical purposes, it is convenient to introduce the dimensionless time variable $x \equiv k\tau$, so that inflation corresponds to $x<0$ and radiation domination corresponds to $x>0$. It is also convenient to work with the dimensionless fields
\begin{align}
\tilde{\gamma}^\lambda_{\bf k} \equiv (\sqrt{2}k^{3/2}/H_{\rm inf})\,\gamma^\lambda_{\bf k} \, ,
\qquad
\tilde{t}^\lambda_{\bf k} \equiv (\sqrt{2}k^{3/2}/H_{\rm inf})\,t^\lambda_{\bf k}.
\end{align}
In the coupled system, however, the operator expansion is in terms of two oscillator sectors
\begin{align}
\hat{\tilde\gamma}_{\bf k}^\lambda(x)
&=\sum_{I=1}^2\left(\tilde U_{\gamma I}^\lambda(x)\,\hat a_I^\lambda({\bf k})+\tilde U_{\gamma I}^{\lambda *}(x)\,\hat a_I^{\lambda\dagger}(-{\bf k})\right),
\\
\hat{\tilde t}_{\bf k}^\lambda(x)
&=\sum_{I=1}^2\left(\tilde U_{t I}^\lambda(x)\,\hat a_I^\lambda({\bf k})+\tilde U_{t I}^{\lambda *}(x)\,\hat a_I^{\lambda\dagger}(-{\bf k})\right),
\end{align}
where the annihilation and creation operators satisfy the commutation relations \eqref{eq-a-ad}. For each helicity $\lambda$, the two columns $(\tilde U_{\gamma 1}^\lambda,\tilde U_{t1}^\lambda)^T$ and $(\tilde U_{\gamma 2}^\lambda,\tilde U_{t2}^\lambda)^T$ are the two independent mode columns of the coupled system.

\subsection{Inflationary era}

During inflation we take $a=-1/(H_{\rm inf}\tau)$, so $x=k\tau<0$. Ignoring the mass term and setting $\kappa=\eta=0$, the equations for the two mode columns become
\begin{align}
    \frac{\D^2\tilde U_{\gamma I}^\lambda}{\D x^2}
    -\frac{2}{x}\frac{\D\tilde U_{\gamma I}^\lambda}{\D x}
    +\tilde U_{\gamma I}^\lambda
    &=
    \frac{\ncpl}{x}\left(
    \frac{\D\tilde U_{tI}^\lambda}{\D x}
    -\frac{3}{x}\tilde U_{tI}^\lambda
    \right),
    \label{eq:appendix-Ugamma-inf}
    \\
    \frac{\D^2\tilde U_{tI}^\lambda}{\D x^2}
    -2\left(\frac{1}{x}-\frac{1}{f}\frac{\D f}{\D x}\right)\frac{\D\tilde U_{tI}^\lambda}{\D x}
    +\left(c_t^2-\frac{2\lambda\pvc}{x}\right)\tilde U_{tI}^\lambda
    &= 
    -\frac{\ncpl}{f^2x}\frac{\D\tilde U_{\gamma I}^\lambda}{\D x},
    \label{eq:appendix-Ut-inf}
\end{align}
for $I=1,2$. These are the coupled equations for the two independent oscillator sectors, written in terms of the dimensionless variables.

As in the main text, we assume that $\ncpl$ and $\pvc$ are constant, while either $f$ or $c_t$ has a localized time dependence. It is then useful to rewrite the profile $F_i(\tau,\tau_\ast)$ in terms of $x$ and the dimensionless momentum ratio $\tilde k\equiv k/k_\ast$, where $k_\ast=-1/\tau_\ast$. One finds
\begin{align}\label{eq:f-x-corrected}
    F_i(x,\tilde k)=c_i-{\cal A}\exp\!\left[-\frac{\big(\ln[-x/\tilde k]\big)^2}{2\tilde\sigma^2}\right],
    \qquad
    c_i=\{1,c_{t,0}\},
    \qquad
    \tilde\sigma\equiv \sigma H_{\rm inf}.
\end{align}
In this form, once the equations are written in terms of $x$, the scale dependence enters only through $\tilde k$. Therefore the mode $k=k_\ast$, corresponding to $\tilde k=1$, is the one most strongly affected around horizon crossing $x=-1$.

Deep inside the horizon, $x\to-\infty$, the mixing terms are negligible and the two sectors reduce to the decoupled adiabatic vacuum. A convenient initialization is to choose the first column to be initially $\gamma$-like and the second column to be initially $t$-like. At leading adiabatic order these initial data are
\begin{align}
\begin{aligned}
    \tilde U_{\gamma 1}^\lambda(x)
    &\simeq
    -x e^{-ix},
    &
    \frac{\D\tilde U_{\gamma 1}^\lambda}{\D x}(x)
    &\simeq
    -(1-ix)e^{-ix},
    \\
    \tilde U_{t1}^\lambda(x)
    &=
    0,
    &
    \frac{\D\tilde U_{t1}^\lambda}{\D x}(x)
    &=
    0,
\end{aligned}
\end{align}
and
\begin{equation}
    \begin{aligned}
        \tilde U_{\gamma 2}^\lambda(x)
        &=
        0,
        \qquad\qquad
        \frac{\D\tilde U_{\gamma 2}^\lambda}{\D x}(x)
        =
        0,
        \\
        \tilde U_{t2}^\lambda(x)
        &=
        -\frac{x}{f(x,\tilde k)\sqrt{c_t(x,\tilde k)}}
        \exp\left[-i\int^x c_t(y,\tilde k)\,dy\right],
        \\
        \frac{\D\tilde U_{t2}^\lambda}{\D x}(x)
        &=
        \tilde U_{t2}^\lambda(x)
        \left(
        \frac{1}{x}
        -\frac{f_x(x,\tilde k)}{f(x,\tilde k)}
        -\frac{c_{t,x}(x,\tilde k)}{2\,c_t(x,\tilde k)}
        -i\,c_t(x,\tilde k)
        \right),
    \end{aligned}
\end{equation}
all evaluated at $x=x_{\rm in}$, with $x_{\rm in}\ll-1$. If $c_t$ is constant at early times and $f\to1$, this reduces to the standard decoupled adiabatic initial conditions.

Starting from these initial data, one solves Eqs.~\eqref{eq:appendix-Ugamma-inf} and \eqref{eq:appendix-Ut-inf} from $x=x_{\rm in}$ up to a late inflationary time $x=x_{\rm end}$, with $|x_{\rm end}|\ll1$, where the mode is already superhorizon and the decaying transients have become negligible. The output of the inflationary evolution is
\begin{align}
    \tilde U_{\gamma I}^{\lambda,{\rm inf}}(x_{\rm end}),\qquad
    \frac{\D\tilde U_{\gamma I}^{\lambda,{\rm inf}}}{\D x}\Big|_{x=x_{\rm end}},\qquad
    \tilde U_{tI}^{\lambda,{\rm inf}}(x_{\rm end}),\qquad
    \frac{\D\tilde U_{tI}^{\lambda,{\rm inf}}}{\D x}\Big|_{x=x_{\rm end}},
    \qquad I=1,2.
\end{align}

At any time during inflation, the helicity-resolved spectra are obtained by summing over the two oscillator sectors,
\begin{align}
    {\cal P}_h^\lambda(k,\tau)
    &=
    \left(\frac{H_{\rm inf}}{\pi M_{\rm Pl}}\right)^2
    \sum_{I=1}^2 \big|\tilde U_{\gamma I}^\lambda(x)\big|^2,
    \\
    {\cal P}_t^\lambda(k,\tau)
    &=
    \left(\frac{H_{\rm inf}}{\pi M_{\rm Pl}}\right)^2
    \sum_{I=1}^2 \big|\tilde U_{tI}^\lambda(x)\big|^2.
\end{align}
In particular, evaluating these expressions at the late inflationary time $x_{\rm end}$ gives the superhorizon spectra at the end of inflation,
\begin{align}
    {\cal P}_{h,{\rm inf}}^\lambda(k)
    &=
    \left(\frac{H_{\rm inf}}{\pi M_{\rm Pl}}\right)^2
    \sum_{I=1}^2 \big|\tilde U_{\gamma I}^{\lambda,{\rm inf}}(x_{\rm end})\big|^2,
    \label{eq:appendix-Ph-inf-super}
    \\
    {\cal P}_{t,{\rm inf}}^\lambda(k)
    &=
    \left(\frac{H_{\rm inf}}{\pi M_{\rm Pl}}\right)^2
    \sum_{I=1}^2 \big|\tilde U_{tI}^{\lambda,{\rm inf}}(x_{\rm end})\big|^2.
    \label{eq:appendix-Pt-inf-super}
\end{align}
Therefore, once the numerical integration reaches a sufficiently late time $x_{\rm end}$ with $|x_{\rm end}|\ll1$, the inflationary superhorizon spectra are read off directly from the values of the two evolved columns. These are the spectra plotted in Figs.~\ref{fig:gamma-PS-f-inf} and~\ref{fig:gamma-PS-ct-inf}, up to the overall normalization factor $(H_{\rm inf}/\pi M_{\rm Pl})^2$.

\subsection{Radiation-dominated era}

After inflation we assume that the mixing switches off, so $\ncpl=0$ during radiation domination. In that case the $\gamma$- and $t$-sectors decouple. Since the GW signal is determined by the metric tensor perturbation, we only need to evolve the $\gamma$-columns during radiation domination.

In the radiation era, with $a\propto\tau$, each column $\tilde U_{\gamma I}^{\lambda,{\rm RD}}$ obeys
\begin{align}\label{eq:appendix-Ugamma-RD}
    \frac{\D^2\tilde U_{\gamma I}^{\lambda,{\rm RD}}}{\D x^2}
    +\frac{2}{x}\frac{\D\tilde U_{\gamma I}^{\lambda,{\rm RD}}}{\D x}
    +\tilde U_{\gamma I}^{\lambda,{\rm RD}}=0,
    \qquad
    I=1,2.
\end{align}

The important point is that the initial conditions for the radiation-era evolution are not chosen independently. They are inherited from the end of the inflationary evolution. More precisely, after solving the coupled system during inflation up to a late time $x=x_{\rm end}$ with $|x_{\rm end}|\ll1$, one takes the values of the inflationary solutions for the $\gamma$-columns and uses them as the initial data for the radiation era. Thus, for each $I=1,2$, we impose
\begin{align}\label{eq:appendix-match-RD}
    \tilde U_{\gamma I}^{\lambda,{\rm RD}}(x_{\rm end})
    &=
    \tilde U_{\gamma I}^{\lambda,{\rm inf}}(x_{\rm end}),
    \qquad
    \frac{\D}{\D x}\tilde U_{\gamma I}^{\lambda,{\rm RD}}(x_{\rm end})
    =
    \frac{\D}{\D x}\tilde U_{\gamma I}^{\lambda,{\rm inf}}\Big|_{x_{\rm end}}.
\end{align}
In other words, the radiation-dominated evolution starts from the superhorizon values produced during inflation. This is just the continuity of the metric tensor perturbation and its first derivative across the transition.

Therefore, the numerical procedure is the following. First, solve the coupled inflationary system for the two columns from $x=x_{\rm in}$ to $x=x_{\rm end}$. This gives $\tilde U_{\gamma I}^{\lambda,{\rm inf}}(x_{\rm end})$ and $\D\tilde U_{\gamma I}^{\lambda,{\rm inf}}/\D x|_{x=x_{\rm end}}$. Second, use Eq.~\eqref{eq:appendix-match-RD} as the initial condition for Eq.~\eqref{eq:appendix-Ugamma-RD}, and evolve the two $\gamma$-columns further during radiation domination.

The corresponding power spectrum at any time during radiation domination is then obtained by summing over the two oscillator sectors,
\begin{align}
    {\cal P}_h^\lambda(k,\tau)
    &=
    \left(\frac{H_{\rm inf}}{\pi M_{\rm Pl}}\right)^2
    \sum_{I=1}^2
    \big|\tilde U_{\gamma I}^{\lambda,{\rm RD}}(x)\big|^2.
    \label{eq:appendix-Ph-RD}
\end{align}
These are the spectra plotted in Fig.~\ref{fig:gamma-PS-f} when evaluated at horizon re-entry, $x=1$, up to the same overall normalization factor.

The spectral GW energy density is defined in Eq.~\eqref{Omega-def}. Using $U_{\gamma I}^{\lambda}=(H_{\rm inf}/\sqrt{2}\,k^{3/2})\,\tilde U_{\gamma I}^{\lambda}$ and $U_{\gamma I}^{\lambda\,\prime}=(H_{\rm inf}/\sqrt{2}\,k^{1/2})\,\D\tilde U_{\gamma I}^{\lambda}/\D x$, and using the fact that during radiation domination one has $a\propto\tau$, it follows that $k/(aH)=k\tau=x$, and we find
\begin{align}\label{eq:appendix-Omega-RD-tilde}
    \Omega^\lambda_{\rm GW}(x,\tilde{k})
    &=
    \left(\frac{H_{\rm inf}}{\pi M_{\rm Pl}}\right)^2
    \frac{x^2}{24}
    \sum_{I=1}^2
    \left[
    \left|\frac{\D\tilde U_{\gamma I}^{\lambda,{\rm RD}}}{\D x}\right|^2
    +
    \left|\tilde U_{\gamma I}^{\lambda,{\rm RD}}\right|^2
    \right].
\end{align}
This is the expression to use for the GW energy density during radiation domination after the two $\gamma$-columns have been evolved from the inflationary initial data.

\subsection{Matching inflation to radiation domination}

As explained above, we solve the system separately during inflation and radiation domination, and then match the solutions to describe the full evolution of modes that start deep inside the horizon during inflation and later re-enter during radiation domination.

At the background level, in order to avoid an artificial discontinuity in the expansion history, one may match the inflationary scale factor $a=-1/(H_{\rm inf}\tau)$, valid for $-\infty<\tau<\tau_{\rm end}<0$, to a radiation-dominated scale factor at $\tau=\tau_{\rm end}$. A convenient choice is $a=(\tau-2\tau_{\rm end})/(H_{\rm inf}\tau_0^2)$ for $\tau_{\rm end}<\tau<\infty$, which gives $a(\tau_{\rm end})=-1/(H_{\rm inf}\tau_{\rm end})$ and also makes $a'(\tau)$ continuous at the transition.\footnote{For simplicity this assumes an instantaneous transition from inflation to radiation domination. In a more realistic treatment, reheating would interpolate smoothly between the two eras.} In terms of the dimensionless variable $x=k\tau$, this introduces a shift $x\to x-2x_{\rm end}$ after inflation, where $x_{\rm end}\equiv k\tau_{\rm end}$. Since in practice the matching is performed for $|x_{\rm end}|\ll1$, this shift is negligible and can be ignored.

At the perturbation level, the inflationary evolution gives, for each helicity $\lambda$ and each oscillator sector $I=1,2$, the values of $\tilde U_{\gamma I}^{\lambda,{\rm inf}}(x_{\rm end})$ and $\D \tilde U_{\gamma I}^{\lambda,{\rm inf}}/\D x|_{x=x_{\rm end}}$ at a late inflationary time $x_{\rm end}<0$ with $|x_{\rm end}|\ll1$. These quantities are then used as the initial conditions for the radiation-dominated evolution. Since the background is matched continuously and there is no singular source at the transition, the matching conditions are simply the continuity relations given in Eq.~\eqref{eq:appendix-match-RD}.

Thus, the superhorizon values found at the end of inflation directly determine the initial conditions for Eq.~\eqref{eq:appendix-Ugamma-RD} during radiation domination. One then evolves Eq.~\eqref{eq:appendix-Ugamma-RD} from $|x_{\rm end}|\ll1$ up to horizon re-entry, $x_{\rm hc}=1$, or beyond, depending on the application. The resulting $\tilde U_{\gamma I}^{\lambda,{\rm RD}}(x)$ are then used to compute the tensor power spectrum through the sum over the two oscillator sectors discussed above.

The full numerical procedure can then be summarized as follows. First, choose the helicity $\lambda$, the momentum ratio $\tilde k=k/k_\ast$, and the inflationary profiles $f(x,\tilde k)$ or $c_t(x,\tilde k)$. Second, impose decoupled adiabatic initial conditions at some early time $x_{\rm in}\ll-1$ for the two independent columns $I=1,2$. Third, evolve the coupled inflationary system $\{\tilde U_{\gamma I}^\lambda,\tilde U_{tI}^\lambda\}$ from $x_{\rm in}$ to a late inflationary time $x_{\rm end}$ with $|x_{\rm end}|\ll1$. Fourth, use $\tilde U_{\gamma I}^{\lambda,{\rm inf}}(x_{\rm end})$ and $\D\tilde U_{\gamma I}^{\lambda,{\rm inf}}/\D x|_{x=x_{\rm end}}$ as the initial conditions for the radiation-dominated evolution of $\tilde U_{\gamma I}^{\lambda,{\rm RD}}$. Finally, compute the power spectrum by summing over the two oscillator sectors, ${\cal P}_h^\lambda \propto \sum_{I=1}^2 |\tilde U_{\gamma I}^{\lambda,{\rm RD}}|^2$, and compute the GW energy density from the same two evolved columns through $\Omega_{\rm GW}^\lambda \propto \sum_{I=1}^2 \left(|\D\tilde U_{\gamma I}^{\lambda,{\rm RD}}/\D x|^2+|\tilde U_{\gamma I}^{\lambda,{\rm RD}}|^2\right)$.

\bibliographystyle{JHEPmod}
\bibliography{ref-new}

\end{document}